\documentclass[fleqn,usenatbib]{mnras}
\usepackage[T1]{fontenc}
\usepackage{newtxtext,newtxmath}
\usepackage[utf8]{inputenc}
\usepackage{natbib}
\usepackage{graphicx}
\usepackage{hyperref}
\usepackage{threeparttable}
\usepackage{placeins}
\usepackage[normalem]{ulem}
\usepackage{pdflscape}
\usepackage{amsmath}
\usepackage{orcidlink}
\usepackage{multirow}
\usepackage{xcolor}

\title[WASP-96b with NIRISS/SOSS]{Awesome SOSS: Transmission Spectroscopy of WASP-96b with NIRISS/SOSS}

\author[Radica et al.]{Michael Radica\orcidlink{0000-0002-3328-1203}$^{1}$\thanks{E-mail: michael.radica@umontreal.ca},
Luis Welbanks\orcidlink{0000-0003-0156-4564}$^{2}$\thanks{NHFP Sagan Fellow},
Néstor Espinoza\orcidlink{0000-0001-9513-1449}$^{3,4}$,
Jake Taylor\orcidlink{0000-0003-4844-9838}$^{1,5}$,
Louis-Philippe Coulombe\orcidlink{0000-0002-2195-735X}$^{1}$,
\newauthor{Adina D.\ Feinstein\orcidlink{0000-0002-9464-8101}$^{6}$\thanks{NSF Graduate Research Fellow},
Jayesh Goyal\orcidlink{0000-0002-8515-7204}$^{7}$,
Nicholas Scarsdale\orcidlink{0000-0003-3623-7280}$^{8}$, 
Loïc Albert\orcidlink{0000-0003-0475-9375}$^{1}$,
Priyanka Baghel\orcidlink{0000-0002-4000-6840}$^{7}$,}
\newauthor{Jacob L.\ Bean\orcidlink{0000-0003-4733-6532}$^{6}$,
Jasmina Blecic\orcidlink{0000-0002-0769-9614}$^{9,10}$,
David Lafrenière\orcidlink{0000-0002-6780-4252}$^{1}$,
Ryan J.\ MacDonald\orcidlink{0000-0003-4816-3469}$^{11}$\thanks{NHFP Sagan Fellow},
Maria Zamyatina\orcidlink{0000-0002-9705-0535}$^{12}$,}
\newauthor{Romain Allart\orcidlink{0000-0002-1199-9759}${1}$\thanks{Trottier Postdoctoral Fellow},
Étienne Artigau\orcidlink{0000-0003-3506-5667}$^{1,13}$,
Natasha E.\ Batalha\orcidlink{0000-0003-1240-6844}$^{14}$,
Neil James Cook\orcidlink{0000-0003-4166-4121}$^{1}$,
Nicolas B.\ Cowan\orcidlink{0000-0001-6129-5699}$^{15,16}$,}
\newauthor{Lisa Dang\orcidlink{0000-0003-4987-6591}$^{1}$\thanks{Banting Postdoctoral Fellow},
René Doyon\orcidlink{0000-0001-5485-4675}$^{1,13}$,
Marylou Fournier-Tondreau\orcidlink{0000-0002-5428-0453}$^{1}$,
Doug Johnstone\orcidlink{0000-0002-6773-459X}$^{17,18}$,
Michael R.\ Line\orcidlink{0000-0001-6247-8323}$^{2}$,}
\newauthor{Sarah E. Moran\orcidlink{0000-0002-6721-3284}$^{19}$,
Sagnick Mukherjee\orcidlink{0000-0003-1622-1302}$^{8}$,
Stefan Pelletier\orcidlink{0000-0002-8573-805X}$^{1}$,
Pierre-Alexis Roy\orcidlink{0000-0001-6809-3520}$^{1}$,
Geert Jan Talens\orcidlink{0000-0003-4787-2335}$^{20}$,}
\newauthor{Joseph Filippazzo\orcidlink{0000-0002-0201-8306}$^{3}$,
Klaus Pontoppidan\orcidlink{0000-0001-7552-1562}$^{3}$,
and Kevin Volk$^{3}$}
\\
$^{1}$Institut Trottier de recherche sur les exoplanètes and Département de Physique, Université de Montréal, 1375 Avenue Thérèse-Lavoie-Roux,\\ Montréal, QC, H2V 0B3, Canada\\
$^{2}$School of Earth and Space Exploration, Arizona State University, 781 Terrace Mall, Tempe, AZ, 85287, USA\\
$^{3}$Space Telescope Science Institute, 3700 San Martin Drive, Baltimore, MD 21218, USA\\
$^{4}$Department of Physics and Astronomy, Johns Hopkins University, 3400 N Charles St, Baltimore, MD 21218, USA\\
$^{5}$Department of Physics (Atmospheric, Oceanic and Planetary Physics), University of Oxford, Parks Rd, Oxford OX1 3PU, UK\\
$^{6}$Department of Astronomy \& Astrophysics, University of Chicago, 5640 S Ellis Ave, Chicago, IL 60637, USA\\
$^{7}$School of Earth and Planetary Sciences (SEPS), National Institute of Science Education and Research (NISER), HBNI, Jatani, Odisha 752050, India\\
$^{8}$Department of Astronomy and Astrophysics, University of California, Santa Cruz, CA 95060, USA\\
$^{9}$Department of Physics, New York University Abu Dhabi, PO Box 129188 Abu Dhabi, UAE\\
$^{10}$Center for Astro, Particle, and Planetary Physics (CAP3), New York University Abu Dhabi, PO Box 129188 Abu Dhabi, UAE\\
$^{11}$Department of Astronomy, University of Michigan, 1085 S. University Ave., Ann Arbor, MI 48109, USA\\
$^{12}$Department of Physics and Astronomy, Faculty of Environment, Science and Economy, University of Exeter, Exeter, EX4 4QL, UK\\
$^{13}$Observatoire du Mont-Mégantic, Université de Montréal, Montréal, QC, H3C 3J7, Canada\\
$^{14}$NASA Ames Research Center, Moffett Field, CA 94035, USA\\
$^{15}$Department of Physics, McGill University, 3600 rue University, Montréal, QC, H3A 2T8, Canada\\
$^{16}$Department of Earth and Planetary Sciences, McGill University, 3600 rue University, Montréal, QC, H3A 2T8, Canada\\
$^{17}$NRC Herzberg Astronomy and Astrophysics, 5071 West Saanich Rd, Victoria, BC, V9E 2E7, Canada\\
$^{18}$Department of Physics and Astronomy, University of Victoria, Victoria, BC, V8P 5C2, Canada\\
$^{19}$Department of Planetary Sciences and Lunar and Planetary Laboratory, University of Arizona, Tuscon, AZ 85721, USA\\
$^{19}$Department of Astrophysical Sciences, Princeton University, 4 Ivy Lane, Princeton, NJ 08544, USA\\
}
\date{Accepted XXX. Received YYY; in original form ZZZ}

\pubyear{2022}

\begin{document}
\label{firstpage}
\pagerange{\pageref{firstpage}--\pageref{lastpage}}
\maketitle

\begin{abstract}
The future is now --- after its long-awaited launch in December 2021, JWST began science operations in July 2022 and is already revolutionizing exoplanet astronomy. The Early Release Observations (ERO) program was designed to provide the first images and spectra from JWST, covering a multitude of science cases and using multiple modes of each on-board instrument. Here, we present transmission spectroscopy observations of the hot-Saturn WASP-96\,b with the Single Object Slitless Spectroscopy (SOSS) mode of the Near Infrared Imager and Slitless Spectrograph, observed as part of the ERO program. As the SOSS mode presents some unique data reduction challenges, we provide an in-depth walk-through of the major steps necessary for the reduction of SOSS data: including background subtraction, correction of 1/$f$ noise, and treatment of the trace order overlap. We furthermore offer potential routes to correct for field star contamination, which can occur due to the SOSS mode's slitless nature. By comparing our extracted transmission spectrum with grids of atmosphere models, we find an atmosphere metallicity between 1$\times$ and 5$\times$ solar, and a solar carbon-to-oxygen ratio. Moreover, our models indicate that no grey cloud deck is required to fit WASP-96\,b's transmission spectrum, but find evidence for a slope shortward of 0.9\,µm, which could either be caused by enhanced Rayleigh scattering or the red wing of a pressure-broadened Na feature. Our work demonstrates the unique capabilities of the SOSS mode for exoplanet transmission spectroscopy and presents a step-by-step reduction guide for this new and exciting instrument.
\end{abstract}




\section{Introduction}
\label{sec: Introduction}

Transiting exoplanets provide astronomers with an ideal opportunity to study the atmospheres of distant worlds \citep{seager_theoretical_2000}. Spectroscopic observations during the transit or eclipse of an exoplanet have revealed the telltale signs of an abundance of molecular and atomic species in the atmospheres of giant exoplanets, both at high \citep[e.g.,][]{brogi_carbon_2014,hoeijmakers_hot_2020, boucher_characterizing_2021, boucher_co_2023} and low \citep[e.g.,][]{charbonneau_detection_2002, kreidberg_detection_2015, evans_detection_2016} spectral resolution, which have provided insights into the formation histories \citep{oberg_effects_2011, madhusudhan_toward_2014, turrini_tracing_2021} as well as the physical and chemical processes governing their atmospheres \citep{moses_disequilibrium_2011, madhusudhan_co_2012, parmentier_3d_2013, wakeford_high-temperature_2017}. Particularly, space-based observations using the Hubble (HST) and Spitzer Space Telescopes have shed light on the population of giant exoplanet atmospheres, revealing near-ubiquitous detections of water as well as the presence of clouds and hazes \citep[e.g.,][]{sing_continuum_2016, Barstow2017, welbanks_massmetallicity_2019, pinhas_h2o_2019}. The spectral signatures of alkali metals such as Na and K are also common, particularly in cloud-free atmospheres \citep[e.g.,][]{welbanks_massmetallicity_2019, Alam2021, nikolov_solar--supersolar_2022}, as 
well as hints of carbon-bearing species such as CO and CO$_2$ \citep[e.g.,][]{kreidberg_global_2018, dragomir_spitzer_2020, spake_abundance_2020}.

However, since neither observatory has observing modes specifically designed with exoplanet observations in mind, atmospheric studies with \textit{HST} and \textit{Spitzer} were far from ideal. Both presented astronomers with a number of technical challenges \citep[e.g.,][]{deming_infrared_2013, zhou_physical_2017}, and moreover, atmospheric inferences from \textit{HST} and \textit{Spitzer} observations, even when used in conjunction, were often limited. The narrow bandwidth of the Wide Field Camera 3 (WFC3) (0.85--1.7\,µm) and Space Telescope Imaging Spectrograph (STIS) (0.525--1.0\,µm) instruments on board \textit{HST} result in atmosphere spectra sensitive primarily to H$_2$O, alkalis, and clouds \citep{benneke_atmospheric_2012, benneke_how_2013}. \textit{Spitzer} provided up to four photometric points at longer wavelengths (3.6, 4.5, 5.8, and 8\,µm, although only the two bluest wavebands remained available after 2009), which could provide hints of the presence of carbon-bearing molecules. However, \textit{Spitzer} photometry precluded any definitive detections, rendering constraints on bulk atmospheric metallicities or carbon-to-oxygen ratios (C/O) challenging \citep[e.g.,][]{kreidberg_detection_2015, spake_abundance_2020}. 

JWST is the natural successor to \textit{HST} and \textit{Spitzer}. Its four instruments each have observing modes specifically tailored to time series observations (TSOs) of transiting exoplanets, which greatly improve the wavelength range (0.6--14\,µm) and spectral resolution (R$\sim$50--3000) with which exoplanet atmospheres can be spectroscopically probed. Indeed, JWST observations of the giant planet WASP-39\,b \citep{faedi_wasp-39b_2011} using the Near Infrared Spectrograph (NIRSpec) PRISM mode \citep{birkmann_near-infrared_2022} have already yielded the first definitive detections of CO$_2$ \citep{jwst_transiting_exoplanet_community_early_release_science_team_identification_2022}, and SO$_2$ \citep{alderson_early_2023, tsai_direct_2022} in an exoplanet atmosphere. 

The Single Object Slitless Spectroscopy (SOSS) mode \citep{Albert2023SOSS} of the Near Infrared Imager and Slitless Spectrograph (NIRISS) instrument \citep{Doyon2023NIRISS} on board JWST is already proving to be one of the workhorse modes for exoplanet atmosphere observations \citep{feinstein_early_2023, fu_water_2022}. NIRISS/SOSS provides medium resolution (R$\sim$700) spectroscopy from 0.6--2.85\,µm, yielding unprecedented coverage of the near-infrared (NIR) H$_2$O bands, as well as coverage of signatures of Na and K at the bluest wavelengths, and potentially CH$_4$, CO, and CO$_2$ at the reddest wavelengths. The full SOSS wavelength range is divided between two spectral orders, with the wavelengths 0.85--2.85\,µm and 0.6--1.0\,µm covered by orders 1 and 2, respectively.

WASP-96\,b \citep{hellier_transiting_2014} is a highly inflated hot-Saturn with a mass of $\rm 0.48 \pm 0.03\,M_J$ and a radius of $\rm 1.20 \pm 0.06\,R_J$. It orbits its G8 host star in 3.42\,d, which results in an equilibrium temperature of 1285$\pm$40\,K, making WASP-96\,b an excellent candidate for transmission spectroscopy. Indeed WASP-96\,b has already been observed in transmission by \citet{nikolov_absolute_2018} with the FOcal Reducer/low dispersion Spectrograph 2 (FORS2; 0.35--0.8\,µm) on the Very Large Telescope (VLT), revealing the clear pressure-broadened profile of Na and suggesting that WASP-96\,b hosts a mostly cloud-free atmosphere. These findings were later confirmed by the retrieval analyses of \citet{welbanks_massmetallicity_2019} and \citet{Alam2021}, the former of which reported a stellar-to-super-stellar Na abundance.  

\citet{yip_compatibility_2021} and \citet{nikolov_solar--supersolar_2022} also presented independent analyses of transmission observations of WASP-96\,b with \textit{HST}/WFC3 using both the G102 (0.8--1.15\,µm) and G141 (1.08--1.7\,µm) grisms, as well as with \textit{Spitzer}/IRAC at 3.6 and 4.5\,µm. Joint retrievals with the VLT observations point to solar-to-super-solar abundances of Na and oxygen. The oxygen abundance in particular was found to be consistent with the oxygen enrichment level of Jupiter in our own solar system --- suggesting a bulk atmospheric metallicity for WASP-96\,b consistent with the solar system mass-metallicity trend \citep[e.g.,][]{thorngren_massmetallicity_2016, welbanks_massmetallicity_2019}. \citet{nikolov_solar--supersolar_2022} also find a sub-solar C/O ratio, under the assumption of chemical equilibrium. Lastly, \citet{mcgruder_access_2022} presented Magellan/IMACS (0.475--0.825\,µm) transmission spectra of WASP-96\,b, independently confirming the presence of the broad Na feature. Their joint retrievals of all available transmission data also point to solar-to-super-solar atmospheric abundances of Na and H$_2$O. 

Recently, \citet{samra_clouds_2023} explored the combined transmission spectrum of WASP-96\,b, as collated by \citet{nikolov_solar--supersolar_2022}, in the context of cloud formation. Although all previous analyses have concluded a clear atmosphere for WASP-96\,b based on the clearly observable pressure-broadened wings of the Na feature, \citet{samra_clouds_2023} point out that the equilibrium temperature of the planet places it in a regime where asymmetric cloud cover of the terminator could be expected \citep{helling_exoplanet_2023}. Moreover, the general circulation models (GCMs) of \citet{samra_clouds_2023} point to homogeneous clouds, dominated by silicates, covering the terminator region of WASP-96\,b. They perform two retrieval analyses on the \citet{nikolov_solar--supersolar_2022} transmission spectrum, with and without the inclusion of clouds, and find that cloudy solutions can accurately reproduce the observed transmission spectrum. \citet{samra_clouds_2023}, furthermore, suggest avenues which could reconcile their GCM models with the previous cloud-free retrieval results. In particular, they indicate that reduced vertical mixing efficiency could cause an optically thick cloud layer to settle below the observable photosphere, or that increasing the porosity of cloud particles could lead to optically thin clouds, even if they remain within the observable atmosphere. 

Here, we present transmission spectroscopy observations of WASP-96\,b with NIRISS/SOSS taken as part of the JWST Early Release Observations (ERO) program \citep{pontoppidan_jwst_2022}. As the SOSS mode presents several unique challenges, this paper undertakes a step-by-step overview of the data reduction procedures, such that the community can understand the critical steps necessary to extract atmosphere spectra from SOSS TSOs. The companion paper, \citet{Taylor_w96_2023}, presents an in-depth exploration of the modelling and retrieval of this transmission spectrum, as well as the particular insights into atmospheric physics and chemistry which can be gained through NIRISS/SOSS observations. This work is organized as follows: Section~\ref{sec: Observations & Data Reductions} presents the observations, as well as the data reduction procedure. Section~\ref{sec: Light Curve Analysis} outlines the light curve fitting, and atmospheric grid modelling methods are explained in Section~\ref{sec: Grid Retrievals}. We present our initial atmospheric inferences in Section~\ref{sec:Grid_Results}, and summarize and discuss our results in Section~\ref{sec: Conclusions}.

\section{Observations and Data Reductions}
\label{sec: Observations & Data Reductions}

\subsection{Outline of the Observations}
\label{sec: Outline of Observations}

WASP-96\,b was observed in transit using the SOSS mode of the NIRISS instrument as part of the JWST ERO program \citep{pontoppidan_jwst_2022}. The TSO started on UTC June 21, 2022 and spanned 6.4\,hr, which covered the 2.4\,hr transit, as well as 2.5\,hr of baseline before the transit, and 1.5\,hr after. It used the standard GR700XD/CLEAR combination, along with the SUBSTRIP256 subarray which captures three diffraction orders of the target on the detector \citep{Albert2023SOSS}. In total, the TSO is composed of 280 individual integrations, each consisting of 14 groups --- yielding an integration time of 76.9\,s per integration, and a median signal-to-noise ratio of 125 per integration for order 1 at 1.5\,µm. An optional second exposure, using the GR700XD grism in combination with the F277W filter was taken after the GR700XD/CLEAR exposure. The F277W filter limits the wavelength range of SOSS to $\lambda \gtrsim2.6$\,µm, and the exposure lasted only 0.25\,hr, using 11 integrations with the same exposure time per integration as the CLEAR. 

\subsection{Data Reduction}
\label{sec: Data Reductions}

The SOSS mode presents a number of particular challenges: the curved nature of the spectral trace, the unique background shape, the potential contamination by field sources, and the overlap between the first and second diffraction orders on the detector to name a few. Here, we outline the major challenges encountered during reduction of SOSS data, and present the \texttt{supreme-SPOON} pipeline (supreme-Steps to Process sOss ObservatioNs) for the reduction of SOSS data. \texttt{supreme-SPOON} is publicly available\footnote{https://github.com/radicamc/supreme-spoon}, and has already been successfully applied to the SOSS TSOs of WASP-39\,b \citep{feinstein_early_2023} and WASP-18\,b \citep{coulombe_broadband_2023} taken as part of the JWST Transiting Exoplanet Community Early Release Science Program \citep{stevenson_transiting_2016, bean_transiting_2018}. \texttt{supreme-SPOON} was briefly outlined in \citet{feinstein_early_2023}, but here we provide a more in-depth discussion of the key steps, as well as introduce new methods to explicitly deal with field star contamination, which was not included in the reduction of the aforementioned TSOs. For additional verification of our results, three other pipelines (\texttt{NAMELESS}, \texttt{transitspectroscopy}, and \texttt{nirHiss}) were used to perform independent reductions. The particulars of these pipelines have also already been outlined in \citet{feinstein_early_2023}, and a brief summary of each, as it pertains to the analysis of this TSO, is provided in Appendix~\ref{sec: Additional Reductions}.

\begin{figure*}
	\centering
	\includegraphics[width=\textwidth]{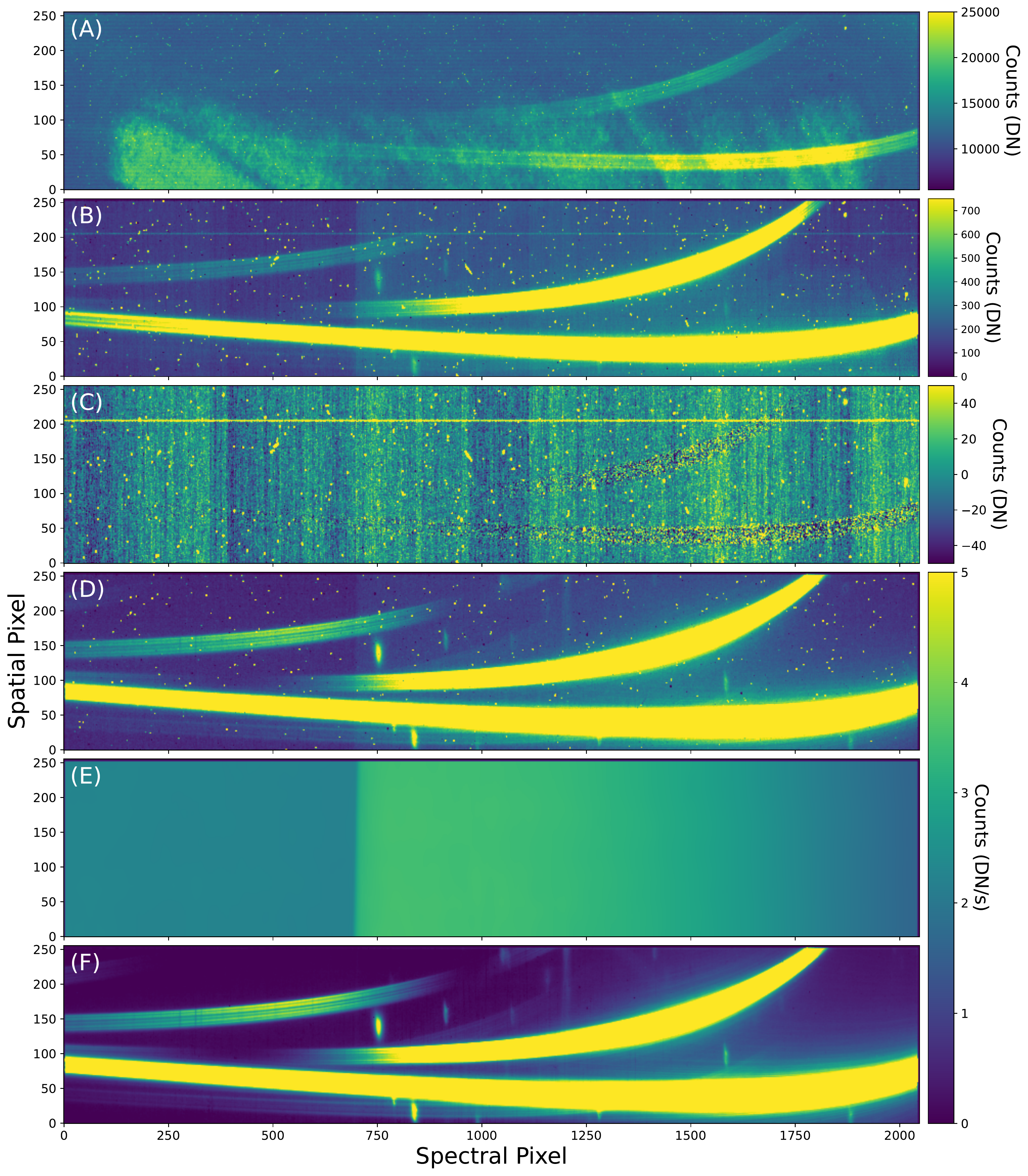}
    \caption{Visualization of the \texttt{supreme-SPOON} data products at several stages of the reduction process.
    \textbf{(A)}: A raw, uncalibrated data frame in data numbers (DN).
    \textbf{(B)}: After superbias subtraction and reference pixel correction. 
    \textbf{(C)}: Frame (B) after the first background subtraction, and subtraction of the scaled median of all integrations to reveal the 1/$f$ noise.
    \textbf{(D)}: Data frame after ramp fitting and flat field correction. 
    \textbf{(E)}: Background model scaled to the flux level of (D).
    \textbf{(F)}: Final calibrated data product.
    The horizontal stripe near row 200 in panels (B) and (C) is a known artifact resulting from FULL frame resets which occur before each subarray exposure.
    \label{fig:Reduction Steps}}
\end{figure*}

\texttt{supreme-SPOON} is composed of four stages, many steps of which are shared with the official \texttt{jwst} data reduction pipeline\footnote{https://jwst-pipeline.readthedocs.io/en/latest/jwst/pipeline/index.html. \texttt{jwst} v1.6.2 is used in this work.}. A summary of the major reduction steps are outlined below, and visualized in Figure~\ref{fig:Reduction Steps}. 

\subsubsection{\texttt{supreme-SPOON} Stage 1 --- Detector Level Calibrations}
\label{sec: Stage 1}

Like the official \texttt{jwst} pipeline, Stage 1 performs the ``detector-level'' calibrations on the full four-dimensional (integrations, groups, spatial pixels, spectral pixels) TSO data cube. The Stage 1 calibrations of \texttt{supreme-SPOON} are separated into 10 main steps, a short description of which is provided below:

\begin{enumerate}
    \item \texttt{GroupScaleStep}: Rescale pixel values to account for on-board averaging of frames. 
    \item \texttt{DQInitStep}: Initialization of data quality flags.
    \item \texttt{SaturationStep}: Flag pixels above the saturation limit. 
    \item \texttt{SuperBiasStep}: Subtraction of the detector bias level.
    \item \texttt{RefPixStep}: Perform initial 1/$f$ noise and odd-even row corrections using reference pixels.
    \item \texttt{BackgroundStep I}: Subtract the background level.
    \item \texttt{OneOverFStep}: Perform further corrections for 1/$f$ noise. Re-add background.
    \item \texttt{LinearityStep}: Correct non-linearity effects.
    \item \texttt{JumpStep}: Flag cosmic ray hits.
    \item \texttt{RampFitStep}: Calculate the mean count rates per pixel by fitting each pixel or each integration ``up-the-ramp'' (that is, along the ``groups" axis).
\end{enumerate}

The initial stages (i--v) are already well handled by the official \texttt{jwst} pipeline, and \texttt{supreme-SPOON} simply provides wrappers for these steps. For more in-depth information about each of these specific steps, please see the \texttt{jwst} documentation\footnote{https://jwst-pipeline.readthedocs.io/en/latest/jwst/pipeline/

calwebb\_detector1.html}. Panel A of Figure~\ref{fig:Reduction Steps} shows a raw, uncalibrated data frame, and panel B shows the same data frame after the completion of these first five steps. 

After the \texttt{RefPixStep}, \texttt{supreme-SPOON} begins to diverge from \texttt{jwst} by more comprehensively treating 1/$f$ noise. 1/$f$ noise is a unique noise source caused by the introduction of a small, arbitrary bias level when pixels are read off of the detector, and is present to varying degrees in all of the NIR JWST instruments \citep[e.g.,][]{ahrer_early_2023, alderson_early_2023, feinstein_early_2023, rustamkulov_early_2023}. The bias level introduced varies over time. As the SOSS detector is read column-by-column, this manifests as column-correlated ``streaks'' (e.g., panel C of Figure~\ref{fig:Reduction Steps}). The 1/$f$ noise level technically also varies along a single column, as pixels at the bottom of the detector are read slightly after those at the top. 

The ostensible purpose of the \texttt{RefPixStep} in the official \texttt{jwst} pipeline is to correct this 1/$f$ noise using non-illuminated pixels on the top edge of the detector. Since these pixels are not light sensitive, whatever counts are measured must be introduced during readout. However, we have found this correction to be inadequate for the complete removal of the 1/$f$ noise. This is likely due to the fact that there are only four rows of reference pixels, and since the 1/$f$ noise also varies in a given column, the 1/$f$ level determined in the top four reference pixels will not necessarily hold for the entire column. 

We, therefore, develop an alternate strategy to handle the 1/$f$ noise and implement it in \texttt{supreme-SPOON}. Since the 1/$f$ noise is introduced during readout, it is one of the, if not the last, noise source injected into the data and should therefore be one of the first to be corrected. The background subtraction and correction of 1/$f$ noise though are necessarily highly coupled, and for reasons that will be described more fully below, the background \textit{must} be subtracted before considering the correction of 1/$f$ noise.
 
The SOSS background has a unique structure due to the combination of the JWST ``pick-off'' mirror and the GR700XD grism \citep{Albert2023SOSS}. The zodiacal background falling off the pick-off mirror creates a step in the background at around column $\sim$750 (e.g., panel E of Figure~\ref{fig:Reduction Steps}). Furthermore, the dispersal of this zodiacal light by the GR700XD grism causes low-frequency variations, which can be seen as a fading of the background level towards the right edge of the detector. 

\begin{figure}
	\centering
	\includegraphics[width=\columnwidth]{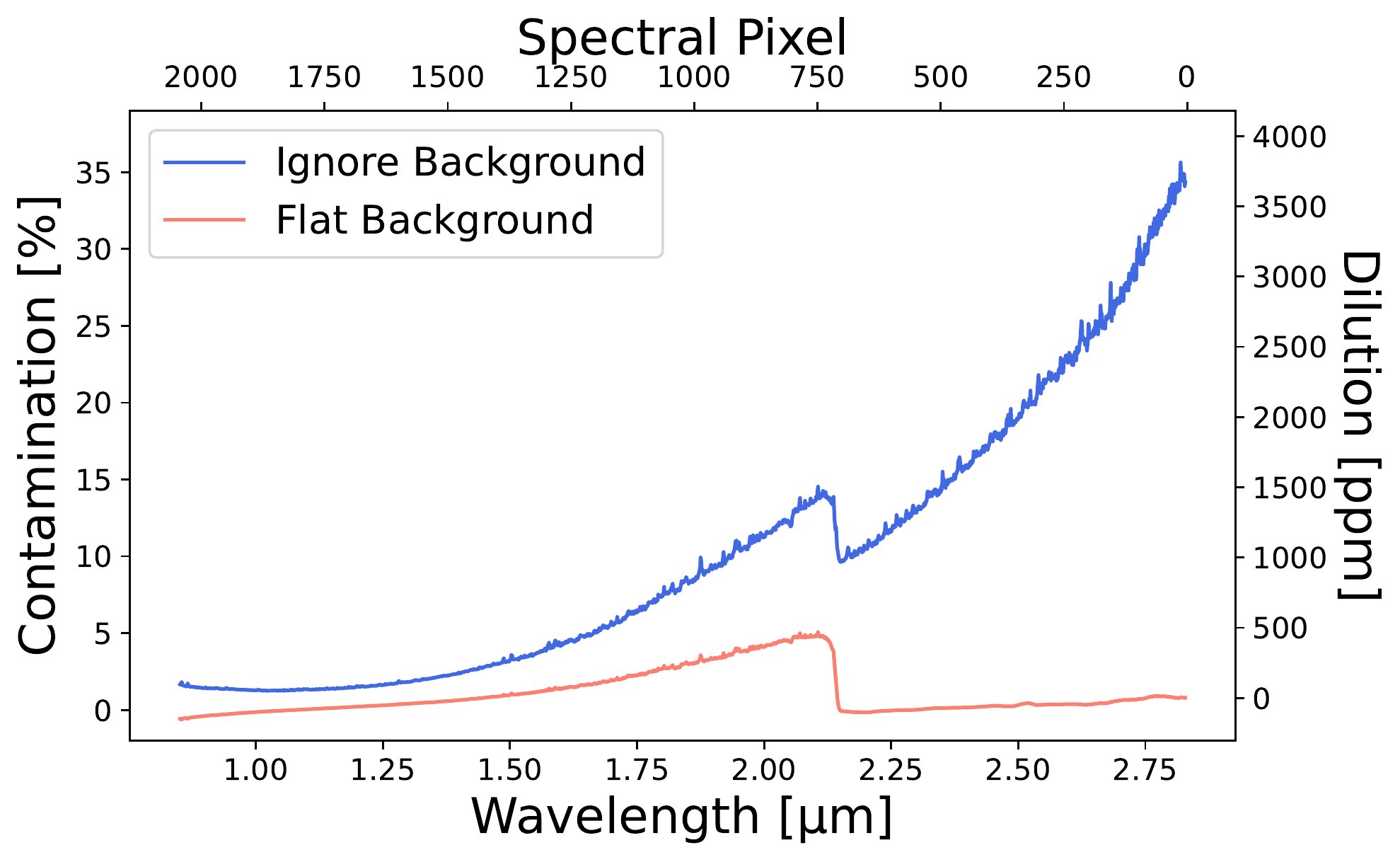}
    \caption{Dilution incurred through improperly treated background subtraction. The blue line shows the dilution to the WASP-96\,b light curves if the background contributions are ignored entirely. In the red is the dilution which would occur if a constant background level is subtracted from the entire frame. Due to the structured nature of the SOSS background, this constant subtraction is ill-advised and can result in a dilution of several hundred ppm. 
    \label{fig:Background Dilution}}
\end{figure}

Due to this unique structure, a constant background subtraction (i.e., estimating the background level based on one region of the detector, and subtracting this everywhere) will not be optimal. Indeed, as the strength of the background changes with wavelength (detector column), this will introduce a significant, wavelength-dependant dilution to transit (or eclipse) depths (e.g., Figure~\ref{fig:Background Dilution}). Instead, \texttt{supreme-SPOON} subtracts a model of the SOSS background, scaled to the flux level of each group. This model, shown in panel E of Figure~\ref{fig:Reduction Steps}, was created during commissioning\footnote{Program 1541} and is available from jdox\footnote{https://jwst-docs.stsci.edu/jwst-calibration-pipeline-caveats/jwst-time-series-observations-pipeline-caveats/niriss-time-series-observation-pipeline-caveats\#NIRISSTimeSeriesObservationPipelineCaveats-SOSSskybackground}. We note that other background models can be swapped into \texttt{supreme-SPOON} by the user if required. First, a median stack of each group is taken across the out-of-transit integrations. Then the background model is scaled to the flux level of each stack using a small region (x$\in$[250, 500], y$\in$[210, 250]) in the upper-left corner of the SUBSTRIP256 detector where the contribution from the target orders is minimal. This region should be adjusted as needed for each individual TSO as background sources may contaminate this particular region. After the model is rescaled, it is then subtracted from each group in each integration.  

We then proceed with the correction of the 1/$f$ noise. The 1/$f$ noise structures are generally quite faint (the streaks are barely visible in panel B of Figure~\ref{fig:Reduction Steps}). We, therefore ``reveal'' them by constructing difference images. To this end, we construct a group-wise median stack using all out-of-transit integrations; that is to say, we create one median stack for each group. For each integration, we then subtract each group frame from that group's median stack to remove the trace of the target star and reveal the 1/$f$ noise structures (panel C of Figure~\ref{fig:Reduction Steps}). A column-wise median is then calculated on this difference image and subtracted from the original data frame. The core region of the target trace, any bad pixels, and bright background contaminants should also be masked during this procedure.

In order to accurately estimate the 1/$f$ level, it is critical to have fully subtracted off the trace of the target star. The wings of the SOSS point spread function (PSF) are extremely broad, extending across virtually the entire detector. If residuals of the wings remain in the difference image when the 1/$f$ noise level is calculated, this will again introduce a wavelength-dependent dilution. To subtract the target trace as fully as possible, it is critical to take into account the fact that it will dim due to the transit (or eclipse) of the planet. Therefore, we rescale the group-wise median to the flux level of each integration before subtracting it, using an estimate of the white light curve (e.g., a factor of 1.0 for out-of-transit frames, and 0.99 for in-transit frames of a 1\% transit). This procedure allows us to completely subtract the target trace, leaving only Gaussian noise in the trace core. In theory, wavelength dependant variations (due to the planet's atmosphere signal, for example) could also change the brightness of the trace within a given frame. However, these will generally be higher order effects and can safely be ignored in most cases. For planets with unusually strong atmospheric signals, or eclipses with large wavelength-dependent changes in brightness, this may become an important effect \citep[e.g.,][]{coulombe_broadband_2023}. 

An additional consideration in this process is that when rescaling the group-wise median frame, one must be careful not to \textit{also rescale the background}. The flux in the target trace will change due to a transit, however, the background level will naturally remain constant. This is why \texttt{supreme-SPOON} treats the background at the group level, and why the background is subtracted before correcting the 1/$f$ noise. It is, furthermore, critical to adequately mask bright background sources (e.g., order 0 contaminants) during the 1/$f$ correction. Since the flux level of these contaminants remains (to first order) constant over the course of the TSO, subtracting the scaled median frame from each integration will leave positive residuals at the locations of order 0 contaminants. These residuals bias the calculated 1/$f$ level in affected columns to larger values --- thereby resulting in the over-subtraction of 1/$f$ noise and anti-dilution in the resulting transit depths.

The final level of complexity in the background-1/$f$ coupling stems from the non-linearity correction. NIR integrations are taken in a non-destructive fashion, so-called ``up-the-ramp'' reads, and final calibrated 2D images are composed of rate quantities (counts per second), obtained via fitting a linear trend to the pixel values of each group. For bright objects and large counts, the detector response begins to become non-linear, where the detector measures fewer counts that are received in reality. For SOSS, non-linearity effects become large around 35\,000 counts \citep{Albert2023SOSS}; for this WASP-96\,b TSO, the peak counts reached is only $\lesssim$20\,000. This detector response correction is performed via the \texttt{LinearityStep}. It is important here to consider the order in which ``noise'' is ``added'' to the observations. The background flux arrives at the detector simultaneously to the photons from the target. The background is therefore subject to any non-linearity effects. However, the 1/$f$ noise is introduced as the detector is read and \textit{is not affected by non-linearity}. Therefore, for the most precise results, the 1/$f$ noise should be treated before the non-linearity correction, and the background after. We therefore re-add the previously subtracted background to the data at the end of the \texttt{OneOverFStep}, and re-subtract it during Stage 2. We note here that performing the 1/$f$ correction at the integration level (that is, after the non-linearity correction) does not lead to any systematic biases in the resulting transmission spectrum, however it does result in overall less precise transit depths and more scatter (see Appendix~\ref{sec: Group vs Integration}). 

After the correction of the 1/$f$ noise, the remaining steps for Stage 1 are well taken care of by the official \texttt{jwst} pipeline, and \texttt{supreme-SPOON} once again simply provides wrappers around the corresponding steps for steps viii--x. We note though that \texttt{supreme-SPOON} skips any dark current subtraction, as the dark reference file provided in the Calibration Reference Data System (CRDS) shows clear signs of superbias residuals and 1/$f$ noise, which actually increases the noise level in the observations when subtracted. We find though that the dark level is in general $\lesssim$25 counts, and can therefore be safely ignored.

\subsubsection{\texttt{supreme-SPOON} Stage 2 --- Spectroscopic Calibrations}
\label{sec: Stage 2}

The second stage of \texttt{supreme-SPOON} performs further, high-level calibrations to SOSS data frames. A brief summary of the four major steps are outlined below:

\begin{enumerate}
    \item \texttt{FlatFieldStep}: Removal of flat field.
    \item \texttt{BackgroundStep II}: Subtract the background level.
    \item \texttt{BadPixStep}: Interpolation of hot, and other persistent bad pixels.
    \item \texttt{TracingStep}: Determine the centroids and stability of the target trace.
\end{enumerate}

The \texttt{FlatFieldStep} is a wrapper around the corresponding step of the official JWST pipeline, and performs the standard correction by dividing the data frames by a flat field reference image. This is the only step in this stage shared with the official pipeline. The \texttt{BackgroundStep} is once again performed, to permanently remove the SOSS background. The process is identical to that described in Section~\ref{sec: Stage 1}, except only one median stack is constructed. The background model is then scaled to the flux level of this median stack, and subtracted from each integration. A single integration, before subtracting the background model is shown in panel D of Figure~\ref{fig:Reduction Steps}, and the scaled background model itself in panel E. 

We skip both of the flux calibration steps, \texttt{pathloss} and \texttt{photom}, of the official \texttt{jwst} pipeline, as exoplanet atmospheric spectroscopy relies on relative measurements (e.g., in- vs. out-of-transit), which renders an absolute flux calibration unnecessary. 

\texttt{supreme-SPOON} then corrects any remaining hot pixels. This is accomplished by first constructing a median stack of all integrations. Pixels in the median stack which deviate by more than a given threshold from the surrounding pixels in time (for this work we use a threshold of 5$\sigma$) are flagged and then corrected in each individual integration using the median of the neighbouring pixels. This is the final reduction stage, and an example of the resulting data frame is shown in panel F of Figure~\ref{fig:Reduction Steps}.

\begin{figure}
	\centering
	\includegraphics[width=\columnwidth]{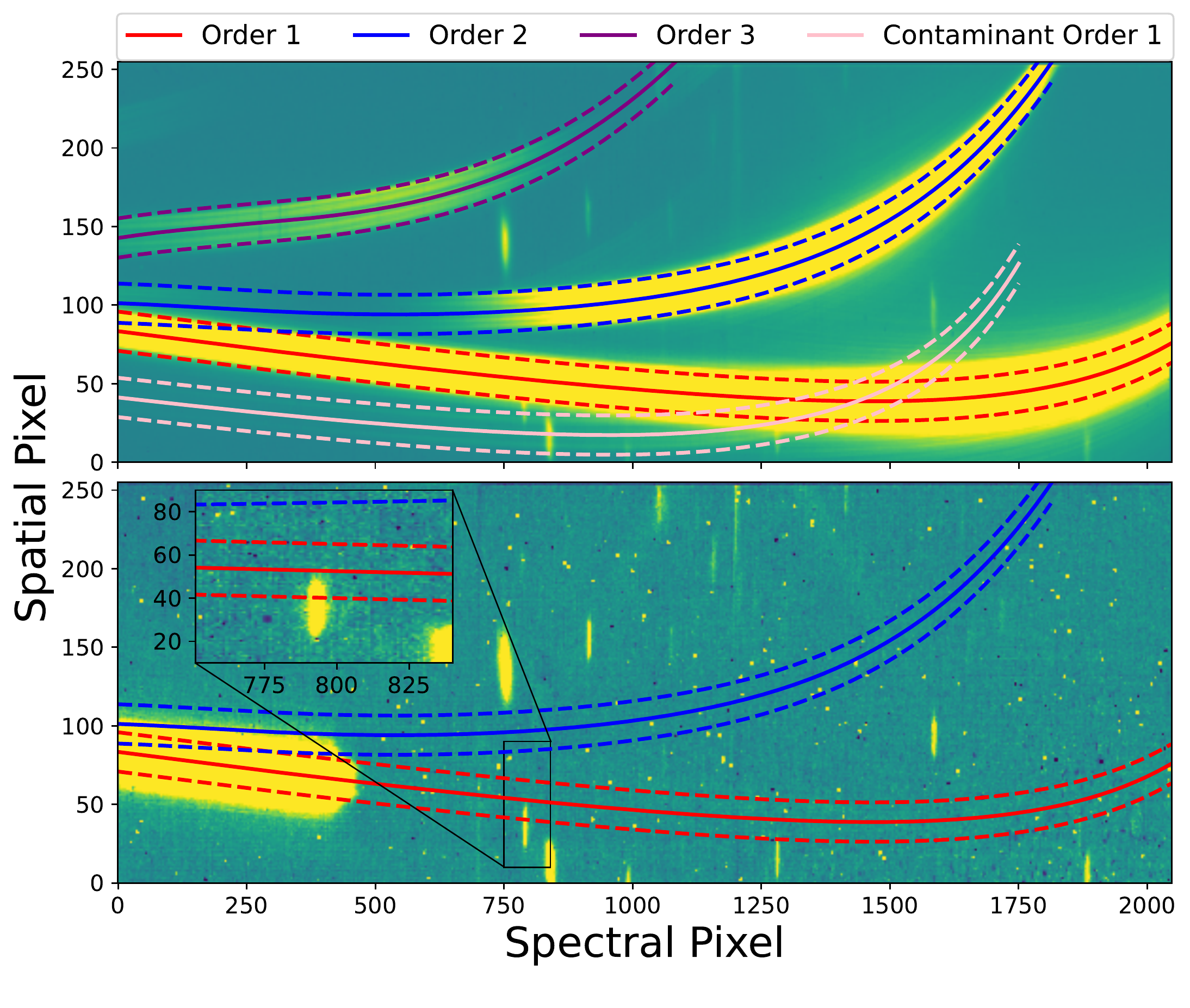}
    \caption{Visualization of field star contaminants in the WASP-96\,b SOSS frame. 
    \emph{Top}: Median stack of the CLEAR exposure. The centroids of the target traces for the first, second, and third orders are indicated via the red, blue, and green solid lines respectively. The 25-pixel extraction boxes are bounded by the dashed lines in each colour (Note that order 3 is not extracted, but included in this plot for completeness). The contaminating order 1 from a field star is indicated in pink and can be clearly seen to intersect the extract box of the target order 1 between pixels $\sim$1250--1600.
    \emph{Bottom}: Median stack of the F277W exposure. The centroids and extraction boxes for orders 1 and 2 are again indicated identically to the above. Multiple field star order 0s are visible in this frame. One, centered at spectral pixel $\sim$790, clearly intersects the order 1 extraction box (see inset). 
    \label{fig:Contaminants}}
\end{figure}

The last step does not perform any additional reductions but is a utility step to aid in 1/$f$ correction as well as eventual spectral extraction and light curve fitting. The \texttt{TracingStep} extracts the centroids of the target traces, and optionally also determines the stability of the target trace over the course of the TSO. Centroids are extracted for each order from a final median stack of all integrations via the ``edgetrigger'' algorithm \citep{radica_applesoss_2022}. The centroids are overplotted on the final median stack for the WASP-96\,b TSO in the top panel of Figure~\ref{fig:Contaminants}. The centroids are necessary to define the extraction box (see Section~\ref{sec: Stage 3}), but also to define a trace mask which can be used to mask the trace core during 1/$f$ correction.

Furthermore, the \texttt{TracingStep} can determine the stability of the target trace throughout the TSO. Pointing jitter, for example, may cause the target trace to shift positions on the detector over the course of a many-hour TSO. The stability is assessed through a cross-correlation function (CCF) analysis. The median stack is shifted by a small fraction of a pixel individually in the vertical and horizontal direction and cross-correlated with the data frame from each integration. This enables the measurement of the position of the target trace to sub-pixel accuracy, relative to the position of the median trace. Furthermore, the full width at half-maximum (FWHM) of the trace is estimated by creating a ``super-profile'' for each integration by stacking the PSFs of the first order near the peak in throughput (detector columns 1500--1750). A Gaussian profile is then fit to this super profile to estimate the FWHM. The change in FWHM, as well as $x$ and $y$ position, are shown in the bottom panel of Figure~\ref{fig:SOSS Stability}. In general, the trace position is remarkably stable, with root mean square (RMS) deviations in $x$ and $y$ of $\sim$2 and $\sim$4 milli-pixels respectively. The FWHM is also extremely stable, except during a mirror tilt event $\sim$1\,hr after the transit midpoint (see Section~\ref{sec: White Light Curve Analysis}).

\subsubsection{\texttt{supreme-SPOON} Stage 3 --- 1D Spectral Extraction}
\label{sec: Stage 3}

Stage 3 of the \texttt{supreme-SPOON} pipeline performs the 1D spectral extraction. The four major steps are listed below.

\begin{enumerate}
    \item \texttt{SpecProfileStep}: Construct a model of the SOSS trace profile for all orders.
    \item \texttt{SossSolverStep}: Determine the correct transform to match the CRDS reference files. 
    \item \texttt{Extract1dStep}: Perform the 1D extraction via either a box extraction or using the \texttt{ATOCA} algorithm.
    \item \texttt{LightCurveStep}: Convert spectra into a convenient data format.
\end{enumerate}

\begin{figure}
	\centering
	\includegraphics[width=\columnwidth]{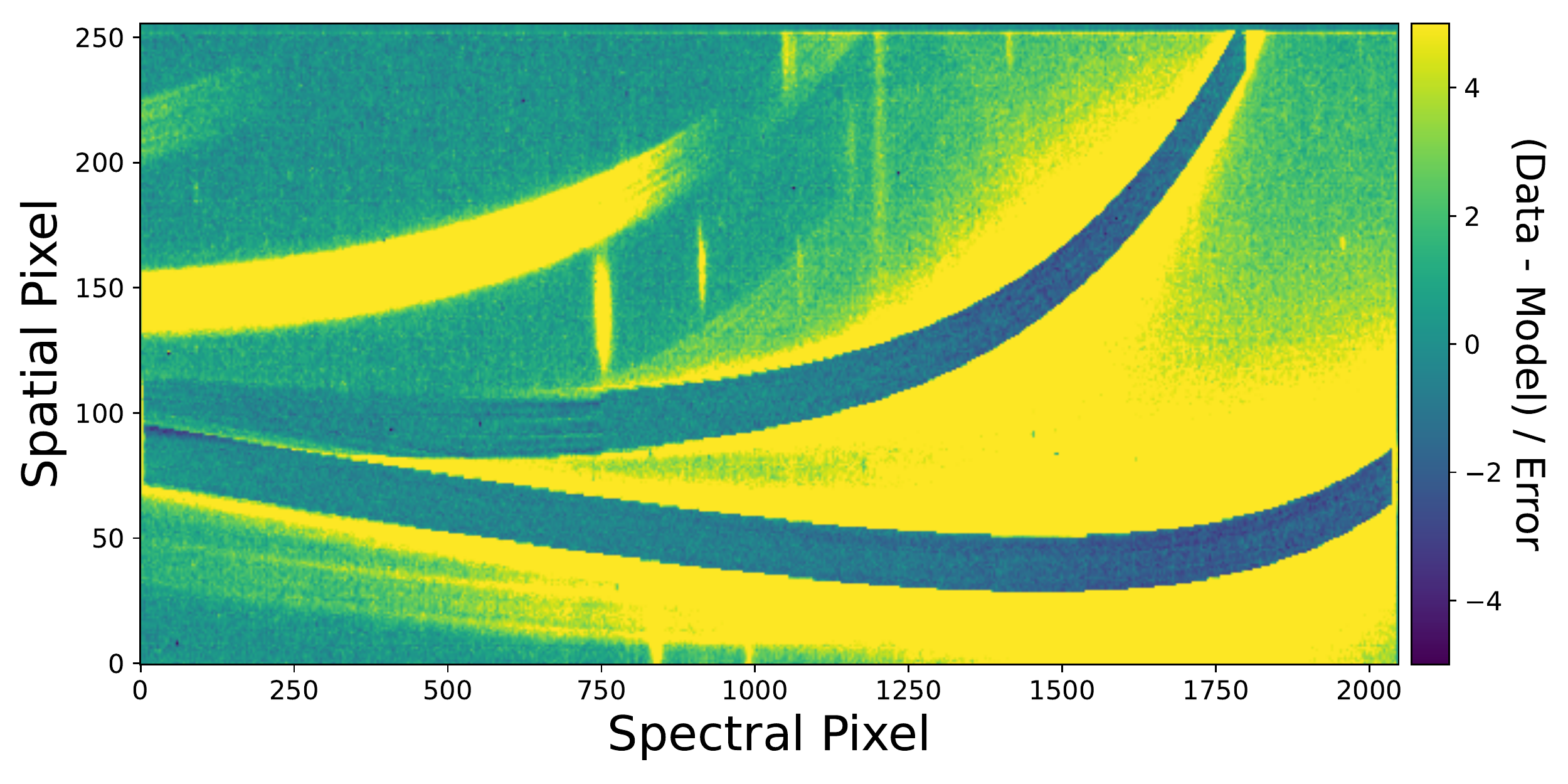}
    \caption{Example \texttt{ATOCA} decontamination results for integration 100. The first and second orders are well modelled across the entire detector, with residuals generally bounded within 1.5$\sigma$. 
    \label{fig:ATOCA Decontamination}}
\end{figure}

Within \texttt{supreme-SPOON}, the 1D spectral extraction can be performed via two different methodologies: either with a simple box aperture extraction or using the \texttt{ATOCA} algorithm \citep{darveau-bernier_atoca_2022}. \texttt{ATOCA} was developed to explicitly treat the contamination resulting from the overlap of the first and second SOSS orders on the detector. Briefly, \texttt{ATOCA} constructs a linear model of every pixel on the detector, including contributions from both orders. It then ``decontaminates'' the detector --- that is, it produces a model of orders 1 and 2 which are then individually subtracted from the data to create decontaminated images of order 2 and 1 respectively. A box extraction can then be safely performed on these decontaminated frames completely free of contamination from the other order. An example of the decontamination for integration 100 of the WASP-96\,b TSO is shown in Figure~\ref{fig:ATOCA Decontamination}.

Although this order contamination can introduce significant biases for absolute flux measurements, for relative flux measurements, like those used in exoplanet transmission and emission spectroscopy, the dilution caused by the contamination is negligible \citep{darveau-bernier_atoca_2022, radica_applesoss_2022}.

The \texttt{SpecProfileStep} is skipped for box extractions, but critical for extractions using \texttt{ATOCA}. This is because it constructs an estimate of the spatial PSF of the target trace upon which \texttt{ATOCA} relies using the \texttt{APPLESOSS} algorithm \citep{radica_applesoss_2022}. A trace profile estimate, which was constructed with \texttt{APPLESOSS} on commissioning data of a quiet A0V star, is provided as part of the \texttt{specprofile} CRDS reference file. However, the fine structure of the SOSS PSF may change subtly over time due to, among other factors, changes in the wavefront caused by small temperature changes, or more drastically due to things such as tilt events. Therefore, it is recommended to construct a trace profile specifically for each TSO instead of relying on the reference profile provided. 
We additionally note that \texttt{APPLESOSS} has been significantly upgraded since the commissioning period to produce profiles which better reproduce the extended wing structure of the SOSS PSF (Fournier-Tondreau et al.~in prep).

Next, the \texttt{SossSolverStep} uses the extracted centroids to calculate the necessary parameters (rotation, $x$-shift, $y$-shift) to transform the reference centroids provided in the \texttt{spectrace} CRDS reference file to match the data. This is performed via a sequential least squares $\chi^2$ minimization implemented in the \texttt{scipy.optimize.minimize} routine. For WASP-96\,b, we find that the centroids are already well-matched to those provided by the CRDS without applying any transform. The required transform is then passed to the \texttt{Extract1dStep} along with the calibrated data cube to do the 1D extraction. This is once again a wrapper around the official JWST pipeline step of the same name. We perform extractions using both the box and \texttt{ATOCA} methods, but since the results are identical (e.g., Figure~\ref{fig:Compare Spectra Contamination}), we only consider the \texttt{ATOCA} results for the remainder of this work. In both cases, we use a box width of 25 pixels as this was found to maximize the signal-to-noise ratio (S/N) of the extracted stellar spectra. The extraction boxes as well as trace centroids for both orders are shown in the top panel of Figure~\ref{fig:Contaminants}.

Lastly, the \texttt{LightCurveStep} interpolates any outliers in time above a user-specified threshold (in this work, we use 5$\sigma$ as we find that sufficient to remove the handful of remaining outlier points) in the extracted stellar spectra, and packages the spectra into an easily portable Flexible Image Transport System (FITS) file format along with all the relevant extraction parameters to ensure reproducibility of the results.

\subsubsection{Spectrum Post-Processing}
\label{sec: Post Processing}

The steps described above will be necessary for all SOSS TSOs, and can be performed, with slight fine-tuning, for any observation. However, the SOSS mode presents a further complexity, namely field star contamination, which must be treated on a case-by-case basis. As SOSS is a slitless mode, any background star lying within the field of view of NIRISS will also be dispersed onto the detector. The amount of field star contamination can be minimized by effectively choosing targets, as well as telescope roll angles. However, occasionally some level of contamination will be impossible to avoid. 

The WASP-96\,b TSO is an example of a case where background contamination was unavoidable. In panel F of Figure~\ref{fig:Reduction Steps}, as well as the top panel of Figure~\ref{fig:Contaminants}, two types of field star contaminants are clearly visible: order 0 contaminants, which appear as bright smudges at various locations, and an order 1 contaminant (whose trace is outlined in pink in the top panel of Figure~\ref{fig:Contaminants}). Below we suggest a method for correcting each type of contaminant and apply them to the WASP-96\,b TSO.

\paragraph{Order 0 Contaminants}

Order 0 contaminants, due to their concentrated brightness, can potentially cause a large amount of dilution if they intersect the target trace. Although most order 0 contaminants are bright enough to be visible in a GR700XD/CLEAR exposure, we make use of the GR700XD/F277W exposure to more efficiently identify potential contaminants. The F277W filter only transmits light redder than $\sim$2.6\,µm, effectively isolating the red end of the order 1 trace. This was initially posited as a key to understanding the extent to which the target orders self-contaminate, which in the end was proven negligible (see Section~\ref{sec: Stage 3}). Instead, this filter does allow for the effective identification of field star contaminants. We process the F277W filter exposures in a similar manner to that of the GR700XD/CLEAR science exposures. The bottom panel of Figure~\ref{fig:Contaminants} shows a median stack of the F277W filter exposure, as well as the extraction boxes for orders 1 and 2. Field star order 0s are clearly visible in the frame, even those which could potentially be hidden behind the target trace. We identify one order 0 (shown zoomed-in in the inset) centered at column $\sim$790 which intersects the target order 1 trace. There is a second contaminant, at pixel $\sim$1300 which appears to graze the order 1 extraction box, however, we were not able to ascertain that this contaminant causes any meaningful dilution, and we, therefore, ignore it.  

Then, we proceed to estimate the dilution introduced by the order 0 contaminant at column $\sim$790. We roughly estimate the extent of the contaminant in the horizontal direction, and again construct a ``super-profile'' by stacking the PSFs of the five columns on each side of the contaminant. We linearly interpolate the two super profiles over the pixels contaminated by the order 0 to construct a roughly ``uncontaminated'' trace, which we subtract off of the detector to reveal the order 0 contaminant. Using the same extraction box as for the target trace, we then extract over the contaminant to estimate the amount of flux, and therefore the dilution factor, it introduces. 

The dilution introduced by this order 0 contaminant is found to be quite significant, at a level of $\sim$750\,ppm, but only over a very small wavelength region ($\sim$15 columns; e.g., Figure~\ref{fig:Compare Spectra Contamination}).


\paragraph{Order 1 Contaminant}

In the top panel of Figure~\ref{fig:Contaminants}, a contaminant first order trace (outlined in pink) can be clearly seen to intersect the first order trace of the target. This will again introduce some dilution to the extracted light curves. In order to estimate the level of dilution from this contaminant, we first subtract off the order 1 trace of the target. We accomplish this by once again using the \texttt{ATOCA} algorithm to construct a model of the first order trace. Instead of using a width of 25 pixels as we did for our initial extraction, we instead select a width of 100 pixels in order to capture the extended wing structure in addition to the profile core. 

After subtraction of the modelled target order 1, we then manually fit the extracted centroids of the target order 1, to the contaminant. We find that only order 1 is able to match the shape of the contaminant trace and that its shape is well represented by a shift of -510 pixels in the horizontal and -21 in the vertical direction from the target order 1. This also allows for the estimation of the wavelength solution for the contaminant via applying the same transform to the wavelength solution of the target. We then define an extraction box and extract the spectrum of the order 1 contaminant over pixel columns 0--500. It is critical that the order 1 of the target be subtracted off before extracting the contaminant spectrum, as the target order 1 wings introduce significant contamination to the spectrum of the contaminant; how the tables turn. 

\begin{figure}
	\centering
	\includegraphics[width=\columnwidth]{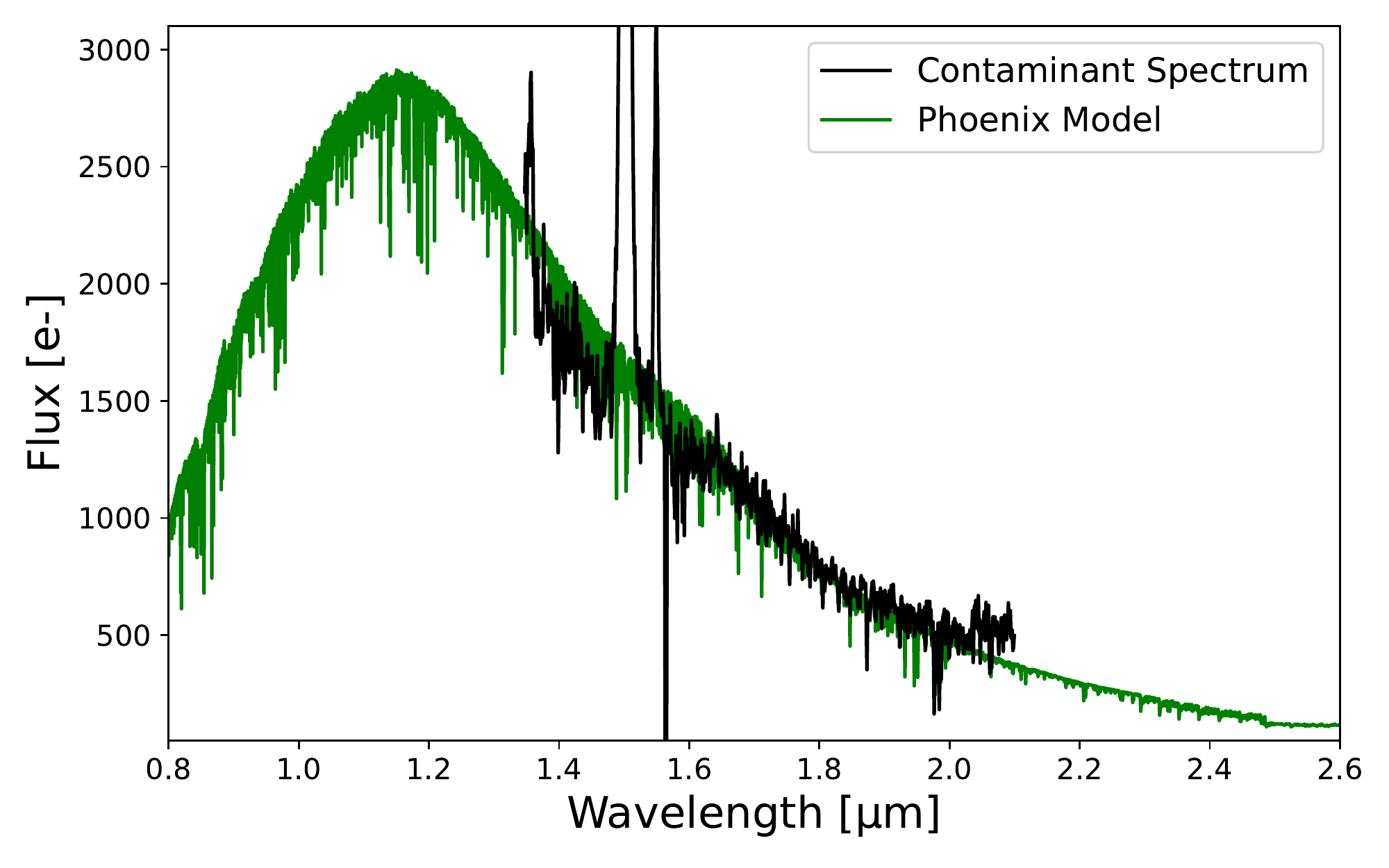}
    \caption{Comparison of the extracted spectrum of the contaminant order 1 field star and the best fitting PHOENIX stellar model. The contaminant spectrum is well fit by a stellar model with parameters $\rm T_{eff} = 3900K$, $\rm \log g = 4.5$, and $\rm Fe/H = 0.5$.
    \label{fig:Contaminant Spectrum}}
\end{figure}

Using a custom SOSS contamination tool\footnote{http://maestria.astro.umontreal.ca/niriss/SOSS\_cont/SOSScontam.php} we were able to identify the contaminant star in the Gaia DR3 catalogue\footnote{Gaia Source ID 4990044874536787328} \citep{2021Gaia}. Cross-referencing with the \textit{Transiting Exoplanet Survey Satellite} \citep{2014RickerTess} Input Catalogue, we retrieve the effective temperature and gravity of the source: $\rm T_{eff}=3900\,K$, $\rm \log g = 4.65$. We then compared the extracted contaminant spectrum to a grid of PHOENIX stellar models \citep{husser_new_2013}, which have been corrected for the instrument throughput, and find the spectrum to be well matched by a model with $\rm T_{eff}=3900\,K$, $\rm \log g = 4.5$, and $\rm Fe/H=0.5$ (Figure~\ref{fig:Contaminant Spectrum}) --- which are consistent with the stellar parameters quoted above. We then smoothed the model stellar spectrum and used it to estimate the level of dilution introduced in the region where it intersects the target trace. 

As the strength of any of the contaminants does not appear to vary in time, we then subtract this estimated contamination from the extracted stellar spectra at each integration to construct contamination-corrected spectra (but c.f., \citealp{fu_water_2022} for a case of time-variable background star contamination). The transmission spectra before and after contamination correction are shown in Figure~\ref{fig:Compare Spectra Contamination}. Compared to the order 0 contaminant, the order 1 contaminant introduces a much lower level of dilution ($\lesssim$100\,ppm vs. $\sim$750\,ppm) but over a much larger wavelength range.

\section{Light Curve Analyses}
\label{sec: Light Curve Analysis}

\subsection{White Light Curve Analysis}
\label{sec: White Light Curve Analysis}

\begin{figure*}
	\centering
	\includegraphics[width=\textwidth]{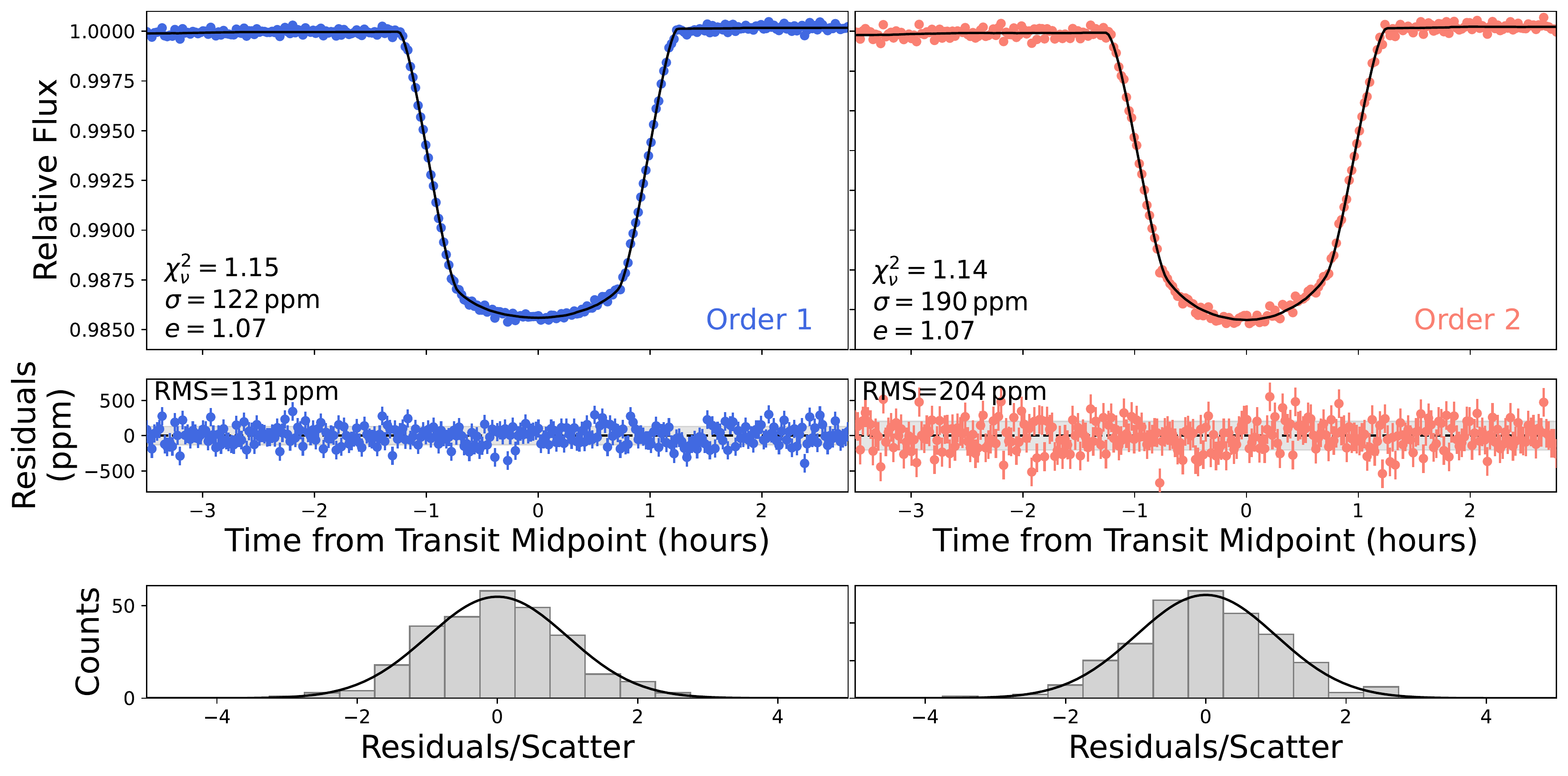}
    \caption{\emph{Top}: White light curves for for order 1 (\emph{left}) and order 2 (\emph{right}). The best-fitting transit model is overplotted in black. The fit statistics are shown for each order: $\chi^2_\nu$, $\sigma$; the average error bar, and $e$; the error multiple to obtain a $\chi^2_\nu$ equal to unity. 
    \emph{Middle}: Residuals to the transit fits. The RMS scatter in the residuals is indicated on each panel.
    \emph{Bottom}: Histogram of residuals. The histograms trace Gaussian distributions indicating that we have well handled systematic trends in the data.
    \label{fig:White Light Curves}}
\end{figure*}

We sum the flux from all wavelengths to create white light curves for each order. We only consider wavelengths $<$0.85\,µm for order 2 as at longer wavelengths the information is redundant with, and at a much lower signal-to-noise level than order 1. The white light curves are shown in Figure~\ref{fig:White Light Curves}.

We fit a \texttt{batman} transit model \citep{kreidberg_batman_2015} to each white light curve using the flexible \texttt{juliet} package \citep{espinoza_juliet_2019}. There appears to be a slight linear trend with time in the light curves for both orders; as well as a small ramp-like effect in the first $\sim$five integrations; but other than that they both appear free of systematic effects. We, therefore, cut these initial five integrations and also test a variety of different detrending techniques to find the optimal combination of transit and systematics models. We perform the \texttt{juliet} fits using its nested sampling capabilities, as implemented via \texttt{dynesty} \citep{speagle_dynesty_2020}, which allows for the estimation of the Bayesian evidence, $\ln \mathcal{Z}$, for each model. We test detrending against linear models with time ($\ln \mathcal{Z}=2012.54$), trace $x$-position ($\ln \mathcal{Z}=2015.99$), trace $y$-position ($\ln \mathcal{Z}=2013.95$), and trace FWHM ($\ln \mathcal{Z}=2015.12$). There is a slight preference \citep[$\sim$2$\sigma$ based on the scale from][]{benneke_how_2013} for detrending against the trace $x$-position. All systematics models are highly preferred ($>$5$\sigma$) over the case with no detrending ($\ln \mathcal{Z}=1992.84$).

There was a so-called ``tilt-event'' which occurred approximately 1.4\,hr after mid-transit (integration $\sim$200) during the WASP-96\,b TSO. Tilt events are abrupt changes in the telescope wavefront potentially due to the thermal relaxation of the primary mirror segments \cite{rigby_commissioning22}. As the SOSS PSF is so highly dispersed, it is incredibly sensitive to wavefront changes, and tilt events can sometimes manifest as large discontinuities in light curves as more (or fewer) photons fall within the extraction aperture \citep[e.g.,][]{coulombe_broadband_2023}. There is no clear evidence of a tilt-event-induced discontinuity in our white light curves. However, the tilt event is clearly visible in the FWHM of the trace (bottom panel of Figure~\ref{fig:SOSS Stability}), as well as the 2D detector images themselves (Figure~\Ref{fig:2D tilt}). This is because the morphological change in the PSF which occurs during the tilt-event is contained entirely within our 25-pixel extraction box, and there is, therefore, no net change in flux within the aperture. However, using instead a 20-pixel aperture does result in a marked discontinuity in the light curve as the tilt-event causes additional flux to fall into the aperture (see Figures~\ref{fig:SOSS Stability} and \ref{fig:2D tilt}). We, therefore, do not explicitly model the tilt event in our light curve fits (e.g., via detrending against an additional ``jump'' parameter in the fits) as there is no evidence it affects the light curve at a level greater than our measurement precision, and a systematics model consisting of a jump term in addition to the trace $x$-position was disfavoured by $\sim$4$\sigma$ over the case with just the trace $x$-position ($\Delta \ln \mathcal{Z}=9.2$).

Our final white light transit model, therefore, has nine parameters: the transit mid-point time, $t_0$, the scaled planet radius, $R_p/R_*$, the transit impact parameter, $b$, the scaled orbital distance, $a/R_*$, the two parameters of the quadratic limb-darkening law, $u_1$ and $u_2$ \citep{claret_new_2000} sampled via the \citet{kipping_efficient_2013} parameterization, a scalar jitter which is added in quadrature to the error bars, $\sigma$, as well as the two parameters of a linear systematics model with $x$-position, $\theta_0$ and $\theta_1$. Wide uninformative flat priors are used for each parameter. We additionally fix the period to 3.4252602\,d \citep{nikolov_solar--supersolar_2022} and assume a circular orbit. The best-fitting transit models are overplotted in black in the top panels of Figure~\ref{fig:White Light Curves}, and the residuals to the fits are shown in the middle and bottom panels. Additionally, the best-fitting transit parameters are shown in Table~\ref{tab: WLC Parameters}, along with those from all other reductions.

\begin{table*}
\caption{Comparison of best-fitting white light curve parameters}
\label{tab: WLC Parameters}
\begin{threeparttable}
    \begin{tabular}{ccccc}
        \hline
        \hline
        Reduction & $\rm t_0$ [BJD - 2400000] & $\rm R_p/R_*$ & $\rm b$ & $\rm a/R_*$ \\
        \hline
        \texttt{supreme-SPOON} & 59751.82467$\pm$3$\times$10$^{-5}$ & 0.1190$\pm$5$\times$10$^{-4}$ & 0.7301$\pm$0.0037 & 8.963$\pm$0.040 \\
        \texttt{nirHiss} & 59751.82471$\pm$5$\times$10$^{-5}$& 0.1197$\pm$9$\times$10$^{-4}$& 0.7276$\pm$0.0055 & 8.988$\pm$0.058\\
        \texttt{transitspectroscopy} & 59751.82464 $\pm$5$\times$10$^{-5}$ & 0.1191$\pm$5$\times$10$^{-4}$ & 0.7243$\pm$0.0045 & 9.011$\pm$0.044\\
        \texttt{NAMELESS} & 59751.82469$\pm$4$\times$10$^{-5}$ & 0.1193$\pm$9$\times$10$^{-4}$ & 0.7287$\pm$0.0051 & 8.978$\pm$0.051\\
        \citet{nikolov_absolute_2018} & -- & -- & 0.749$\pm$0.020 & 8.84$\pm$0.10 \\
        \hline
    \end{tabular}
    \begin{tablenotes}
        \small
        \item \textbf{Notes}: Values listed above are for order 1 white light curves. 
    \end{tablenotes}
\end{threeparttable}
\end{table*}

We assess the quality of the transit model fits to each order via the reduced chi-squared ($\chi^2_\nu$) metric. This value, along with the size of the average error bar, $\sigma$, and $e$; the error multiple necessary to reach a $\chi^2_\nu$ equal to unity are listed for each order in the top panel of Figure~\ref{fig:White Light Curves}. For both orders, we find a very favourable $\chi^2_\nu$ metric (1.15, and 1.14 for orders 1 and 2 respectively) indicating that the assumed model is an excellent fit to the data. The residuals are furthermore highly Gaussian, and bin down as would be expected from pure photon noise (e.g., Figure~\ref{fig:Allan Variance}).

\subsection{Spectrophotometric Light Curve Analysis}
\label{sec: Spectroscopic Light Curve Analysis}

\begin{figure}
	\centering
	\includegraphics[width=\columnwidth]{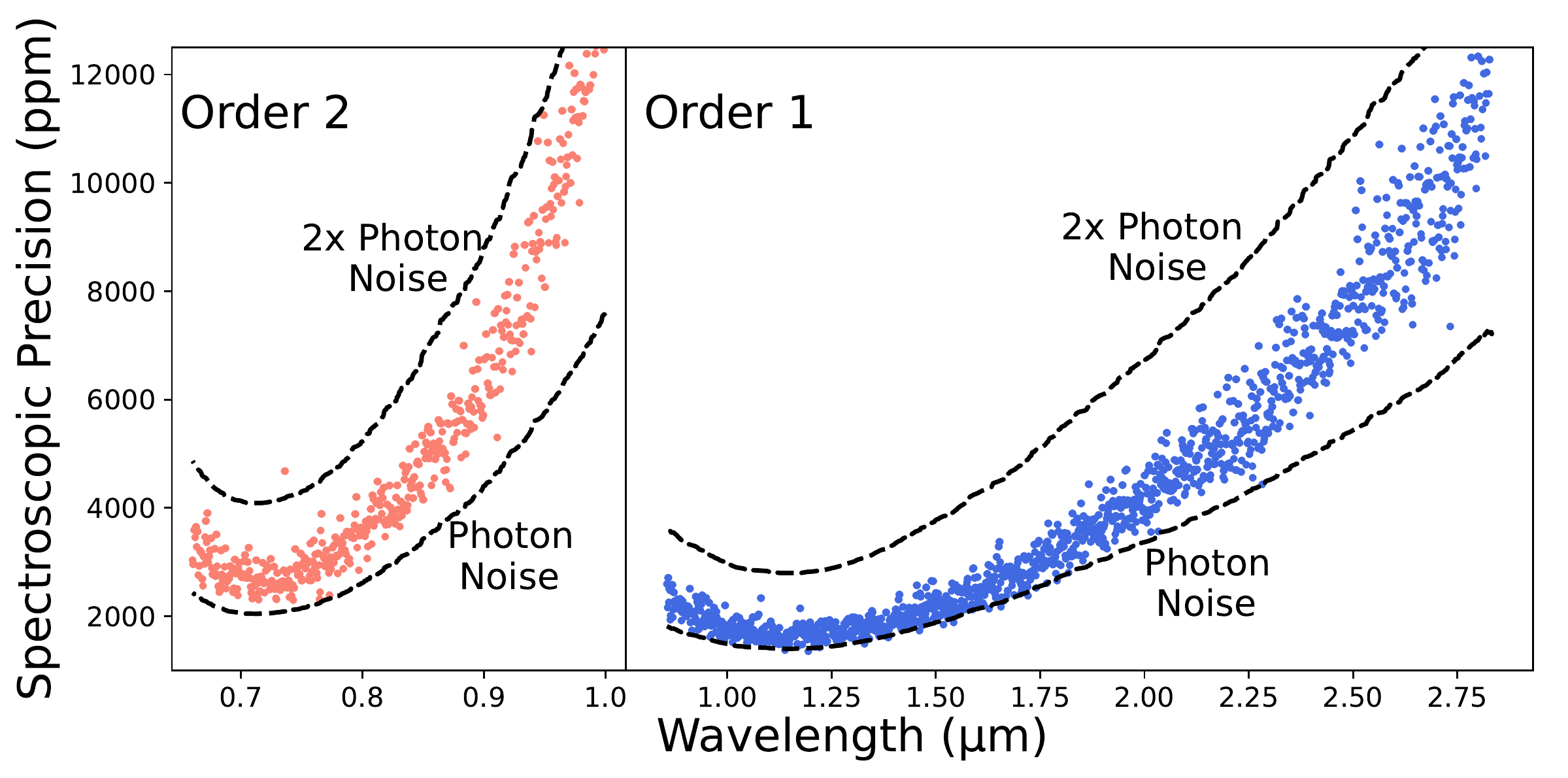}
    \caption{Spectroscopic precision per bin obtained in this NIRISS/SOSS TSO of WASP-96\,b at the pixel level (bin width of one pixel) compared to the predicted photon noise limits (dashed black lines) for order 1 (right) and order 2 (left). The estimated read noise ($\sim$21\,ppm) is removed here from the precision calculated in each bin. Our \texttt{supreme-SPOON} reduction reaches an average precision of $\sim$1.2$\times$ and $\sim$1.4$\times$ the predicted photon noise for order 1 and order 2 respectively. 
    \label{fig:Spectroscopic Precision}}
\end{figure}

To obtain the final transmission spectrum, we proceed to fit the above transit model to light curves at the pixel level (that is one light curve per pixel column). Studies during commissioning have shown that binning detector columns after spectral extraction, but before light curve fitting, results in higher than expected levels of scatter in the out-of-transit light curve baseline. This has been attributed to unaccounted-for covariance between detector columns \citep[see e.g., Fig.~8 in][for an example with NIRSpec BOTS]{espinoza_spectroscopic_2023}. \citet{espinoza_spectroscopic_2023} in particular, conclude that the optimal way to work with NIR spectra from JWST is to extract spectra and fit light curves at the pixel level, and bin observables, such as transit depths, at a later stage. We thus adopt this strategy, which results in 2040 bins for order 1 (2048 pixel columns minus eight reference pixel columns) and 567 bins for order 2, where we only consider wavelengths $<$0.85\,µm. In each bin our \texttt{supreme-SPOON} reduction reached an average precision of 1.2 and 1.4$\times$ the photon noise for orders 1 and 2 respectively (e.g., Figure~\ref{fig:Spectroscopic Precision}).

\begin{figure*}
	\centering
	\includegraphics[width=\textwidth]{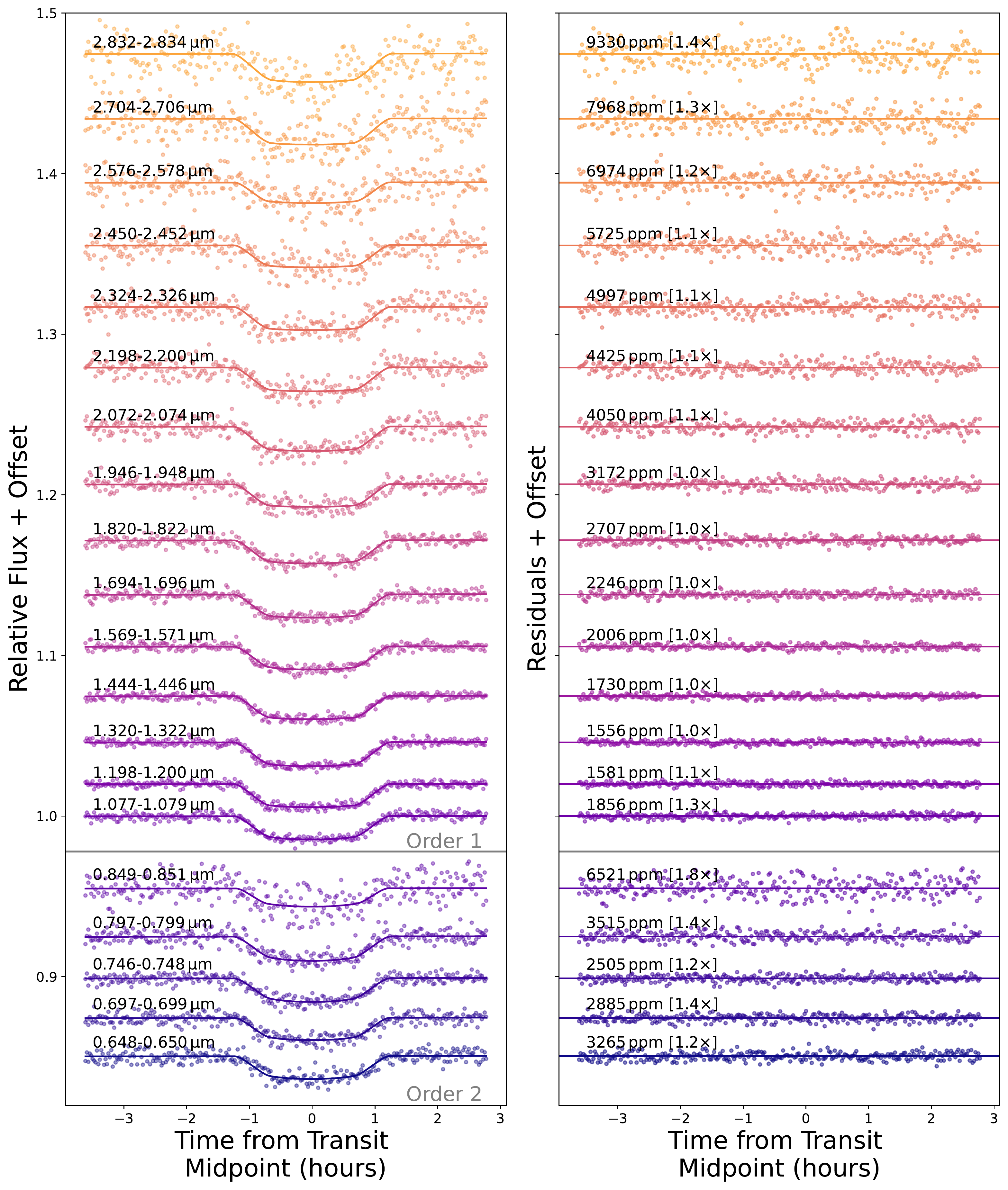}
    \caption{Spectrophotometric light curves and transit fits for the NIRISS/SOSS TSO of WASP-96\,b. \emph{Left}: Normalized spectrophotometric light curves at pixel resolution. The extent of each bin is indicated above each light curve. Overplotted on each is its best fitting transit model. No systematic trends have been removed from the data. \emph{Right}: Residuals for the light curve fit in each bin. The RMS scatter of the residuals in each bin are indicated as well as the ratio to the predicted photon noise in brackets. 
    \label{fig:SpectroPhotometric Lightcurves}}
\end{figure*}

\begin{figure*}
	\centering
	\includegraphics[width=\textwidth]{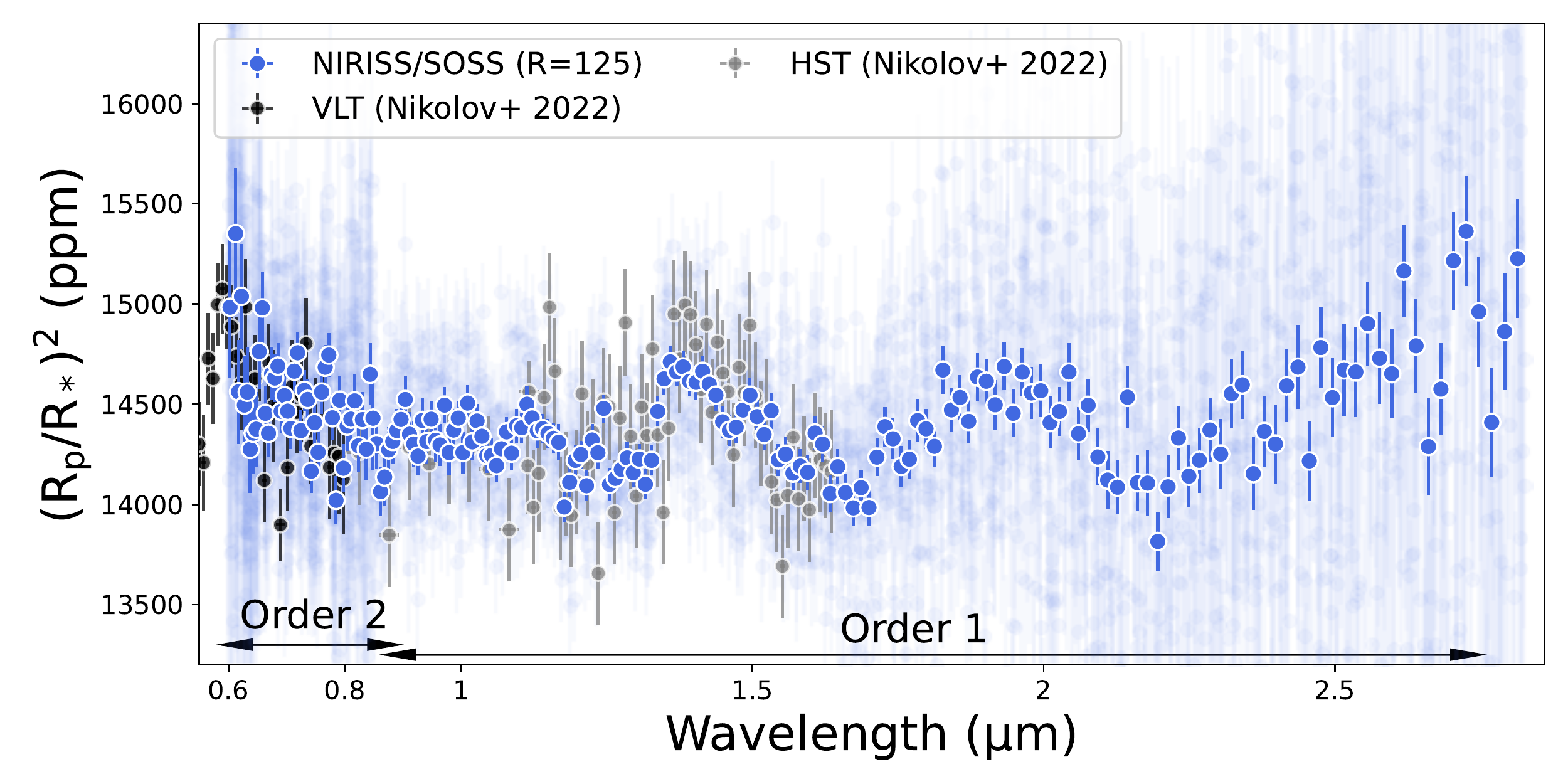}
    \caption{NIRISS/SOSS transmission spectrum of WASP-96\,b at pixel resolution (faded blue points) and binned to a resolution of R$\sim$125 (blue points). Also shown are the VLT (black points) and \textit{HST}/WFC3 G102 and G141 (grey points) transmission spectra from \citet{nikolov_solar--supersolar_2022}. Offsets of 400 and 200\,ppm have been applied to the VLT and \textit{HST} data respectively (see text).    
    \label{fig:Transmission Spectrum}}
\end{figure*}

Instead of fixing the orbital parameters of WASP-96\,b to the best-fitting values from the white light curve, we instead used the $b$, and $a/R_*$ provided by \citet{nikolov_solar--supersolar_2022} (last row of Table~\ref{tab: WLC Parameters}). This was chosen, in order to maintain a constant set of orbital parameters for the four different reductions. The transit mid-point time however, was fixed to the best-fitting value from the order 1 white light curve. We note here that what follows pertains only to the \texttt{supreme-SPOON} reduction, and details on the light curve fits for the other three reductions can be found in Appendix~\ref{sec: Additional Reductions}. We placed Gaussian priors on both of the limb-darkening coefficients, centered on predictions from the \texttt{ExoTiC-LD} package \citep{sing_stellar_2010, laginja_exotic-ism_2020} using the 3D stellar models of \citet{magic_stagger-grid_2015}. In order to estimate both the position and width of the Gaussian prior, we calculated limb-darkening coefficients over a three-dimensional grid in the stellar effective temperature ($\rm T_{eff}$), metallicity ([Fe/H]), and gravity ($\log g$) parameter space. The grid had five nodes, equally spaced in each dimension, and the extent of the grid was determined by the published uncertainties in each parameter. For all three stellar parameters, we used the following values and uncertainties from \citet{hellier_transiting_2014}: $\rm T_{eff}=5400\pm140\,K$, $\rm [Fe/H]=0.14\pm0.19\,dex$, and $\rm \log g=4.42\pm0.02\,g/cm^3$. The position of the Gaussian prior for the limb-darkening parameters at each wavelength was taken to be the mean values of the grid, and the width of the prior as the standard deviation across the grid. We therefore fit seven parameters to each bin. The light curves for several representative bins, as well as their best fitting transit models are shown in the left panel of Figure~\ref{fig:SpectroPhotometric Lightcurves}, and the fit residuals on the right. The resulting transmission spectrum is shown in Figure~\ref{fig:Transmission Spectrum} at both the pixel resolution (faded points) and binned to a constant R=125 for visual clarity.

Figure~\ref{fig:Transmission Spectrum} also includes the VLT and \textit{HST}/WFC3 transmission spectra as published by \citet{nikolov_solar--supersolar_2022}. We find offsets of 400 and 200\, ppm for the VLT and \textit{HST} transmission spectra relative to our SOSS spectrum. All of \citet{yip_compatibility_2021}, \citet{nikolov_solar--supersolar_2022}, and \citet{mcgruder_access_2022} also find offsets between their individual reductions of the VLT and \textit{HST} spectra, concluding that these offsets are linked to common-mode corrections applied to each dataset in order to correct systematics in the binned light curves. Indeed, \citet{mcgruder_access_2022} explicitly explored the effects of common-mode corrections on retrieved transit depths, and find that although information regarding relative depths is preserved, absolute transit depths are poorly reproduced. Moving forward in the JWST era, we should therefore expect to find offsets between existing transmission spectra which require common-mode corrections (e.g., \textit{HST}/WFC3), and those from JWST on which no common-mode corrections are applied and absolute transit depths are reliable.

\section{Grid Model Fits}
\label{sec: Grid Retrievals}

We interpret the NIRISS/SOSS observations by comparing them to a series of self-consistent radiative-convective equilibrium atmosphere models. We follow the procedure established by the JWST Transiting Exoplanet Community Early Release Science Team \citep{bean_transiting_2018} and perform an initial atmospheric assessment using grids of models \citep[e.g.,][]{ahrer_early_2023,alderson_early_2023,feinstein_early_2023,rustamkulov_early_2023}. We defer detailed atmospheric retrieval analyses of these observations to the subsequent companion paper, \citet{Taylor_w96_2023}. Below we describe the three grids of models used to interpret our WASP-96\,b observations. Since all grids converge to broadly consistent results, we select the ScCHIMERA grid, which produces the overall best fit, as the reference for our discussion and conclusions. 

We note that the model fits were performed on the observations at the different resolutions of R=125, 250, 500, and pixel-level. The results for each grid are generally independent of the resolution of the observations, and we defer further analysis into any resolution-dependence of atmosphere inferences to \citet{Taylor_w96_2023}. The best fitting parameters for each model grid are included in Table~\ref{tab: Model results}.

\begin{table*}
\caption{Comparison of best-fitting model grid parameters}
\label{tab: Model results}
\begin{threeparttable}
    \begin{tabular}{cccccc}
        \hline
        \hline
        Model & Metallicity [$\times$Solar] & C/O [$\times$Solar] & Redistribution Factor & Cloud Parameters & Haze Scattering Parameters\\
        \hline
        PICASO & 10 & 1.5 & 0.5 (full redist.) & $\rm f_{sed}$=1, $\rm \log_{10} K_{zz}$=7 & ---\\
        ATMO & 5 & 1 & 0.5 (full redist.) & 0$\times$ grey cloud & 150$\times$ enhanced Rayleigh\\
        ScCHIMERA & 1 & 1 & 1.319 (partial redist.) & $\rm P_{cloud}$=1\,bar & $a$=1.78, $\gamma$=4\\
        \hline
    \end{tabular}
    \begin{tablenotes}
        \small
        \item \textbf{Notes}: Results are from fits to R=125 data. --- indicates that this aspect was not considered in the grid. 
        The ScCHIMERA results refer to model 2 described in the text. 
    \end{tablenotes}
\end{threeparttable}
\end{table*}

\subsection{PICASO}
\label{sec: picaso}

We use the modelling tool PICASO 3.0 \citep{batalha_exoplanet_2019,mukherjee_picasso22} to create a grid of model atmospheres in 1D radiative-convective equilibrium for WASP-96\,b. PICASO is based on the legacy "Extrasolar Giant Planet (EGP)" code \citep{Marley_2021_sonora, Fortney_2005, Marley_1999} and uses the opacities listed in \cite{Marley_2021_sonora}. We generate two sets of models: one cloud-free and one with clouds. Both sets contain grids of models with atmospheric metallicities in the range (0.1, 0.3, 1, 3.1, 10, 31, 50, 100) times solar, and C/O ratios in the range (0.5, 1, 1.5, 2) times solar. We fix the energy redistribution between the day- and nightsides of the planet at 0.5, which represents full day-to-night heat redistribution in PICASO. Metallicity is computed by multiplying the abundances of elements heavier than hydrogen/helium by the appropriate factor. (C+O)/H ratio is held constant at a given metallicity so that changing the C/O ratio does not also change the metallicity. The chemistry grid for these metallicities and C/O ratios is computed using the thermocheical equilibrium models presented in \citet{gordon1994,Visscher2010} and also presented in \citet{Marley_2021_sonora}. We use 0.458 for the solar C/O value (\citealp{Lodders_2009_solar_co}). The cloudy grid additionally uses the Virga \citep{Rooney_2022_virgaupdate} implementation of the Eddysed \citep{Ackerman&Marley01} framework, with clouds parameterized by a vertical mixing coefficient $\log_{10}$K$_\textrm{zz}$ in the range (7, 9, 11; cgs units) and a sedimentation parameter $f_{\textrm{sed}}$ in the range (0.6, 1, 3, 6). This excludes grid edge points $\log_{10}$K$_\textrm{zz}=5$ and $f_{\textrm{sed}}=10$ used by the ERS collaboration previously in \citet{feinstein_early_2023}. A $\log_{10}$K$_\textrm{zz}=5$ is unphysically small at the expected temperatures of WASP-96\,b \citep[see Fig. 2 in][as well as references therein]{moses2022}, while $f_{\textrm{sed}}=10$ produces a compact, vertically thin cloud deck at pressures below the transmission contribution of our observations. Therefore, omitting these points does not affect the overall interpretation of WASP-96b. Both cloud parameters are taken to be constant with altitude. We account for cloud species MnS, MgSiO$_3$, Cr, Mg$_2$SiO$_4$, and Fe. Our grids use the orbital parameters described in \citet{hellier_transiting_2014} which yield an equilibrium temperature of 1291 K (for zero albedo). The best-fit internal temperature is 200 K, yielding a planetary effective temperature of 1491 K.

\subsection{ATMO}
\label{sec: atmo}

We also use a grid of self-consistent model atmospheres presented in \citet{Goyal2020} and previously applied to the WASP-96\,b VLT and \textit{HST} observations from \citet{nikolov_solar--supersolar_2022}. This model grid has been generated using the 1D-2D planetary atmosphere model ATMO \citep{Tremblin2015,Drummond2016,Goyal2018}. In this grid, model transmission spectra are generated using radiative-convective equilibrium pressure-temperature (P-T) profiles consistent with equilibrium chemistry. These spectra are generated at R$\sim$1000 for all the opacity species listed in Table 1 of the supplementary material of \citet{Goyal2020} using the correlated-\textit{k} methodology with random overlap \citep{Amundsen2017}. For this work, model transmission spectra are generated for a range of heat re-distribution parameters (0.25, 0.50, 0.75, 1.0), metallicities (0.1, 1, 3, 5, 10, 50, 100, 200$\times$ solar), C/O ratios (0.35, 0.55, 0.7, 0.75, 0.90, 1.0), haze scattering parameters (1, 10, 150, 550, 1100$\times$ multi-gas Rayleigh scattering) and grey cloud parameters (0, 0.5, 1.0, 5.0$\times$ H$_2$ Rayleigh scattering cross-section at 350\,nm). Here, a heat re-distribution parameter of 0.5 corresponds to efficient heat re-distribution. Radiative-convective equilibrium P-T profiles with four different re-distribution factors cover a range from hottest (1) to coolest (0.25) P-T profiles that could be encountered in the planet's atmosphere. The solar C/O ratio is 0.55, and a grey cloud parameter of 0 implies cloud free model spectra. The metallicity is perturbed in the model by multiplying the abundances of all the elements heavier than H and He by the appropriate metallicity factor. C/O ratio is varied in the model via the O/H ratio.

\subsection{ScCHIMERA}
\label{sec: chimera}

Detailed descriptions of the ScCHIMERA framework are included in the previous works of \citet{Arcangeli2018}, \citet{Piskorz2018}, and \citet{mansfield21}, and more recently in applications to JWST data in \citet{jwst_transiting_exoplanet_community_early_release_science_team_identification_2022} and \citet{feinstein_early_2023}. Generally, for a set of planetary parameters, this method pre-computes the P-T structure of the planet and the gas mixing ratio profiles under thermochemical equilibrium. The temperature pressure profile is computed under the assumption of radiative-convective equilibrium. The layer net fluxes are computed using the \citet{Toon1989} two stream source function technique and a Newton-Raphson iteration \citep{McKay1989} is used to march to convergence. Opacities are derived from correlated-K tables mixed on-the-fly using the random-overlap `resort-rebin' procedure \citep[e.g.,][]{Amundsen2016}. The gas mixing ratios used to weigh the mixed correlated-K opacities are derived under the assumption of thermochemical equilibrium using the NASA CEA2 routine \citep{gordon1994}. While the calculation of the grid models fixes the internal temperature to 150K, we find that this value has no significant impact on the resulting P-T profile given the assumption of chemical equilibrium \citep[e.g.,][]{Fortney2020}. The computations are performed on a grid of atmospheric metallicity ([M/H], i.e., $\log_{10}$ enrichment relative to solar; \citet{lodders2009}) spaced at 0.125 dex values between -1 and 2.5 (i.e., 0.1 to 316$\times$ solar) and C/O at values of 0.1, 0.2, 0.35, 0.45, 0.55, 0.65, 0.7, 0.725, 0.75, 0.775, 0.8, 0.85, 0.9, 0.95 (where 0.55 is solar). The grid also explores the energy redistribution ($f$) between the day and night sides of the planet \citep{fortney05}, with values of 0.657, 0.721, 0.791, 0.865, 1.0, 1.03, 1.12, 1.217, 1.319 in our grid, where $f$=1.0 corresponds to full day-to-night heat redistribution and $f$=2.0 corresponds to dayside only redistribution.

The P-T structure and volume mixing ratio profiles of the different chemical species are then used to compute the transmission spectrum of the planet with CHIMERA \citep{Line2013,iyer&line2020,Mai&Line19} using correlated-$k$ tables at a resolution of R=3000. The computed spectra are then compared to the observations using the Bayesian inference framework MultiNest \citep{MULTINEST} through its Python implementation PyMultiNest \citep{PyMultiNest} to obtain an optimal (e.g., best-fit) solution for the atmospheric metallicity, C/O ratio, and heat redistribution. We also account for slight changes in the planetary radius by fitting for the planet radius at 1 bar, an arbitrary pressure with no direct impact on the inferred atmospheric properties \citep[e.g.,][]{Welbanks2019}. Our atmospheric model considers the opacity sources expected to affect gas giant planets \citep[e.g.,][]{Madhusudhan2019}, including H$_2$-H$_2$ and H$_2$-He CIA \cite[][]{Richard2012} alongside H$_2$O \citep{Polyansky+18, Freedman2014}, CO$_2$ \citep[][]{Freedman2014}, CO \citep{Rothman+10}, CH$_4$ \citep{Rothman+10}, H$_2$S \citep{Azzam2016}, HCN \citep{Barber2014},  Na \cite[]{Kramida2018, Allard2019}, and K \citep[]{Kramida2018, Allard+16}. The opacities were computed following the methods described in \cite{Gharib2021, Grimm2021}.

Our atmospheric models also consider different cloud and haze treatments to explore the degree of cloudiness of WASP-96\,b. We follow the model framework in \citet{feinstein_early_2023} and consider three base models:
(1) a vertically distributed cloud opacity that is spatially uniform and gray (i.e., a single parameter $\kappa_{\rm{cloud}}$ which describes the opacity at all wavelengths).
(2) A parameterization for scattering hazes following \citet{Lecavelier+08}. That is, one parameter, $\gamma$, for the scattering slope and another, $a$, for the Rayleigh enhancement factor such that $\sigma_{\mathrm{hazes}}=a\sigma_0(\lambda/\lambda_0)^\gamma$. Here, $\sigma_0$ is the H$_2$ Rayleigh cross section at $\lambda_0$, given by 2.3$\times$10$^{-27}$ cm$^2$ and 430\,nm respectively. This model also includes an optically thick cloud deck at a given atmospheric pressure, $\mathrm{P}_\mathrm{cloud}$ \citep{macdonald_hd_2017,Welbanks2021}. 
(3) A physically motivated droplet sedimentation model assuming enstatite grains with parameters to capture the behaviour of the eddy diffusion coefficient and the ratio of sedimentation velocity to the vertical mixing velocity \citep{Ackerman&Marley01} following the prescription of  \citet{Mai&Line19}. This case has a total of four parameters: P$_\mathrm{base}$ the cloud-base pressure, $f_\mathrm{cond.}$ for the condensate mixing ratio at the cloud base, $K_{zz}$ for the Eddy diffusion coefficient, and $f_\mathrm{sed}$ for the ratio of sedimentation velocity to characteristic vertical mixing velocity. 
Furthermore, models 2 and 3 are considered as either fully cloudy or with inhomogeneous cloud cover by using the linear combination approach of \citet{line16}.

\section{Atmospheric Inference Results}
\label{sec:Grid_Results}

\begin{figure*}
	\centering
	\includegraphics[width=\textwidth]{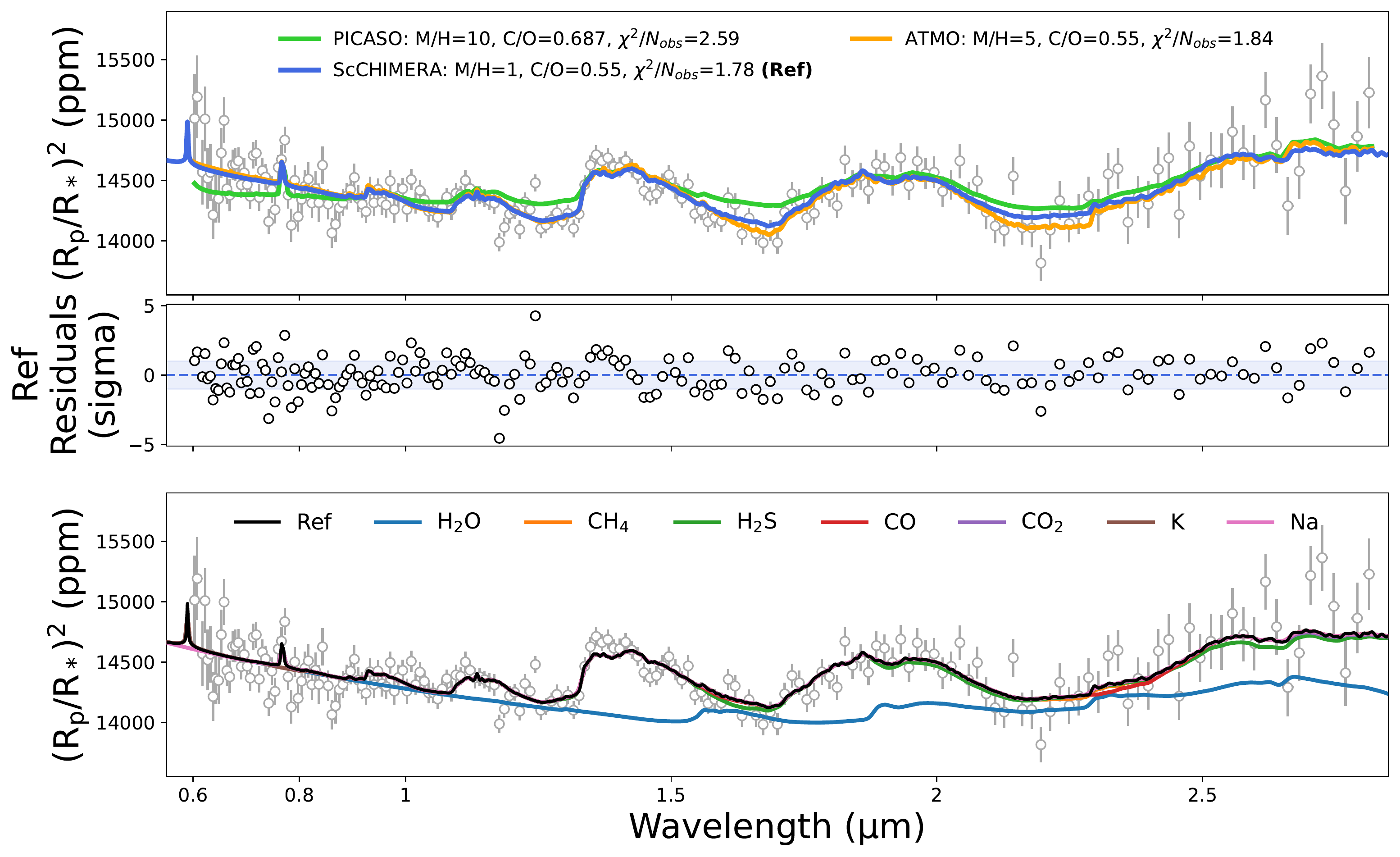}
    \caption{Grid modelling results of the WASP-96\,b spectrum at R=125. \emph{Top}: Best fitting models from the PICASO, ATMO, and ScCHIMERA grids overplotted with the R=125 transmission spectrum. The ATMO and ScCHIMERA models prefer cloud-free solar-to-super-solar metallicity transmission spectra, with a solar C/O ratio and an enhanced Rayleigh scattering haze. The PICASO model prefers a super-solar metallicity and C/O ratio, as well as some grey cloud opacity --- however this is likely do be due to the fact that an enhanced scattering parameter was not included in the PICASO grids.  
    \emph{Middle}: Residuals (as (data - model) / error) to the reference ScCHIMERA model.
    \emph{Bottom}: Impact on the transmission spectrum of removing the contributions of individual chemical species. H$_2$O opacity dominates the spectrum over nearly the entire wavelength range, with a model preference \citep[e.g., Bayesian evidence comparison, also known as `detection significance'][]{benneke_how_2013, Welbanks2021} of $\sim$17$\sigma$. The alkalis Na and K are marginally detected at $\sim$2$\sigma$ and $\sim$3$\sigma$, respectively. There are also hints of contributions from CO, CO$_2$ and H$_2$S with marginal model preference of $\sim$2$\sigma$. 
    \label{fig:Best Fit Models}}
\end{figure*}

The best fit atmosphere models to the NIRISS/SOSS WASP-96\,b transmission spectrum at R=125 are shown in Figure~\ref{fig:Best Fit Models}. The best fit solution found by PICASO ($\chi^2_\nu = 2.59$) suggests the atmosphere of WASP-96\,b is best explained by a 10$\times$ solar metallicity atmosphere, a slightly super-solar C/O ratio of 0.687, and a preference for cloudy models in agreement with \cite{samra_clouds_2023}. However, this preference for clouds over a cloud-free transmission spectrum may be driven by the lack of enhanced scattering opacity included in this particular grid of PICASO models. This grid only included self-consistent parameterized clouds from Virga with physically driven optical properties and particle sizes. As a result, the cloudy models within the PICASO grid that generated sufficient scattering to match the short-wavelength slope from NIRISS/SOSS also overly suppressed the strength of molecular features at longer wavelengths. Therefore, the best-fitting model from the PICASO grid results in a model that compromises between the short and long wavelengths while fitting neither, resulting is a poorer fit, suggesting scattering particles not captured by the Virga clouds need to be introduced to explain the spectrum. On account of this, we do not consider the PICASO results in the remainder of our discussion.

The results from the ATMO grid provide a comparatively better fit ($\chi^2_\nu = 1.84$). The ATMO best-fit suggests a metallicity of 5$\times$ solar and a solar C/O ratio, and also requires haze scattering as well as very low cloud opacity to explain the observations. The best fit ($\chi^2_\nu = 1.78$) across all grids was obtained by ScCHIMERA, which suggests a 1$\times$ solar metallicity, and solar C/O ratio. Across all resolutions, the ScCHIMERA grid finds that the best-fit solutions are largely consistent with a solar composition, and correspond to the class of models with a power-law for the scattering hazes and an optically thick cloud deck (i.e., model 2 described above), which are strongly preferred ($\sim$12$\sigma$) over the uniform vertically distributed cloud deck model (i.e., model 1). When using this model for clouds and hazes, only the scattering slope of the hazes has an impact on the spectrum and the optically thick deck cloud is placed deep in the atmosphere where it does not significantly affect the spectrum. That is, we find that the power law model for hazes best explains the spectrum, with no need for an optically thick cloud deck or inhomogeneous clouds. The median retrieved haze parameters are $\log_{10}(a)=1.78$ and $\gamma=4$, corresponding to enhanced Rayleigh-like scattering. The cloud deck pressure is constrained to pressures below the photosphere (e.g., P$\rm _{cloud}>1$~bar). This will be referred to as our ``reference model'' for some further tests described below. Overall, these results suggest that WASP-96\,b's terminator region is mostly cloud-free at the pressures probed here with possible enhanced scattering due to small particle hazes. Furthermore, the consideration of inhomogeneous clouds and hazes does not significantly impact our results.

Assuming that both ScCHIMERA and ATMO provide an equally plausible scenario (given their similar $\chi^2_{\nu}$ values) for the atmospheric composition of WASP-96\,b, our analysis suggests an atmospheric metallicity of 1--5$\times$ solar and a solar C/O ratio. Across both models, we find that there is a need for enhanced Rayleigh scattering to explain the blueward transit depths. However, we do not find strong evidence for optically thick clouds.

Our inferences on the metallicity and C/O ratio of WASP-96\,b are enabled by the large spectral features present in the observations. The bottom panel of Figure~\ref{fig:Best Fit Models} shows the contribution to the spectrum by different molecules and atoms. The main absorber in our WASP-96\,b transmission spectrum is H$_2$O, showing three clearly visible molecular bands. There are also signatures of absorption due to K near the known doublet peaks at $\sim$0.76\,µm. The Na absorption feature seen by \citet{nikolov_absolute_2018} is not resolved in these observations, due to lack of wavelength coverage, but, as mentioned previously, we see a slope in the bluest wavelengths. Our models fit this slope with enhanced Rayleigh scattering, though we note that this could indeed be the red wing of a highly-broadened Na feature. 

We therefore investigated the robustness of our inferences against this blueward slope. Given that these observations alone cannot robustly identify the nature of this slope, we aim to establish whether it has an impact on our inferred metallicity. We refit the R=125 observations using the ScCHIMERA grid without the data shorter than 1\,µm, finding a consistent inference of a 1$\times$ solar metallicity with a solar C/O ratio. We repeat this exercise for all other resolutions and find relatively consistent results, with an average best-fit metallicity across all five model configurations with ScCHIMERA (i.e., three cloud/haze models + two inhomogeneous cloud/haze models) and the four resolutions tested (i.e., R=125, 250, 500, pixel-level) of 2$\times$ solar with a standard deviation of 1$\times$ solar. The larger average is driven by the inferences at the pixel-level, which prefer metallicities of $\sim$4$\times$ solar across most model configurations. For the same model/resolution combinations, when not considering the observations blue-wards of 1\,µm, the average C/O is 0.46 with a standard deviation of 0.17; values consistent with solar expectations. This degeneracy will be further explored in the retrieval analyses of \citet{Taylor_w96_2023}, but we note that jointly fitting NIRISS/SOSS transmission spectra with ground based transmission measurements, such as in this case the existing VLT/FORS2 observations, may be necessary to fully constrain Na and haze scattering properties. 

Finally, some unresolved hints of carbon- (e.g., CO and CO$_2$ both with a model preference of $\sim$2$\sigma$) and/or sulfur- (e.g., H$_2$S with a model preference of $\sim$2$\sigma$) bearing species may be present near the 2.5\,µm feature in the spectrum (under the assumption of chemical equilibrium). 

\begin{figure*}
	\centering
	\includegraphics[width=\textwidth]{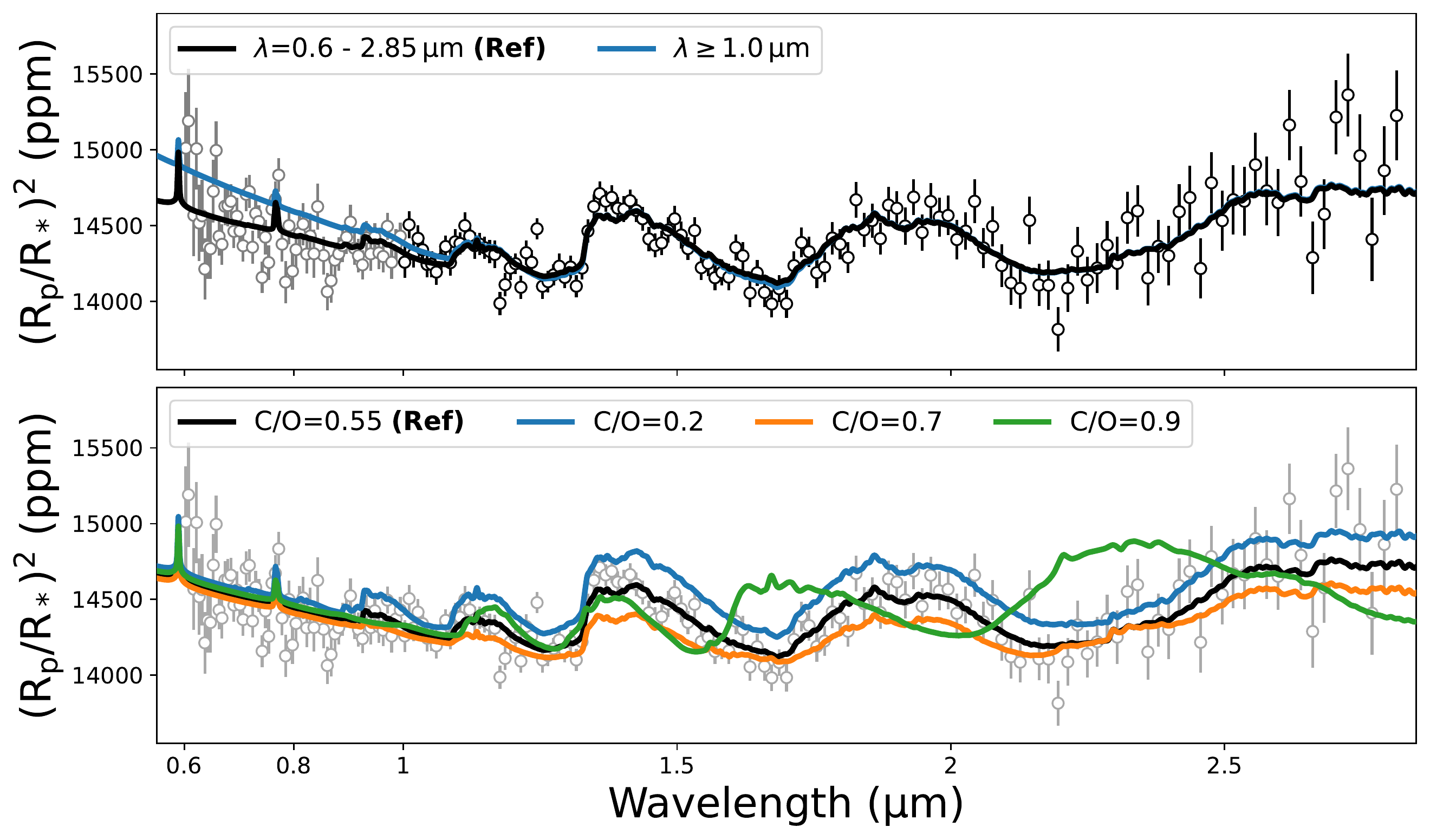}
    \caption{\emph{Top}: Comparison of model fits to different wavelength ranges. A fit to the data for wavelengths $\lambda \gtrsim1.0$\,µm (blue) is compared to the reference ScCHIMERA model (black) fit to the entire wavelength range. At wavelengths $\lambda \gtrsim1.0$\,µm, the two fits are entirely consistent and yield identical C/) and metallicity, showing that the haze slope detected at shorter wavelengths is not biasing the other atmospheric inferences.
    \emph{Bottom}: Investigation of the effects of C/O on the transmission spectrum. All model parameters are fixed to those from the reference ScCHIMERA model, except for the C/O ratio, which is allowed to vary. Four cases are compared here: 0.2, 0.55 (solar: best fit), 0.7, and 0.9. Super-solar C/O ratios result in large changes to the shape of the transmission spectrum (e.g., the emergence of CH$_4$, which we do not see in the data), allowing us to strongly rule out a super-solar C/O ratio for WASP-96\,b.
    \label{fig:CO Variation}}
\end{figure*}

Additionally, we explored the reliability of our solar C/O ratio inference. We fixed the best-fit parameters from ScCHIMERA to the R=125 observations (with the exception of the C/O ratio) and investigated the impact of changing the best-fit C/O ratio to sub-solar and super-solar values, while fixing the atmospheric metallicity and cloud/haze properties to their best-fitting values. We find that super-solar C/O ratios (i.e., 0.9) are incompatible with existing observations due to the expected signatures of carbon-bearing species, such as CH$_4$, which are not seen with NIRISS/SOSS. Sub-solar C/O ratios provide a worse fit by increasing the size of the observed H$_2$O features. A more robust statistical inference on the atmospheric C/O ratio and metallicity will be possible with the more detailed retrieval study that follows our investigation \citep{Taylor_w96_2023}. Figure \ref{fig:CO Variation} shows the results of our C/O ratio investigation and the reliability of our metallicity inferences for a partial fit to the observations explained above.

Our grid models find an atmospheric composition in good agreement with previous retrieval studies. The chemical equilibrium retrieval of \citet{mcgruder_access_2022}, using the full ensemble of all pre-JWST data, found a metallicity of $Z/Z_\odot = 0.32^{+2.91}_{-0.20}$, which is consistent with our best fitting metallicity at the 1$\sigma$ level. The free- and equilibrium chemistry retrievals of \citet{yip_compatibility_2021} and \citet{nikolov_solar--supersolar_2022} respectively, both point to solar-to-super-solar abundances of Na and O which, if extrapolated to a full atmosphere metallicity, are again consistent with our findings.  

\begin{figure}
	\centering
	\includegraphics[width=\columnwidth]{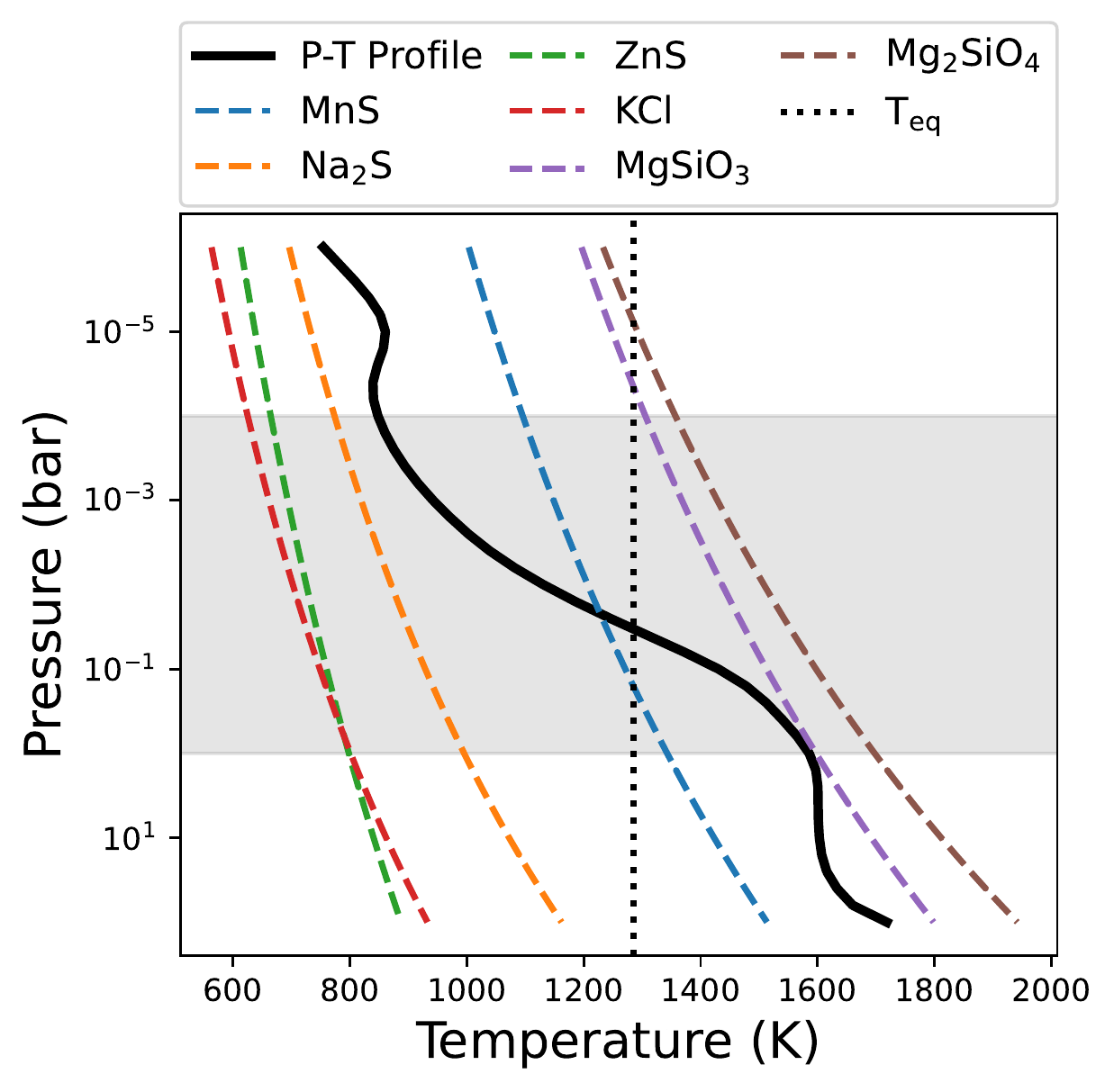}
    \caption{P-T profile from the ScCHIMERA grid (black) along with condensation curves for prominent cloud species from \citet{Visscher2010} and \citet{Morley_cloud_2012}. The shaded gray region shows the approximate pressure region of the photosphere \citep[see e.g.,][]{welbanks_massmetallicity_2019}. Our best-fitting grid models find no evidence for an optically thick grey cloud deck in the observable photosphere of WASP-96\,b despite the temperature structure not precluding conditions favourable for cloud condensation.
    \label{fig:PT Profiles}}
\end{figure}

Moreover, previous observations of this target have pointed to an atmosphere free from optically thick clouds \citep[e.g.,][]{nikolov_absolute_2018, yip_compatibility_2021, nikolov_solar--supersolar_2022, mcgruder_access_2022} --- in large part due to the pressure-broadened Na wings visible at optical wavelengths. The microphysical cloud models of \citet{samra_clouds_2023}, however, predict a homogeneous cloud coverage of WASP-96\,b's terminator. In Figure~\ref{fig:PT Profiles}, we show the P-T profile associated with the reference ScCHIMERA model, along with condensation curves for several prominent cloud condensates \citep{Visscher2010, Morley_cloud_2012}. The shaded region shows the pressure range generally probed by the photosphere \citep[e.g.,][]{welbanks_massmetallicity_2019, Welbanks2022}. The best-fit model has a P-T profile consistent with the planet's equilibrium temperature of $\sim1285$~K. Comparing the P-T profile to the condensation curves from \citep{Visscher2010, Morley_cloud_2012}, our results do not preclude the formation of some cloud species. However, with the expanded wavelength range of NIRISS/SOSS, our transmission spectrum still shows no strong preference for optically thick clouds in the observable photosphere. We do, though, find a slope at the bluest wavelengths, which our models fit with an enhanced Rayleigh scattering haze, though we note that such a slope could also potentially be the red wing of the highly-broadened Na feature. Without sampling the wavelengths corresponding to the peak of the Na feature, these two scenarios are difficult to disentangle with SOSS observations alone, and jointly fitting the NIRISS/SOSS and VLT/FORS2 spectra will likely be a fruitful avenue to break this degeneracy. In this initial work, we also do not explicitly explore microphysical clouds models, which may potentially be able to help explain the blue-wavelength slope while having minimal opacity at redder wavelengths. 

\section{Summary and Discussion}
\label{sec: Conclusions}

In this work, we have presented transmission spectroscopy observations of the hot-Saturn WASP-96\,b, taken with NIRISS/SOSS as part of the JWST Early Release Observations Program. As this is one of the very first SOSS datasets to be observed, we present a detailed walkthrough of the reductions for this new instrument, paying special attention to background subtraction and the correction of 1/$f$ noise. We further suggest and implement strategies to mitigate contamination due to order 0 and order 1 traces of field stars. 

We compare our transmission spectrum to grid models generated with the PICASO, ATMO, and ScCHIMERA codes. Overall, our grid fits suggest an atmosphere metallicity in the range of 1 -- 5$\times$ solar, a solar C/O ratio, and a cloud-free upper atmosphere in the terminator region of WASP-96\,b, which is in agreement with previous works using \textit{HST} and ground-based observations. We also identify a strong slope towards bluer wavelengths, which may be either the red wing of a highly broadened Na feature or enhanced Rayleigh scattering from small particles in the atmosphere. This serves as a preliminary glimpse into the atmosphere of WASP-96\,b, whose nature will be explored in more depth in \citet{Taylor_w96_2023}. 

The SOSS mode is slated to observe numerous exoplanet targets in Cycle 1, and will certainly continue to be a workhorse instrument, especially for observations of small, temperate worlds, throughout Cycle 2 and beyond. We therefore hope that this work will provide a useful reference to the community and aid in understanding this novel and awesome observing mode.

\section*{Acknowledgements}
MR would like to acknowledge funding from the National Sciences and Research Council of Canada (NSERC), the Fonds de Recherche du Qu\'ebec - Nature et Technologies (FRQNT), and the Institut Trottier de recherche sur les exoplanètes (iREx). He would also like to thank A, T, U, and Th for numerous helpful conversations and songs. DJ is supported by NRC Canada and by an NSERC Discovery Grant. LD acknowledges support from the Banting Postdoctoral Fellowship program, administered by the Government of Canada. RA is a Trottier Postdoctoral Fellow and acknowledges support from the Trottier Family Foundation. LW and RJM acknowledge support provided by NASA through NASA Hubble Fellowships awarded by the Space Telescope Science Institute, which is operated by the Association of Universities for Research in Astronomy, Inc., for NASA, under contract NAS5-26555. This project was undertaken with the financial support of the Canadian Space Agency. This work is based on observations made with the NASA/ESA/CSA JWST. The data were obtained from the Mikulski Archive for Space Telescopes at the Space Telescope Science Institute, which is operated by the Association of Universities for Research in Astronomy, Inc., under NASA contract NAS 5-03127 for JWST.

\section*{Software}
\begin{itemize}
    \renewcommand\labelitemi{--}
    \item \texttt{astropy}; \citet{astropy:2013, astropy:2018}
    \item \texttt{batman}; \citet{kreidberg_batman_2015}
    \item \texttt{dynesty}; \citet{speagle_dynesty_2020}
    \item \texttt{ExoTiC-LD}; \citet{laginja_exotic-ism_2020}
    \item \texttt{ipython}; \citet{PER-GRA:2007}
    \item \texttt{juliet}; \citet{espinoza_juliet_2019}
    \item \texttt{matplotlib}; \citet{Hunter:2007}
    \item \texttt{numpy}; \citet{harris2020array}
    \item \texttt{PyMultiNest}; \citet{PyMultiNest}
    \item \texttt{scipy}; \citet{2020SciPy-NMeth}
\end{itemize}

\section*{Data Availability}
All data used in this study is publicly available from the Barbara A. Mikulski Archive for Space Telescopes\footnote{https://mas    t.stsci.edu/portal/Mashup/Clients/Mast/Portal.html}.

\bibliography{W96b.bib}

\begin{thebibliography}{}
\expandafter\ifx\csname natexlab\endcsname\relax\def\natexlab#1{#1}\fi
\providecommand{\url}[1]{\href{#1}{#1}}
\providecommand{\dodoi}[1]{doi:~\href{http://doi.org/#1}{\nolinkurl{#1}}}
\providecommand{\doeprint}[1]{\href{http://ascl.net/#1}{\nolinkurl{http://ascl.net/#1}}}
\providecommand{\doarXiv}[1]{\href{https://arxiv.org/abs/#1}{\nolinkurl{https://arxiv.org/abs/#1}}}

\bibitem[{{Ackerman} \& {Marley}(2001)}]{Ackerman&Marley01}
{Ackerman}, A.~S., \& {Marley}, M.~S. 2001, \apj, 556, 872,
  \dodoi{10.1086/321540}

\bibitem[{Ahrer {et~al.}(2023)Ahrer, Stevenson, Mansfield, Moran, Brande,
  Morello, Murray, Nikolov, Petit dit de~la Roche, Schlawin, Wheatley, Zieba,
  Batalha, Damiano, Goyal, Lendl, Lothringer, Mukherjee, Ohno, Batalha,
  Battley, Bean, Beatty, Benneke, Berta-Thompson, Carter, Cubillos, Daylan,
  Espinoza, Gao, Gibson, Gill, Harrington, Hu, Kreidberg, Lewis, Line,
  López-Morales, Parmentier, Powell, Sing, Tsai, Wakeford, Welbanks, Alam,
  Alderson, Allen, Anderson, Barstow, Bayliss, Bell, Blecic, Bryant, Burleigh,
  Carone, Casewell, Changeat, Chubb, Crossfield, Crouzet, Decin, Désert,
  Feinstein, Flagg, Fortney, Gizis, Heng, Iro, Kempton, Kendrew, Kirk, Knutson,
  Komacek, Lagage, Leconte, Lustig-Yaeger, MacDonald, Mancini, May, Mayne,
  Miguel, Mikal-Evans, Molaverdikhani, Palle, Piaulet, Rackham, Redfield,
  Rogers, Roy, Rustamkulov, Shkolnik, Sotzen, Taylor, Tremblin, Tucker, Turner,
  de~Val-Borro, Venot, \& Zhang}]{ahrer_early_2023}
Ahrer, E.-M., Stevenson, K.~B., Mansfield, M., {et~al.} 2023, Nature,
  \dodoi{10.1038/s41586-022-05590-4}

\bibitem[{{Alam} {et~al.}(2021){Alam}, {L{\'o}pez-Morales}, {MacDonald},
  {Nikolov}, {Kirk}, {Goyal}, {Sing}, {Wakeford}, {Rathcke}, {Deming},
  {Sanz-Forcada}, {Lewis}, {Barstow}, {Mikal-Evans}, \& {Buchhave}}]{Alam2021}
{Alam}, M.~K., {L{\'o}pez-Morales}, M., {MacDonald}, R.~J., {et~al.} 2021,
  \apjl, 906, L10, \dodoi{10.3847/2041-8213/abd18e}

\bibitem[{{Albert} {et~al.}(2023){Albert}, {Lafreniere}, {Doyon}, {Artigau},
  {Volk}, {Goudfrooij}, {Martel}, {Radica}, {Rowe}, {Espinoza}, {Roy},
  {Filippazzo}, {Darveau-Bernier}, {Talens}, {Sivaramakrishnan}, {Willott},
  {Fullerton}, {LaMassa}, {Hutchings}, {Rowlands}, {Begona Vila}, {Zhou},
  {Aldridge}, {Maszkiewicz}, {Beaulieu}, {Cook}, {Piaulet}, {Roy},
  {Lamontagne}, {Morel}, {Frost}, {Salhi}, {Coulombe}, {Benneke}, {MacDonald},
  {Johnstone}, {Turner}, {Fournier-Tondreau}, {Allart}, \&
  {Kaltenegger}}]{Albert2023SOSS}
{Albert}, L., {Lafreniere}, D., {Doyon}, R., {et~al.} 2023, arXiv e-prints,
  arXiv:2306.04572, \dodoi{10.48550/arXiv.2306.04572}

\bibitem[{Alderson {et~al.}(2023)Alderson, Wakeford, Alam, Batalha, Lothringer,
  Redai, Barat, Brande, Damiano, Daylan, Espinoza, Flagg, Goyal, Grant, Hu,
  Inglis, Lee, Mikal-Evans, Ramos-Rosado, Roy, Wallack, Batalha, Bean, Benneke,
  Berta-Thompson, Carter, Changeat, Colón, Crossfield, Désert,
  Foreman-Mackey, Gibson, Kreidberg, Line, López-Morales, Molaverdikhani,
  Moran, Morello, Moses, Mukherjee, Schlawin, Sing, Stevenson, Taylor,
  Aggarwal, Ahrer, Allen, Barstow, Bell, Blecic, Casewell, Chubb, Crouzet,
  Cubillos, Decin, Feinstein, Fortney, Harrington, Heng, Iro, Kempton, Kirk,
  Knutson, Krick, Leconte, Lendl, MacDonald, Mancini, Mansfield, May, Mayne,
  Miguel, Nikolov, Ohno, Palle, Parmentier, Petit dit de~la Roche, Piaulet,
  Powell, Rackham, Redfield, Rogers, Rustamkulov, Tan, Tremblin, Tsai, Turner,
  de~Val-Borro, Venot, Welbanks, Wheatley, \& Zhang}]{alderson_early_2023}
Alderson, L., Wakeford, H.~R., Alam, M.~K., {et~al.} 2023, Nature,
  \dodoi{10.1038/s41586-022-05591-3}

\bibitem[{{Allard} {et~al.}(2016){Allard}, {Spiegelman}, \&
  {Kielkopf}}]{Allard+16}
{Allard}, N.~F., {Spiegelman}, F., \& {Kielkopf}, J.~F. 2016, \aap, 589, A21,
  \dodoi{10.1051/0004-6361/201628270}

\bibitem[{{Allard} {et~al.}(2019){Allard}, {Spiegelman}, {Leininger}, \&
  {Molliere}}]{Allard2019}
{Allard}, N.~F., {Spiegelman}, F., {Leininger}, T., \& {Molliere}, P. 2019,
  \aap, 628, A120, \dodoi{10.1051/0004-6361/201935593}

\bibitem[{{Amundsen} {et~al.}(2017){Amundsen}, {Tremblin}, {Manners},
  {Baraffe}, \& {Mayne}}]{Amundsen2017}
{Amundsen}, D.~S., {Tremblin}, P., {Manners}, J., {Baraffe}, I., \& {Mayne},
  N.~J. 2017, \aap, 598, A97, \dodoi{10.1051/0004-6361/201629322}

\bibitem[{{Amundsen} {et~al.}(2016){Amundsen}, {Mayne}, {Baraffe}, {Manners},
  {Tremblin}, {Drummond}, {Smith}, {Acreman}, \& {Homeier}}]{Amundsen2016}
{Amundsen}, D.~S., {Mayne}, N.~J., {Baraffe}, I., {et~al.} 2016, \aap, 595,
  A36, \dodoi{10.1051/0004-6361/201629183}

\bibitem[{{Arcangeli} {et~al.}(2018){Arcangeli}, {D{\'e}sert}, {Line}, {Bean},
  {Parmentier}, {Stevenson}, {Kreidberg}, {Fortney}, {Mansfield}, \&
  {Showman}}]{Arcangeli2018}
{Arcangeli}, J., {D{\'e}sert}, J.-M., {Line}, M.~R., {et~al.} 2018, \apjl, 855,
  L30, \dodoi{10.3847/2041-8213/aab272}

\bibitem[{{Astropy Collaboration} {et~al.}(2013){Astropy Collaboration},
  {Robitaille}, {Tollerud}, {Greenfield}, {Droettboom}, {Bray}, {Aldcroft},
  {Davis}, {Ginsburg}, {Price-Whelan}, {Kerzendorf}, {Conley}, {Crighton},
  {Barbary}, {Muna}, {Ferguson}, {Grollier}, {Parikh}, {Nair}, {Unther},
  {Deil}, {Woillez}, {Conseil}, {Kramer}, {Turner}, {Singer}, {Fox}, {Weaver},
  {Zabalza}, {Edwards}, {Azalee Bostroem}, {Burke}, {Casey}, {Crawford},
  {Dencheva}, {Ely}, {Jenness}, {Labrie}, {Lim}, {Pierfederici}, {Pontzen},
  {Ptak}, {Refsdal}, {Servillat}, \& {Streicher}}]{astropy:2013}
{Astropy Collaboration}, {Robitaille}, T.~P., {Tollerud}, E.~J., {et~al.} 2013,
  \aap, 558, A33, \dodoi{10.1051/0004-6361/201322068}

\bibitem[{{Astropy Collaboration} {et~al.}(2018){Astropy Collaboration},
  {Price-Whelan}, {Sip{\H{o}}cz}, {G{\"u}nther}, {Lim}, {Crawford}, {Conseil},
  {Shupe}, {Craig}, {Dencheva}, {Ginsburg}, {Vand erPlas}, {Bradley},
  {P{\'e}rez-Su{\'a}rez}, {de Val-Borro}, {Aldcroft}, {Cruz}, {Robitaille},
  {Tollerud}, {Ardelean}, {Babej}, {Bach}, {Bachetti}, {Bakanov}, {Bamford},
  {Barentsen}, {Barmby}, {Baumbach}, {Berry}, {Biscani}, {Boquien}, {Bostroem},
  {Bouma}, {Brammer}, {Bray}, {Breytenbach}, {Buddelmeijer}, {Burke},
  {Calderone}, {Cano Rodr{\'\i}guez}, {Cara}, {Cardoso}, {Cheedella}, {Copin},
  {Corrales}, {Crichton}, {D'Avella}, {Deil}, {Depagne}, {Dietrich}, {Donath},
  {Droettboom}, {Earl}, {Erben}, {Fabbro}, {Ferreira}, {Finethy}, {Fox},
  {Garrison}, {Gibbons}, {Goldstein}, {Gommers}, {Greco}, {Greenfield},
  {Groener}, {Grollier}, {Hagen}, {Hirst}, {Homeier}, {Horton}, {Hosseinzadeh},
  {Hu}, {Hunkeler}, {Ivezi{\'c}}, {Jain}, {Jenness}, {Kanarek}, {Kendrew},
  {Kern}, {Kerzendorf}, {Khvalko}, {King}, {Kirkby}, {Kulkarni}, {Kumar},
  {Lee}, {Lenz}, {Littlefair}, {Ma}, {Macleod}, {Mastropietro}, {McCully},
  {Montagnac}, {Morris}, {Mueller}, {Mumford}, {Muna}, {Murphy}, {Nelson},
  {Nguyen}, {Ninan}, {N{\"o}the}, {Ogaz}, {Oh}, {Parejko}, {Parley}, {Pascual},
  {Patil}, {Patil}, {Plunkett}, {Prochaska}, {Rastogi}, {Reddy Janga},
  {Sabater}, {Sakurikar}, {Seifert}, {Sherbert}, {Sherwood-Taylor}, {Shih},
  {Sick}, {Silbiger}, {Singanamalla}, {Singer}, {Sladen}, {Sooley},
  {Sornarajah}, {Streicher}, {Teuben}, {Thomas}, {Tremblay}, {Turner},
  {Terr{\'o}n}, {van Kerkwijk}, {de la Vega}, {Watkins}, {Weaver}, {Whitmore},
  {Woillez}, {Zabalza}, \& {Astropy Contributors}}]{astropy:2018}
{Astropy Collaboration}, {Price-Whelan}, A.~M., {Sip{\H{o}}cz}, B.~M., {et~al.}
  2018, \aj, 156, 123, \dodoi{10.3847/1538-3881/aabc4f}

\bibitem[{{Azzam} {et~al.}(2016){Azzam}, {Tennyson}, {Yurchenko}, \&
  {Naumenko}}]{Azzam2016}
{Azzam}, A. A.~A., {Tennyson}, J., {Yurchenko}, S.~N., \& {Naumenko}, O.~V.
  2016, \mnras, 460, 4063, \dodoi{10.1093/mnras/stw1133}

\bibitem[{{Barber} {et~al.}(2014){Barber}, {Strange}, {Hill}, {Polyansky},
  {Mellau}, {Yurchenko}, \& {Tennyson}}]{Barber2014}
{Barber}, R.~J., {Strange}, J.~K., {Hill}, C., {et~al.} 2014, \mnras, 437,
  1828, \dodoi{10.1093/mnras/stt2011}

\bibitem[{{Barstow} {et~al.}(2017){Barstow}, {Aigrain}, {Irwin}, \&
  {Sing}}]{Barstow2017}
{Barstow}, J.~K., {Aigrain}, S., {Irwin}, P.~G.~J., \& {Sing}, D.~K. 2017,
  \apj, 834, 50, \dodoi{10.3847/1538-4357/834/1/50}

\bibitem[{Batalha {et~al.}(2019)Batalha, Marley, Lewis, \&
  Fortney}]{batalha_exoplanet_2019}
Batalha, N.~E., Marley, M.~S., Lewis, N.~K., \& Fortney, J.~J. 2019, ApJ, 878,
  70, \dodoi{10.3847/1538-4357/ab1b51}

\bibitem[{Bean {et~al.}(2018)Bean, Stevenson, Batalha, Berta-Thompson,
  Kreidberg, Crouzet, Benneke, Line, Sing, Wakeford, Knutson, Kempton, Désert,
  Crossfield, Batalha, Wit, Parmentier, Harrington, Moses, Lopez-Morales, Alam,
  Blecic, Bruno, Carter, Chapman, Decin, Dragomir, Evans, Fortney, Fraine, Gao,
  Muñoz, Gibson, Goyal, Heng, Hu, Kendrew, Kilpatrick, Krick, Lagage, Lendl,
  Louden, Madhusudhan, Mandell, Mansfield, May, Morello, Morley, Nikolov,
  Redfield, Roberts, Schlawin, Spake, Todorov, Tsiaras, Venot, Waalkes,
  Wheatley, Zellem, Angerhausen, Barrado, Carone, Casewell, Cubillos, Damiano,
  Val-Borro, Drummond, Edwards, Endl, Espinoza, France, Gizis, Greene, Henning,
  Hong, Ingalls, Iro, Irwin, Kataria, Lahuis, Leconte, Lillo-Box, Lines,
  Lothringer, Mancini, Marchis, Mayne, Palle, Rauscher, Roudier, Shkolnik,
  Southworth, Swain, Taylor, Teske, Tinetti, Tremblin, Tucker, Boekel,
  Waldmann, Weaver, \& Zingales}]{bean_transiting_2018}
Bean, J.~L., Stevenson, K.~B., Batalha, N.~M., {et~al.} 2018, PASP, 130,
  114402, \dodoi{10.1088/1538-3873/aadbf3}

\bibitem[{Bell {et~al.}(2022)Bell, Ahrer, Brande, Carter, Feinstein, Caloca,
  Mansfield, Zieba, Piaulet, Benneke, Filippazzo, May, Roy, Kreidberg, \&
  Stevenson}]{bell_eureka_2022}
Bell, T.~J., Ahrer, E.-M., Brande, J., {et~al.} 2022, JOSS, 7, 4503,
  \dodoi{10.21105/joss.04503}

\bibitem[{Benneke(2015)}]{benneke_strict_2015}
Benneke, B. 2015, Strict {Upper} {Limits} on the {Carbon}-to-{Oxygen} {Ratios}
  of {Eight} {Hot} {Jupiters} from {Self}-{Consistent} {Atmospheric}
  {Retrieval},  arXiv.
\newblock \url{http://arxiv.org/abs/1504.07655}

\bibitem[{Benneke \& Seager(2012)}]{benneke_atmospheric_2012}
Benneke, B., \& Seager, S. 2012, ApJ, 753, 100,
  \dodoi{10.1088/0004-637X/753/2/100}

\bibitem[{Benneke \& Seager(2013)}]{benneke_how_2013}
---. 2013, ApJ, 778, 153, \dodoi{10.1088/0004-637X/778/2/153}

\bibitem[{Benneke {et~al.}(2019)Benneke, Wong, Piaulet, Knutson, Lothringer,
  Morley, Crossfield, Gao, Greene, Dressing, Dragomir, Howard, McCullough,
  Kempton, Fortney, \& Fraine}]{benneke_water_2019}
Benneke, B., Wong, I., Piaulet, C., {et~al.} 2019, ApJ, 887, L14,
  \dodoi{10.3847/2041-8213/ab59dc}

\bibitem[{Birkmann {et~al.}(2022)Birkmann, Ferruit, Giardino, Nielsen,
  García~Muñoz, Kendrew, Rauscher, Beck, Keyes, Valenti, Jakobsen, Dorner,
  Alves~de Oliveira, Arribas, Böker, Bunker, Charlot, de~Marchi, Kumari,
  López-Caniego, Lützgendorf, Maiolino, Manjavacas, Marston, Moseley,
  Prizkal, Proffitt, Rawle, Rix, te~Plate, Sabbi, Sirianni, Willott, \&
  Zeidler}]{birkmann_near-infrared_2022}
Birkmann, S.~M., Ferruit, P., Giardino, G., {et~al.} 2022, A\&A, 661, A83,
  \dodoi{10.1051/0004-6361/202142592}

\bibitem[{Boucher {et~al.}(2021)Boucher, Darveau-Bernier, Pelletier,
  Lafrenière, Artigau, Cook, Allart, Radica, Doyon, Benneke, Arnold, Bonfils,
  Bourrier, Cloutier, Gomes~da Silva, Deibert, Delfosse, Donati, Ehrenreich,
  Figueira, Forveille, Fouqué, Gagné, Gaidos, Hébrard, Jayawardhana, Klein,
  Lovis, Martins, Martioli, Moutou, \& Santos}]{boucher_characterizing_2021}
Boucher, A., Darveau-Bernier, A., Pelletier, S., {et~al.} 2021, AJ, 162, 233,
  \dodoi{10.3847/1538-3881/ac1f8e}

\bibitem[{{Boucher} {et~al.}(2023){Boucher}, {Lafreni{\'e}re}, {Pelletier},
  {Darveau-Bernier}, {Radica}, {Allart}, {Artigau}, {Cook}, {Debras}, {Doyon},
  {Gaidos}, {Benneke}, {Cadieux}, {Carmona}, {Cloutier}, {Cort{\'e}s-Zuleta},
  {Cowan}, {Delfosse}, {Donati}, {Fouqu{\'e}}, {Forveille}, {Grankin},
  {H{\'e}brard}, {Martins}, {Martioli}, {Masson}, \&
  {Vinatier}}]{boucher_co_2023}
{Boucher}, A., {Lafreni{\'e}re}, D., {Pelletier}, S., {et~al.} 2023, \mnras,
  522, 5062, \dodoi{10.1093/mnras/stad1247}

\bibitem[{Brogi {et~al.}(2014)Brogi, de~Kok, Birkby, Schwarz, \&
  Snellen}]{brogi_carbon_2014}
Brogi, M., de~Kok, R.~J., Birkby, J.~L., Schwarz, H., \& Snellen, I. A.~G.
  2014, A\&A, 565, A124, \dodoi{10.1051/0004-6361/201423537}

\bibitem[{{Buchner} {et~al.}(2014){Buchner}, {Georgakakis}, {Nandra}, {Hsu},
  {Rangel}, {Brightman}, {Merloni}, {Salvato}, {Donley}, \&
  {Kocevski}}]{PyMultiNest}
{Buchner}, J., {Georgakakis}, A., {Nandra}, K., {et~al.} 2014, \aap, 564, A125,
  \dodoi{10.1051/0004-6361/201322971}

\bibitem[{Charbonneau {et~al.}(2002)Charbonneau, Brown, Noyes, \&
  Gilliland}]{charbonneau_detection_2002}
Charbonneau, D., Brown, T.~M., Noyes, R.~W., \& Gilliland, R.~L. 2002, ApJ,
  568, 377, \dodoi{10.1086/338770}

\bibitem[{Claret(2000)}]{claret_new_2000}
Claret, A. 2000, A\&A, 363, 10

\bibitem[{Coulombe {et~al.}(2023)Coulombe, Benneke, Challener, Piette, Wiser,
  Mansfield, MacDonald, Beltz, Feinstein, Radica, Savel, Santos, Bean,
  Parmentier, Wong, Rauscher, Komacek, Kempton, Tan, Hammond, Lewis, Line, Lee,
  Shivkumar, Crossfield, Nixon, Rackham, Wakeford, Welbanks, Zhang, Batalha,
  Berta-Thompson, Changeat, Désert, Espinoza, Goyal, Harrington, Knutson,
  Kreidberg, López-Morales, Shporer, Sing, Stevenson, Aggarwal, Ahrer, Alam,
  Bell, Blecic, Caceres, Carter, Casewell, Crouzet, Cubillos, Decin, Fortney,
  Gibson, Heng, Henning, Iro, Kendrew, Lagage, Leconte, Lendl, Lothringer,
  Mancini, Mikal-Evans, Molaverdikhani, Nikolov, Ohno, Palle, Piaulet,
  Redfield, Roy, Tsai, Venot, \& Wheatley}]{coulombe_broadband_2023}
Coulombe, L.-P., Benneke, B., Challener, R., {et~al.} 2023, A broadband thermal
  emission spectrum of the ultra-hot {Jupiter} {WASP}-18b,  arXiv.
\newblock \url{http://arxiv.org/abs/2301.08192}

\bibitem[{Darveau-Bernier {et~al.}(2022)Darveau-Bernier, Albert, Talens,
  Lafrenière, Radica, Doyon, Cook, Rowe, Allart, Artigau, Benneke, Cowan,
  Dang, Espinoza, Johnstone, Kaltenegger, Lim, Pauly, Pelletier, Piaulet, Roy,
  Roy, Splinter, Taylor, \& Turner}]{darveau-bernier_atoca_2022}
Darveau-Bernier, A., Albert, L., Talens, G.~J., {et~al.} 2022, PASP, 134,
  094502, \dodoi{10.1088/1538-3873/ac8a77}

\bibitem[{Deming {et~al.}(2013)Deming, Wilkins, McCullough, Burrows, Fortney,
  Agol, Dobbs-Dixon, Madhusudhan, Crouzet, Desert, Gilliland, Haynes, Knutson,
  Line, Magic, Mandell, Ranjan, Charbonneau, Clampin, Seager, \&
  Showman}]{deming_infrared_2013}
Deming, D., Wilkins, A., McCullough, P., {et~al.} 2013, ApJ, 774, 17,
  \dodoi{10.1088/0004-637X/774/2/95}

\bibitem[{{Doyon} {et~al.}(2023){Doyon}, {Willott}, {Hutchings},
  {Sivaramakrishnan}, {Albert}, {Lafreniere}, {Rowlands}, {Begona Vila},
  {Martel}, {LaMassa}, {Aldridge}, {Artigau}, {Cameron}, {Chayer}, {Cook},
  {Cooper}, {Darveau-Bernier}, {Dupuis}, {Earnshaw}, {Espinoza}, {Filippazzo},
  {Fullerton}, {Gaudreau}, {Gawlik}, {Goudfrooij}, {Haley}, {Kammerer},
  {Kendall}, {Lambros}, {Ilinca Ignat}, {Maszkiewicz}, {McColgan}, {Morishita},
  {Ouellette}, {Pacifici}, {Philippi}, {Radica}, {Ravindranath}, {Rowe}, {Roy},
  {Saad}, {Sohn}, {Talens}, {Thatte}, {Taylor}, {Vandal}, {Volk}, {Wander},
  {Warner}, {Zheng}, {Zhou}, {Abraham}, {Beaulieu}, {Benneke}, {Ferrarese},
  {Johnstone}, {Kaltenegger}, {Meyer}, {Pipher}, {Rameau}, {Rieke}, {Salhi}, \&
  {Sawicki}}]{Doyon2023NIRISS}
{Doyon}, R., {Willott}, C.~J., {Hutchings}, J.~B., {et~al.} 2023, arXiv
  e-prints, arXiv:2306.03277.
\newblock \doarXiv{2306.03277}

\bibitem[{Dragomir {et~al.}(2020)Dragomir, Crossfield, Benneke, Wong, Daylan,
  Diaz, Deming, Molliere, Kreidberg, Jenkins, Berardo, Christiansen, Dressing,
  Gorjian, Kane, Mikal-Evans, Morales, Werner, Ricker, Vanderspek, Seager,
  Winn, Jenkins, Colón, Fong, Guerrero, Hesse, Osborn, E.~Rose, Smith, \&
  Ting}]{dragomir_spitzer_2020}
Dragomir, D., Crossfield, I. J.~M., Benneke, B., {et~al.} 2020, ApJ, 903, L6,
  \dodoi{10.3847/2041-8213/abbc70}

\bibitem[{{Drummond} {et~al.}(2016){Drummond}, {Tremblin}, {Baraffe},
  {Amundsen}, {Mayne}, {Venot}, \& {Goyal}}]{Drummond2016}
{Drummond}, B., {Tremblin}, P., {Baraffe}, I., {et~al.} 2016, \aap, 594, A69,
  \dodoi{10.1051/0004-6361/201628799}

\bibitem[{Espinoza {et~al.}(2019)Espinoza, Kossakowski, \&
  Brahm}]{espinoza_juliet_2019}
Espinoza, N., Kossakowski, D., \& Brahm, R. 2019, Monthly Notices of the Royal
  Astronomical Society, 490, 2262, \dodoi{10.1093/mnras/stz2688}

\bibitem[{Espinoza {et~al.}(2023)Espinoza, Úbeda, Birkmann, Ferruit, Valenti,
  Sing, Rustamkulov, Regan, Kendrew, Sabbi, Schlawin, Beatty, Albert, Greene,
  Nikolov, Karakla, Keyes, Alves~de Oliveira, Böker, Pena-Guerrero, Giardino,
  Kumari, Manjavacas, Proffitt, \& Rawle}]{espinoza_spectroscopic_2023}
Espinoza, N., Úbeda, L., Birkmann, S.~M., {et~al.} 2023, PASP, 135, 018002,
  \dodoi{10.1088/1538-3873/aca3d3}

\bibitem[{Evans {et~al.}(2016)Evans, Sing, Wakeford, Nikolov, Ballester,
  Drummond, Kataria, Gibson, Amundsen, \& Spake}]{evans_detection_2016}
Evans, T.~M., Sing, D.~K., Wakeford, H.~R., {et~al.} 2016, ApJ, 822, L4,
  \dodoi{10.3847/2041-8205/822/1/L4}

\bibitem[{Faedi {et~al.}(2011)Faedi, Barros, Anderson, Brown, Collier~Cameron,
  Pollacco, Boisse, Hébrard, Lendl, Lister, Smalley, Street, Triaud, Bento,
  Bouchy, Butters, Enoch, Haswell, Hellier, Keenan, Miller, Moulds, Moutou,
  Norton, Queloz, Santerne, Simpson, Skillen, Smith, Udry, Watson, West, \&
  Wheatley}]{faedi_wasp-39b_2011}
Faedi, F., Barros, S. C.~C., Anderson, D.~R., {et~al.} 2011, A\&A, 531, A40,
  \dodoi{10.1051/0004-6361/201116671}

\bibitem[{Feinstein {et~al.}(2023)Feinstein, Radica, Welbanks, Murray, Ohno,
  Coulombe, Espinoza, Bean, Teske, Benneke, Line, Rustamkulov, Saba, Tsiaras,
  Barstow, Fortney, Gao, Knutson, MacDonald, Mikal-Evans, Rackham, Taylor,
  Parmentier, Batalha, Berta-Thompson, Carter, Changeat, dos Santos, Gibson,
  Goyal, Kreidberg, López-Morales, Lothringer, Miguel, Molaverdikhani, Moran,
  Morello, Mukherjee, Sing, Stevenson, Wakeford, Ahrer, Alam, Alderson, Allen,
  Batalha, Bell, Blecic, Brande, Caceres, Casewell, Chubb, Crossfield, Crouzet,
  Cubillos, Decin, Désert, Harrington, Heng, Henning, Iro, Kempton, Kendrew,
  Kirk, Krick, Lagage, Lendl, Mancini, Mansfield, May, Mayne, Nikolov, Palle,
  Petit dit de~la Roche, Piaulet, Powell, Redfield, Rogers, Roman, Roy, Nixon,
  Schlawin, Tan, Tremblin, Turner, Venot, Waalkes, Wheatley, \&
  Zhang}]{feinstein_early_2023}
Feinstein, A.~D., Radica, M., Welbanks, L., {et~al.} 2023, Nature,
  \dodoi{10.1038/s41586-022-05674-1}

\bibitem[{{Feroz} {et~al.}(2009){Feroz}, {Hobson}, \& {Bridges}}]{MULTINEST}
{Feroz}, F., {Hobson}, M.~P., \& {Bridges}, M. 2009, \mnras, 398, 1601,
  \dodoi{10.1111/j.1365-2966.2009.14548.x}

\bibitem[{{Fortney}(2005{\natexlab{a}})}]{Fortney_2005}
{Fortney}, J.~J. 2005{\natexlab{a}}, \mnras, 364, 649,
  \dodoi{10.1111/j.1365-2966.2005.09587.x}

\bibitem[{{Fortney}(2005{\natexlab{b}})}]{fortney05}
---. 2005{\natexlab{b}}, \mnras, 364, 649,
  \dodoi{10.1111/j.1365-2966.2005.09587.x}

\bibitem[{{Fortney} {et~al.}(2020){Fortney}, {Visscher}, {Marley}, {Hood},
  {Line}, {Thorngren}, {Freedman}, \& {Lupu}}]{Fortney2020}
{Fortney}, J.~J., {Visscher}, C., {Marley}, M.~S., {et~al.} 2020, \aj, 160,
  288, \dodoi{10.3847/1538-3881/abc5bd}

\bibitem[{{Freedman} {et~al.}(2014){Freedman}, {Lustig-Yaeger}, {Fortney},
  {Lupu}, {Marley}, \& {Lodders}}]{Freedman2014}
{Freedman}, R.~S., {Lustig-Yaeger}, J., {Fortney}, J.~J., {et~al.} 2014, \apjs,
  214, 25, \dodoi{10.1088/0067-0049/214/2/25}

\bibitem[{Fu {et~al.}(2022)Fu, Espinoza, Sing, Lothringer, Dos~Santos,
  Rustamkulov, Deming, Kempton, Komacek, Knutson, Albert, Pontoppidan, Volk, \&
  Filippazzo}]{fu_water_2022}
Fu, G., Espinoza, N., Sing, D.~K., {et~al.} 2022, ApJL, 940, L35,
  \dodoi{10.3847/2041-8213/ac9977}

\bibitem[{{Gaia Collaboration} {et~al.}(2021){Gaia Collaboration}, {Brown},
  {Vallenari}, {Prusti}, {de Bruijne}, {Babusiaux}, {Biermann}, {Creevey},
  {Evans}, {Eyer}, {Hutton}, {Jansen}, {Jordi}, {Klioner}, {Lammers},
  {Lindegren}, {Luri}, {Mignard}, {Panem}, {Pourbaix}, {Randich}, {Sartoretti},
  {Soubiran}, {Walton}, {Arenou}, {Bailer-Jones}, {Bastian}, {Cropper},
  {Drimmel}, {Katz}, {Lattanzi}, {van Leeuwen}, {Bakker}, {Cacciari},
  {Casta{\~n}eda}, {De Angeli}, {Ducourant}, {Fabricius}, {Fouesneau},
  {Fr{\'e}mat}, {Guerra}, {Guerrier}, {Guiraud}, {Jean-Antoine Piccolo},
  {Masana}, {Messineo}, {Mowlavi}, {Nicolas}, {Nienartowicz}, {Pailler},
  {Panuzzo}, {Riclet}, {Roux}, {Seabroke}, {Sordo}, {Tanga}, {Th{\'e}venin},
  {Gracia-Abril}, {Portell}, {Teyssier}, {Altmann}, {Andrae}, {Bellas-Velidis},
  {Benson}, {Berthier}, {Blomme}, {Brugaletta}, {Burgess}, {Busso}, {Carry},
  {Cellino}, {Cheek}, {Clementini}, {Damerdji}, {Davidson}, {Delchambre},
  {Dell'Oro}, {Fern{\'a}ndez-Hern{\'a}ndez}, {Galluccio}, {Garc{\'\i}a-Lario},
  {Garcia-Reinaldos}, {Gonz{\'a}lez-N{\'u}{\~n}ez}, {Gosset}, {Haigron},
  {Halbwachs}, {Hambly}, {Harrison}, {Hatzidimitriou}, {Heiter},
  {Hern{\'a}ndez}, {Hestroffer}, {Hodgkin}, {Holl}, {Jan{\ss}en}, {Jevardat de
  Fombelle}, {Jordan}, {Krone-Martins}, {Lanzafame}, {L{\"o}ffler}, {Lorca},
  {Manteiga}, {Marchal}, {Marrese}, {Moitinho}, {Mora}, {Muinonen}, {Osborne},
  {Pancino}, {Pauwels}, {Petit}, {Recio-Blanco}, {Richards}, {Riello},
  {Rimoldini}, {Robin}, {Roegiers}, {Rybizki}, {Sarro}, {Siopis}, {Smith},
  {Sozzetti}, {Ulla}, {Utrilla}, {van Leeuwen}, {van Reeven}, {Abbas}, {Abreu
  Aramburu}, {Accart}, {Aerts}, {Aguado}, {Ajaj}, {Altavilla}, {{\'A}lvarez},
  {{\'A}lvarez Cid-Fuentes}, {Alves}, {Anderson}, {Anglada Varela}, {Antoja},
  {Audard}, {Baines}, {Baker}, {Balaguer-N{\'u}{\~n}ez}, {Balbinot}, {Balog},
  {Barache}, {Barbato}, {Barros}, {Barstow}, {Bartolom{\'e}}, {Bassilana},
  {Bauchet}, {Baudesson-Stella}, {Becciani}, {Bellazzini}, {Bernet}, {Bertone},
  {Bianchi}, {Blanco-Cuaresma}, {Boch}, {Bombrun}, {Bossini}, {Bouquillon},
  {Bragaglia}, {Bramante}, {Breedt}, {Bressan}, {Brouillet}, {Bucciarelli},
  {Burlacu}, {Busonero}, {Butkevich}, {Buzzi}, {Caffau}, {Cancelliere},
  {C{\'a}novas}, {Cantat-Gaudin}, {Carballo}, {Carlucci}, {Carnerero},
  {Carrasco}, {Casamiquela}, {Castellani}, {Castro-Ginard}, {Castro Sampol},
  {Chaoul}, {Charlot}, {Chemin}, {Chiavassa}, {Cioni}, {Comoretto}, {Cooper},
  {Cornez}, {Cowell}, {Crifo}, {Crosta}, {Crowley}, {Dafonte}, {Dapergolas},
  {David}, {David}, {de Laverny}, {De Luise}, {De March}, {De Ridder}, {de
  Souza}, {de Teodoro}, {de Torres}, {del Peloso}, {del Pozo}, {Delbo},
  {Delgado}, {Delgado}, {Delisle}, {Di Matteo}, {Diakite}, {Diener},
  {Distefano}, {Dolding}, {Eappachen}, {Edvardsson}, {Enke}, {Esquej}, {Fabre},
  {Fabrizio}, {Faigler}, {Fedorets}, {Fernique}, {Fienga}, {Figueras},
  {Fouron}, {Fragkoudi}, {Fraile}, {Franke}, {Gai}, {Garabato},
  {Garcia-Gutierrez}, {Garc{\'\i}a-Torres}, {Garofalo}, {Gavras}, {Gerlach},
  {Geyer}, {Giacobbe}, {Gilmore}, {Girona}, {Giuffrida}, {Gomel}, {Gomez},
  {Gonzalez-Santamaria}, {Gonz{\'a}lez-Vidal}, {Granvik},
  {Guti{\'e}rrez-S{\'a}nchez}, {Guy}, {Hauser}, {Haywood}, {Helmi}, {Hidalgo},
  {Hilger}, {H{\l}adczuk}, {Hobbs}, {Holland}, {Huckle}, {Jasniewicz},
  {Jonker}, {Juaristi Campillo}, {Julbe}, {Karbevska}, {Kervella}, {Khanna},
  {Kochoska}, {Kontizas}, {Kordopatis}, {Korn}, {Kostrzewa-Rutkowska},
  {Kruszy{\'n}ska}, {Lambert}, {Lanza}, {Lasne}, {Le Campion}, {Le Fustec},
  {Lebreton}, {Lebzelter}, {Leccia}, {Leclerc}, {Lecoeur-Taibi}, {Liao},
  {Licata}, {Lindstr{\o}m}, {Lister}, {Livanou}, {Lobel}, {Madrero Pardo},
  {Managau}, {Mann}, {Marchant}, {Marconi}, {Marcos Santos}, {Marinoni},
  {Marocco}, {Marshall}, {Martin Polo}, {Mart{\'\i}n-Fleitas}, {Masip},
  {Massari}, {Mastrobuono-Battisti}, {Mazeh}, {McMillan}, {Messina},
  {Michalik}, {Millar}, {Mints}, {Molina}, {Molinaro}, {Moln{\'a}r},
  {Montegriffo}, {Mor}, {Morbidelli}, {Morel}, {Morris}, {Mulone}, {Munoz},
  {Muraveva}, {Murphy}, {Musella}, {Noval}, {Ord{\'e}novic}, {Orr{\`u}},
  {Osinde}, {Pagani}, {Pagano}, {Palaversa}, {Palicio}, {Panahi}, {Pawlak},
  {Pe{\~n}alosa Esteller}, {Penttil{\"a}}, {Piersimoni}, {Pineau}, {Plachy},
  {Plum}, {Poggio}, {Poretti}, {Poujoulet}, {Pr{\v{s}}a}, {Pulone}, {Racero},
  {Ragaini}, {Rainer}, {Raiteri}, {Rambaux}, {Ramos}, {Ramos-Lerate}, {Re
  Fiorentin}, {Regibo}, {Reyl{\'e}}, {Ripepi}, {Riva}, {Rixon}, {Robichon},
  {Robin}, {Roelens}, {Rohrbasser}, {Romero-G{\'o}mez}, {Rowell}, {Royer},
  {Rybicki}, {Sadowski}, {Sagrist{\`a} Sell{\'e}s}, {Sahlmann}, {Salgado},
  {Salguero}, {Samaras}, {Sanchez Gimenez}, {Sanna}, {Santove{\~n}a},
  {Sarasso}, {Schultheis}, {Sciacca}, {Segol}, {Segovia}, {S{\'e}gransan},
  {Semeux}, {Shahaf}, {Siddiqui}, {Siebert}, {Siltala}, {Slezak}, {Smart},
  {Solano}, {Solitro}, {Souami}, {Souchay}, {Spagna}, {Spoto}, {Steele},
  {Steidelm{\"u}ller}, {Stephenson}, {S{\"u}veges}, {Szabados}, {Szegedi-Elek},
  {Taris}, {Tauran}, {Taylor}, {Teixeira}, {Thuillot}, {Tonello}, {Torra},
  {Torra}, {Turon}, {Unger}, {Vaillant}, {van Dillen}, {Vanel}, {Vecchiato},
  {Viala}, {Vicente}, {Voutsinas}, {Weiler}, {Wevers}, {Wyrzykowski}, {Yoldas},
  {Yvard}, {Zhao}, {Zorec}, {Zucker}, {Zurbach}, \& {Zwitter}}]{2021Gaia}
{Gaia Collaboration}, {Brown}, A.~G.~A., {Vallenari}, A., {et~al.} 2021, \aap,
  649, A1, \dodoi{10.1051/0004-6361/202039657}

\bibitem[{{Gharib-Nezhad} {et~al.}(2021){Gharib-Nezhad}, {Iyer}, {Line},
  {Freedman}, {Marley}, \& {Batalha}}]{Gharib2021}
{Gharib-Nezhad}, E., {Iyer}, A.~R., {Line}, M.~R., {et~al.} 2021, \apjs, 254,
  34, \dodoi{10.3847/1538-4365/abf504}

\bibitem[{Gordon \& McBride(1994)}]{gordon1994}
Gordon, S., \& McBride, B.~J. 1994, Computer program for calculation of complex
  chemical equilibrium compositions and applications. Part 1: Analysis, Tech.
  rep., NASA

\bibitem[{{Goyal} {et~al.}(2018){Goyal}, {Mayne}, {Sing}, {Drummond},
  {Tremblin}, {Amundsen}, {Evans}, {Carter}, {Spake}, {Baraffe}, {Nikolov},
  {Manners}, {Chabrier}, \& {Hebrard}}]{Goyal2018}
{Goyal}, J.~M., {Mayne}, N., {Sing}, D.~K., {et~al.} 2018, \mnras, 474, 5158,
  \dodoi{10.1093/mnras/stx3015}

\bibitem[{{Goyal} {et~al.}(2020){Goyal}, {Mayne}, {Drummond}, {Sing},
  {H{\'e}brard}, {Lewis}, {Tremblin}, {Phillips}, {Mikal-Evans}, \&
  {Wakeford}}]{Goyal2020}
{Goyal}, J.~M., {Mayne}, N., {Drummond}, B., {et~al.} 2020, \mnras, 498, 4680,
  \dodoi{10.1093/mnras/staa2300}

\bibitem[{{Grimm} {et~al.}(2021){Grimm}, {Malik}, {Kitzmann},
  {Guzm{\'a}n-Mesa}, {Hoeijmakers}, {Fisher}, {Mendon{\c{c}}a}, {Yurchenko},
  {Tennyson}, {Alesina}, {Buchschacher}, {Burnier}, {Segransan}, {Kurucz}, \&
  {Heng}}]{Grimm2021}
{Grimm}, S.~L., {Malik}, M., {Kitzmann}, D., {et~al.} 2021, \apjs, 253, 30,
  \dodoi{10.3847/1538-4365/abd773}

\bibitem[{Harris {et~al.}(2020)Harris, Millman, van~der Walt, Gommers,
  Virtanen, Cournapeau, Wieser, Taylor, Berg, Smith, Kern, Picus, Hoyer, van
  Kerkwijk, Brett, Haldane, del R{\'{i}}o, Wiebe, Peterson,
  G{\'{e}}rard-Marchant, Sheppard, Reddy, Weckesser, Abbasi, Gohlke, \&
  Oliphant}]{harris2020array}
Harris, C.~R., Millman, K.~J., van~der Walt, S.~J., {et~al.} 2020, Nature, 585,
  357, \dodoi{10.1038/s41586-020-2649-2}

\bibitem[{Hellier {et~al.}(2014)Hellier, Anderson, Cameron, Delrez, Gillon,
  Jehin, Lendl, Maxted, Pepe, Pollacco, Queloz, Ségransan, Smalley, Smith,
  Southworth, Triaud, Udry, \& West}]{hellier_transiting_2014}
Hellier, C., Anderson, D.~R., Cameron, A.~C., {et~al.} 2014, Monthly Notices of
  the Royal Astronomical Society, 440, 1982, \dodoi{10.1093/mnras/stu410}

\bibitem[{Helling {et~al.}(2023)Helling, Samra, Lewis, Calder, Hirst, Woitke,
  Baeyens, Carone, Herbort, \& Chubb}]{helling_exoplanet_2023}
Helling, C., Samra, D., Lewis, D., {et~al.} 2023, A\&A, 671, A122,
  \dodoi{10.1051/0004-6361/202243956}

\bibitem[{Hoeijmakers {et~al.}(2020)Hoeijmakers, Seidel, Pino, Kitzmann,
  Sindel, Ehrenreich, Oza, Bourrier, Allart, Gebek, Lovis, Yurchenko,
  Astudillo-Defru, Bayliss, Cegla, Lavie, Lendl, Melo, Murgas, Nascimbeni,
  Pepe, Ségransan, Udry, Wyttenbach, \& Heng}]{hoeijmakers_hot_2020}
Hoeijmakers, H.~J., Seidel, J.~V., Pino, L., {et~al.} 2020, A\&A, 641, A123,
  \dodoi{10.1051/0004-6361/202038365}

\bibitem[{Hunter(2007)}]{Hunter:2007}
Hunter, J.~D. 2007, Computing in Science \& Engineering, 9, 90,
  \dodoi{10.1109/MCSE.2007.55}

\bibitem[{Husser {et~al.}(2013)Husser, Wende-von Berg, Dreizler, Homeier,
  Reiners, Barman, \& Hauschildt}]{husser_new_2013}
Husser, T.-O., Wende-von Berg, S., Dreizler, S., {et~al.} 2013, A\&A, 553, A6,
  \dodoi{10.1051/0004-6361/201219058}

\bibitem[{{Iyer} \& {Line}(2020)}]{iyer&line2020}
{Iyer}, A.~R., \& {Line}, M.~R. 2020, \apj, 889, 78,
  \dodoi{10.3847/1538-4357/ab612e}

\bibitem[{{JWST Transiting Exoplanet Community Early Release Science Team}
  {et~al.}(2022){JWST Transiting Exoplanet Community Early Release Science
  Team}, Ahrer, Alderson, Batalha, Batalha, Bean, Beatty, Bell, Benneke,
  Berta-Thompson, Carter, Crossfield, Espinoza, Feinstein, Fortney, Gibson,
  Goyal, Kempton, Kirk, Kreidberg, López-Morales, Line, Lothringer, Moran,
  Mukherjee, Ohno, Parmentier, Piaulet, Rustamkulov, Schlawin, Sing, Stevenson,
  Wakeford, Allen, Birkmann, Brande, Crouzet, Cubillos, Damiano, Désert, Gao,
  Harrington, Hu, Kendrew, Knutson, Lagage, Leconte, Lendl, MacDonald, May,
  Miguel, Molaverdikhani, Moses, Murray, Nehring, Nikolov, Petit dit de~la
  Roche, Radica, Roy, Stassun, Taylor, Waalkes, Wachiraphan, Welbanks,
  Wheatley, Aggarwal, Alam, Banerjee, Barstow, Blecic, Casewell, Changeat,
  Chubb, Colón, Coulombe, Daylan, de~Val-Borro, Decin, Dos~Santos, Flagg,
  France, Fu, García~Muñoz, Gizis, Glidden, Grant, Heng, Henning, Hong,
  Inglis, Iro, Kataria, Komacek, Krick, Lee, Lewis, Lillo-Box, Lustig-Yaeger,
  Mancini, Mandell, Mansfield, Marley, Mikal-Evans, Morello, Nixon,
  Ortiz~Ceballos, Piette, Powell, Rackham, Ramos-Rosado, Rauscher, Redfield,
  Rogers, Roman, Roudier, Scarsdale, Shkolnik, Southworth, Spake, Steinrueck,
  Tan, Teske, Tremblin, Tsai, Tucker, Turner, Valenti, Venot, Waldmann,
  Wallack, Zhang, \&
  Zieba}]{jwst_transiting_exoplanet_community_early_release_science_team_identification_2022}
{JWST Transiting Exoplanet Community Early Release Science Team}, Ahrer, E.-M.,
  Alderson, L., {et~al.} 2022, Nature, \dodoi{10.1038/s41586-022-05269-w}

\bibitem[{Kipping(2013)}]{kipping_efficient_2013}
Kipping, D.~M. 2013, Monthly Notices of the Royal Astronomical Society, 435,
  2152, \dodoi{10.1093/mnras/stt1435}

\bibitem[{Kramida {et~al.}(2018)Kramida, {Yu.~Ralchenko}, Reader, \& {and NIST
  ASD Team}}]{Kramida2018}
Kramida, A., {Yu.~Ralchenko}, Reader, J., \& {and NIST ASD Team}. 2018, {NIST
  Atomic Spectra Database (ver. 5.6.1), [Online]. Available:
  {\tt{https://physics.nist.gov/asd}} [2019, February 6]. National Institute of
  Standards and Technology, Gaithersburg, MD.}

\bibitem[{Kreidberg(2015)}]{kreidberg_batman_2015}
Kreidberg, L. 2015, Publications of the Astronomical Society of the Pacific,
  127, 1161, \dodoi{10.1086/683602}

\bibitem[{Kreidberg {et~al.}(2015)Kreidberg, Line, Bean, Stevenson, Désert,
  Madhusudhan, Fortney, Barstow, Henry, Williamson, \&
  Showman}]{kreidberg_detection_2015}
Kreidberg, L., Line, M.~R., Bean, J.~L., {et~al.} 2015, ApJ, 814, 66,
  \dodoi{10.1088/0004-637X/814/1/66}

\bibitem[{Kreidberg {et~al.}(2018)Kreidberg, Line, Parmentier, Stevenson,
  Louden, Bonnefoy, Faherty, Henry, Williamson, Stassun, Beatty, Bean, Fortney,
  Showman, Désert, \& Arcangeli}]{kreidberg_global_2018}
Kreidberg, L., Line, M.~R., Parmentier, V., {et~al.} 2018, AJ, 156, 17,
  \dodoi{10.3847/1538-3881/aac3df}

\bibitem[{Laginja \& Wakeford(2020)}]{laginja_exotic-ism_2020}
Laginja, I., \& Wakeford, H. 2020, JOSS, 5, 2281, \dodoi{10.21105/joss.02281}

\bibitem[{{Lecavelier Des Etangs} {et~al.}(2008){Lecavelier Des Etangs},
  {Pont}, {Vidal-Madjar}, \& {Sing}}]{Lecavelier+08}
{Lecavelier Des Etangs}, A., {Pont}, F., {Vidal-Madjar}, A., \& {Sing}, D.
  2008, \aap, 481, L83, \dodoi{10.1051/0004-6361:200809388}

\bibitem[{{Line} \& {Parmentier}(2016)}]{line16}
{Line}, M.~R., \& {Parmentier}, V. 2016, \apj, 820, 78,
  \dodoi{10.3847/0004-637X/820/1/78}

\bibitem[{{Line} {et~al.}(2013){Line}, {Wolf}, {Zhang}, {Knutson}, {Kammer},
  {Ellison}, {Deroo}, {Crisp}, \& {Yung}}]{Line2013}
{Line}, M.~R., {Wolf}, A.~S., {Zhang}, X., {et~al.} 2013, \apj, 775, 137,
  \dodoi{10.1088/0004-637X/775/2/137}

\bibitem[{{Lodders} {et~al.}(2009{\natexlab{a}}){Lodders}, {Palme}, \&
  {Gail}}]{Lodders_2009_solar_co}
{Lodders}, K., {Palme}, H., \& {Gail}, H.~P. 2009{\natexlab{a}}, Landolt
  B\&ouml;rnstein, 4B, 712, \dodoi{10.1007/978-3-540-88055-4_34}

\bibitem[{{Lodders} {et~al.}(2009{\natexlab{b}}){Lodders}, {Palme}, \&
  {Gail}}]{lodders2009}
---. 2009{\natexlab{b}}, Landolt B\&ouml;rnstein, 4B, 712,
  \dodoi{10.1007/978-3-540-88055-4_34}

\bibitem[{MacDonald \& Madhusudhan(2017)}]{macdonald_hd_2017}
MacDonald, R.~J., \& Madhusudhan, N. 2017, Monthly Notices of the Royal
  Astronomical Society, 469, 1979, \dodoi{10.1093/mnras/stx804}

\bibitem[{Madhusudhan(2012)}]{madhusudhan_co_2012}
Madhusudhan, N. 2012, ApJ, 758, 36, \dodoi{10.1088/0004-637X/758/1/36}

\bibitem[{{Madhusudhan}(2019)}]{Madhusudhan2019}
{Madhusudhan}, N. 2019, \araa, 57, 617,
  \dodoi{10.1146/annurev-astro-081817-051846}

\bibitem[{Madhusudhan {et~al.}(2014)Madhusudhan, Amin, \&
  Kennedy}]{madhusudhan_toward_2014}
Madhusudhan, N., Amin, M.~A., \& Kennedy, G.~M. 2014, ApJ, 794, L12,
  \dodoi{10.1088/2041-8205/794/1/L12}

\bibitem[{Magic {et~al.}(2015)Magic, Chiavassa, Collet, \&
  Asplund}]{magic_stagger-grid_2015}
Magic, Z., Chiavassa, A., Collet, R., \& Asplund, M. 2015, A\&A, 573, A90,
  \dodoi{10.1051/0004-6361/201423804}

\bibitem[{{Mai} \& {Line}(2019)}]{Mai&Line19}
{Mai}, C., \& {Line}, M.~R. 2019, \apj, 883, 144,
  \dodoi{10.3847/1538-4357/ab3e6d}

\bibitem[{Mandel \& Agol(2002)}]{mandel_analytic_2002}
Mandel, K., \& Agol, E. 2002, The Astrophysical Journal, 580, L171,
  \dodoi{10.1086/345520}

\bibitem[{{Mansfield} {et~al.}(2021){Mansfield}, {Line}, {Bean}, {Fortney},
  {Parmentier}, {Wiser}, {Kempton}, {Gharib-Nezhad}, {Sing},
  {L{\'o}pez-Morales}, {Baxter}, {D{\'e}sert}, {Swain}, \&
  {Roudier}}]{mansfield21}
{Mansfield}, M., {Line}, M.~R., {Bean}, J.~L., {et~al.} 2021, Nature Astronomy,
  5, 1224, \dodoi{10.1038/s41550-021-01455-4}

\bibitem[{{Marley}(1999)}]{Marley_1999}
{Marley}, M. 1999, \pasp, 111, 1591, \dodoi{10.1086/316463}

\bibitem[{{Marley} {et~al.}(2021){Marley}, {Saumon}, {Visscher}, {Lupu},
  {Freedman}, {Morley}, {Fortney}, {Seay}, {Smith}, {Teal}, \&
  {Wang}}]{Marley_2021_sonora}
{Marley}, M.~S., {Saumon}, D., {Visscher}, C., {et~al.} 2021, \apj, 920, 85,
  \dodoi{10.3847/1538-4357/ac141d}

\bibitem[{McGruder {et~al.}(2022)McGruder, López-Morales, Kirk, Espinoza,
  Rackham, Alam, Allen, Nikolov, Weaver, Ortiz~Ceballos, Osip, Apai, Jordán,
  \& Fortney}]{mcgruder_access_2022}
McGruder, C.~D., López-Morales, M., Kirk, J., {et~al.} 2022, AJ, 164, 134,
  \dodoi{10.3847/1538-3881/ac7f2e}

\bibitem[{{McKay} {et~al.}(1989){McKay}, {Pollack}, \& {Courtin}}]{McKay1989}
{McKay}, C.~P., {Pollack}, J.~B., \& {Courtin}, R. 1989, \icarus, 80, 23,
  \dodoi{10.1016/0019-1035(89)90160-7}

\bibitem[{{Morley} {et~al.}(2012){Morley}, {Fortney}, {Marley}, {Visscher},
  {Saumon}, \& {Leggett}}]{Morley_cloud_2012}
{Morley}, C.~V., {Fortney}, J.~J., {Marley}, M.~S., {et~al.} 2012, \apj, 756,
  172, \dodoi{10.1088/0004-637X/756/2/172}

\bibitem[{{Moses} {et~al.}(2022){Moses}, {Tremblin}, {Venot}, \&
  {Miguel}}]{moses2022}
{Moses}, J.~I., {Tremblin}, P., {Venot}, O., \& {Miguel}, Y. 2022, Experimental
  Astronomy, 53, 279, \dodoi{10.1007/s10686-021-09749-1}

\bibitem[{Moses {et~al.}(2011)Moses, Visscher, Fortney, Showman, Lewis,
  Griffith, Klippenstein, Shabram, Friedson, Marley, \&
  Freedman}]{moses_disequilibrium_2011}
Moses, J.~I., Visscher, C., Fortney, J.~J., {et~al.} 2011, ApJ, 737, 15,
  \dodoi{10.1088/0004-637X/737/1/15}

\bibitem[{{Mukherjee} {et~al.}(2022){Mukherjee}, {Batalha}, {Fortney}, \&
  {Marley}}]{mukherjee_picasso22}
{Mukherjee}, S., {Batalha}, N.~E., {Fortney}, J.~J., \& {Marley}, M.~S. 2022,
  arXiv e-prints, arXiv:2208.07836.
\newblock \doarXiv{2208.07836}

\bibitem[{Nikolov {et~al.}(2018)Nikolov, Sing, Fortney, Goyal, Drummond, Evans,
  Gibson, De~Mooij, Rustamkulov, Wakeford, Smalley, Burgasser, Hellier,
  Helling, Mayne, Madhusudhan, Kataria, Baines, Carter, Ballester, Barstow,
  McCleery, \& Spake}]{nikolov_absolute_2018}
Nikolov, N., Sing, D.~K., Fortney, J.~J., {et~al.} 2018, Nature, 557, 526,
  \dodoi{10.1038/s41586-018-0101-7}

\bibitem[{Nikolov {et~al.}(2022)Nikolov, Sing, Spake, Smalley, Goyal,
  Mikal-Evans, Wakeford, Rustamkulov, Deming, Fortney, Carter, Gibson, \&
  Mayne}]{nikolov_solar--supersolar_2022}
Nikolov, N.~K., Sing, D.~K., Spake, J.~J., {et~al.} 2022, Monthly Notices of
  the Royal Astronomical Society, 515, 3037, \dodoi{10.1093/mnras/stac1530}

\bibitem[{Parmentier {et~al.}(2013)Parmentier, Showman, \&
  Lian}]{parmentier_3d_2013}
Parmentier, V., Showman, A.~P., \& Lian, Y. 2013, A\&A, 558, A91,
  \dodoi{10.1051/0004-6361/201321132}

\bibitem[{P\'erez \& Granger(2007)}]{PER-GRA:2007}
P\'erez, F., \& Granger, B.~E. 2007, Computing in Science and Engineering, 9,
  21, \dodoi{10.1109/MCSE.2007.53}

\bibitem[{Pinhas {et~al.}(2019)Pinhas, Madhusudhan, Gandhi, \&
  MacDonald}]{pinhas_h2o_2019}
Pinhas, A., Madhusudhan, N., Gandhi, S., \& MacDonald, R. 2019, Monthly Notices
  of the Royal Astronomical Society, 482, 1485, \dodoi{10.1093/mnras/sty2544}

\bibitem[{{Piskorz} {et~al.}(2018){Piskorz}, {Buzard}, {Line}, {Knutson},
  {Benneke}, {Crockett}, {Lockwood}, {Blake}, {Barman}, {Bender}, {Deming}, \&
  {Johnson}}]{Piskorz2018}
{Piskorz}, D., {Buzard}, C., {Line}, M.~R., {et~al.} 2018, \aj, 156, 133,
  \dodoi{10.3847/1538-3881/aad781}

\bibitem[{{Polyansky} {et~al.}(2018){Polyansky}, {Kyuberis}, {Zobov},
  {Tennyson}, {Yurchenko}, \& {Lodi}}]{Polyansky+18}
{Polyansky}, O.~L., {Kyuberis}, A.~A., {Zobov}, N.~F., {et~al.} 2018, \mnras,
  480, 2597, \dodoi{10.1093/mnras/sty1877}

\bibitem[{Pontoppidan {et~al.}(2022)Pontoppidan, Barrientes, Blome, Braun,
  Brown, Carruthers, Coe, DePasquale, Espinoza, Marin, Gordon, Henry, Hustak,
  James, Jenkins, Koekemoer, LaMassa, Law, Lockwood, Moro-Martin, Mullally,
  Pagan, Player, Proffitt, Pulliam, Ramsay, Ravindranath, Reid, Robberto,
  Sabbi, Ubeda, Balogh, Flanagan, Gardner, Hasan, Meinke, \&
  Nota}]{pontoppidan_jwst_2022}
Pontoppidan, K.~M., Barrientes, J., Blome, C., {et~al.} 2022, ApJL, 936, L14,
  \dodoi{10.3847/2041-8213/ac8a4e}

\bibitem[{Radica {et~al.}(2022)Radica, Albert, Taylor, Lafrenière, Coulombe,
  Darveau-Bernier, Doyon, Cook, Cowan, Espinoza, Johnstone, Kaltenegger,
  Piaulet, Roy, \& Talens}]{radica_applesoss_2022}
Radica, M., Albert, L., Taylor, J., {et~al.} 2022, PASP, 134, 104502,
  \dodoi{10.1088/1538-3873/ac9430}

\bibitem[{{Richard} {et~al.}(2012){Richard}, {Gordon}, {Rothman}, {Abel},
  {Frommhold}, {Gustafsson}, {Hartmann}, {Hermans}, {Lafferty}, {Orton},
  {Smith}, \& {Tran}}]{Richard2012}
{Richard}, C., {Gordon}, I.~E., {Rothman}, L.~S., {et~al.} 2012, \jqsrt, 113,
  1276, \dodoi{10.1016/j.jqsrt.2011.11.004}

\bibitem[{{Ricker} {et~al.}(2014){Ricker}, {Winn}, {Vanderspek}, {Latham},
  {Bakos}, {Bean}, {Berta-Thompson}, {Brown}, {Buchhave}, {Butler}, {Butler},
  {Chaplin}, {Charbonneau}, {Christensen-Dalsgaard}, {Clampin}, {Deming},
  {Doty}, {De Lee}, {Dressing}, {Dunham}, {Endl}, {Fressin}, {Ge}, {Henning},
  {Holman}, {Howard}, {Ida}, {Jenkins}, {Jernigan}, {Johnson}, {Kaltenegger},
  {Kawai}, {Kjeldsen}, {Laughlin}, {Levine}, {Lin}, {Lissauer}, {MacQueen},
  {Marcy}, {McCullough}, {Morton}, {Narita}, {Paegert}, {Palle}, {Pepe},
  {Pepper}, {Quirrenbach}, {Rinehart}, {Sasselov}, {Sato}, {Seager},
  {Sozzetti}, {Stassun}, {Sullivan}, {Szentgyorgyi}, {Torres}, {Udry}, \&
  {Villasenor}}]{2014RickerTess}
{Ricker}, G.~R., {Winn}, J.~N., {Vanderspek}, R., {et~al.} 2014, in Society of
  Photo-Optical Instrumentation Engineers (SPIE) Conference Series, Vol. 9143,
  Space Telescopes and Instrumentation 2014: Optical, Infrared, and Millimeter
  Wave, ed. J.~{Oschmann}, Jacobus~M., M.~{Clampin}, G.~G. {Fazio}, \& H.~A.
  {MacEwen}, 914320, \dodoi{10.1117/12.2063489}

\bibitem[{{Rigby} {et~al.}(2022){Rigby}, {Perrin}, {McElwain}, {Kimble},
  {Friedman}, {Lallo}, {Doyon}, {Feinberg}, {Ferruit}, {Glasse}, {Rieke},
  {Rieke}, {Wright}, {Willott}, {Colon}, {Milam}, {Neff}, {Stark}, {Valenti},
  {Abell}, {Abney}, {Abul-Huda}, {Acton}, {Adams}, {Adler}, {Aguilar}, {Ahmed},
  {Albert}, {Alberts}, {Aldridge}, {Allen}, {Altenburg}, {Alvarez Marquez},
  {Alves de Oliveira}, {Andersen}, {Anderson}, {Anderson}, {Argyriou},
  {Armstrong}, {Arribas}, {Artigau}, {Arvai}, {Atkinson}, {Bacon}, {Bair},
  {Banks}, {Barrientes}, {Barringer}, {Bartosik}, {Bast}, {Baudoz}, {Beatty},
  {Bechtold}, {Beck}, {Bergeron}, {Bergkoetter}, {Bhatawdekar}, {Birkmann},
  {Blazek}, {Blome}, {Boccaletti}, {Boeker}, {Boia}, {Bonaventura}, {Bond},
  {Bosley}, {Boucarut}, {Bourque}, {Bouwman}, {Bower}, {Bowers}, {Boyer},
  {Bradley}, {Brady}, {Braun}, {Breda}, {Bresnahan}, {Bright}, {Britt},
  {Bromenschenkel}, {Brooks}, {Brooks}, {Brown}, {Brown}, {Brown}, {Bunker},
  {Burger}, {Bushouse}, {Cale}, {Cameron}, {Cameron}, {Canipe}, {Caplinger},
  {Caputo}, {Cara}, {Carey}, {Carniani}, {Carrasquilla}, {Carruthers}, {Case},
  {Catherine}, {Chance}, {Chapman}, {Charlot}, {Charlow}, {Chayer}, {Chen},
  {Cherinka}, {Chichester}, {Chilton}, {Chonis}, {Clampin}, {Clark}, {Clark},
  {Coe}, {Coleman}, {Comber}, {Comeau}, {Connolly}, {Cooper}, {Cooper},
  {Coppock}, {Correnti}, {Cossou}, {Coulais}, {Coyle}, {Cracraft}, {Curti},
  {Cuturic}, {Davis}, {Davis}, {Dean}, {DeLisa}, {deMeester}, {Dencheva},
  {Dencheva}, {DePasquale}, {Deschenes}, {Hunor Detre}, {Diaz}, {Dicken},
  {DiFelice}, {Dillman}, {Dixon}, {Doggett}, {Donaldson}, {Douglas}, {DuPrie},
  {Dupuis}, {Durning}, {Easmin}, {Eck}, {Edeani}, {Egami}, {Ehrenwinkler},
  {Eisenhamer}, {Eisenhower}, {Elie}, {Elliott}, {Elliott}, {Ellis},
  {Engesser}, {Espinoza}, {Etienne}, {Etxaluze}, {Falini}, {Feeney}, {Ferry},
  {Filippazzo}, {Fincham}, {Fix}, {Flagey}, {Florian}, {Flynn}, {Fontanella},
  {Ford}, {Forshay}, {Fox}, {Franz}, {Fu}, {Fullerton}, {Galkin}, {Galyer},
  {Garcia Marin}, {Gardner}, {Gardner}, {Garland}, {Garrett}, {Gasman},
  {Gaspar}, {Gaudreau}, {Gauthier}, {Geers}, {Geithner}, {Gennaro}, {Giardino},
  {Girard}, {Giuliano}, {Glassmire}, {Glauser}, {Glazer}, {Godfrey},
  {Golimowski}, {Gollnitz}, {Gong}, {Gonzaga}, {Gordon}, {Gordon},
  {Goudfrooij}, {Greene}, {Greenhouse}, {Grimaldi}, {Groebner}, {Grundy},
  {Guillard}, {Gutman}, {Ha}, {Haderlein}, {Hagedorn}, {Hainline}, {Haley},
  {Hami}, {Hamilton}, {Hammel}, {Hansen}, {Harkins}, {Harr}, {Hart}, {Hart},
  {Hartig}, {Hashimoto}, {Haskins}, {Hathaway}, {Havey}, {Hayden}, {Hecht},
  {Heller-Boyer}, {Henriques}, {Henry}, {Hermann}, {Hernandez}, {Hesman},
  {Hicks}, {Hilbert}, {Hines}, {Hoffman}, {Holfeltz}, {Holler}, {Hoppa},
  {Hott}, {Howard}, {Howard}, {Hunter}, {Hunter}, {Hurst}, {Husemann},
  {Hustak}, {Ilinca Ignat}, {Illingworth}, {Irish}, {Jackson}, {Jahromi},
  {Jakobsen}, {James}, {James}, {Januszewski}, {Jenkins}, {Jirdeh}, {Johnson},
  {Johnson}, {Jones}, {Jones}, {Jones}, {Jones}, {Jordan}, {Jordan}, {Jurczyk},
  {Jurling}, {Kaleida}, {Kalmanson}, {Kammerer}, {Kang}, {Kao}, {Karakla},
  {Kavanagh}, {Kelly}, {Kendrew}, {Kennedy}, {Kenny}, {Keski-kuha}, {Keyes},
  {Kidwell}, {Kinzel}, {Kirk}, {Kirkpatrick}, {Kirshenblat}, {Klaassen},
  {Knapp}, {Knight}, {Knollenberg}, {Koehler}, {Koekemoer}, {Kovacs}, {Kulp},
  {Kumari}, {Kyprianou}, {La Massa}, {Labador}, {Labiano Ortega}, {Lagage},
  {Lajoie}, {Lallo}, {Lam}, {Lamb}, {Lambros}, {Lampenfield}, {Langston},
  {Larson}, {Law}, {Lawrence}, {Lee}, {Leisenring}, {Lepo}, {Leveille},
  {Levenson}, {Levine}, {Levy}, {Lewis}, {Lewis}, {Libralato}, {Lightsey},
  {Link}, {Liu}, {Lo}, {Lockwood}, {Logue}, {Long}, {Long}, {Loomis},
  {Lopez-Caniego}, {Alvarez}, {Love-Pruitt}, {Lucy}, {Luetzgendorf}, {Maghami},
  {Maiolino}, {Major}, {Malla}, {Malumuth}, {Manjavacas}, {Mannfolk},
  {Marrione}, {Marston}, {Martel}, {Maschmann}, {Masci}, {Masciarelli},
  {Maszkiewicz}, {Mather}, {McKenzie}, {McLean}, {McMaster}, {Melbourne},
  {Mel{\'e}ndez}, {Menzel}, {Merz}, {Meyett}, {Meza}, {Miskey}, {Misselt},
  {Moller}, {Morrison}, {Morse}, {Moseley}, {Mosier}, {Mountain}, {Mueckay},
  {Mueller}, {Mullally}, {Murphy}, {Murray}, {Murray}, {Mustelier},
  {Muzerolle}, {Mycroft}, {Myers}, {Myrick}, {Nanavati}, {Nance}, {Nayak},
  {Naylor}, {Nelan}, {Nickson}, {Nielson}, {Nieto-Santisteban}, {Nikolov},
  {Noriega-Crespo}, {O'Shaughnessy}, {O'Sullivan}, {Ochs}, {Ogle}, {Oleszczuk},
  {Olmsted}, {Osborne}, {Ottens}, {Owens}, {Pacifici}, {Pagan}, {Page}, {Park},
  {Parrish}, {Patapis}, {Paul}, {Pauly}, {Pavlovsky}, {Pedder}, {Peek},
  {Pena-Guerrero}, {Pennanen}, {Perez}, {Perna}, {Perriello}, {Phillips},
  {Pietraszkiewicz}, {Pinaud}, {Pirzkal}, {Pitman}, {Piwowar}, {Platais},
  {Player}, {Plesha}, {Pollizi}, {Polster}, {Pontoppidan}, {Porterfield},
  {Proffitt}, {Pueyo}, {Pulliam}, {Quirt}, {Quispe Neira}, {Ramos Alarcon},
  {Ramsay}, {Rapp}, {Rapp}, {Rauscher}, {Ravindranath}, {Rawle}, {Regan},
  {Reichard}, {Reis}, {Ressler}, {Rest}, {Reynolds}, {Rhue}, {Richon},
  {Rickman}, {Ridgaway}, {Ritchie}, {Rix}, {Robberto}, {Robinson}, {Robinson},
  {Robinson}, {Rock}, {Rodriguez}, {Rodriguez Del Pino}, {Roellig}, {Rohrbach},
  {Roman}, {Romelfanger}, {Rose}, {Roteliuk}, {Roth}, {Rothwell}, {Rowlands},
  {Roy}, {Royer}, {Royle}, {Rui}, {Rumler}, {Runnels}, {Russ}, {Rustamkulov},
  {Ryden}, {Ryer}, {Sabata}, {Sabatke}, {Sabbi}, {Samuelson}, {Sapp},
  {Sappington}, {Sargent}, {Sauer}, {Scheithauer}, {Schlawin}, {Schlitz},
  {Schmitz}, {Schneider}, {Schreiber}, {Schulze}, {Schwab}, {Scott}, {Sembach},
  {Shanahan}, {Shaughnessy}, {Shaw}, {Shawger}, {Shay}, {Sheehan}, {Shen},
  {Sherman}, {Shiao}, {Shih}, {Shivaei}, {Sienkiewicz}, {Sing}, {Sirianni},
  {Sivaramakrishnan}, {Skipper}, {Sloan}, {Slocum}, {Slowinski}, {Smith},
  {Smith}, {Smith}, {Smith}, {Snyder}, {Soh}, {Sohn}, {Soto}, {Spencer},
  {Stallcup}, {Stansberry}, {Starr}, {Starr}, {Stewart}, {Stiavelli},
  {Straughn}, {Strickland}, {Stys}, {Summers}, {Sun}, {Sunnquist}, {Swade},
  {Swam}, {Swaters}, {Swoish}, {Taylor}, {Taylor}, {Te Plate}, {Tea}, {Teague},
  {Telfer}, {Temim}, {Thatte}, {Thompson}, {Thompson}, {Thomson}, {Tikkanen},
  {Tippet}, {Todd}, {Toolan}, {Tran}, {Trejo}, {Truong}, {Tsukamoto},
  {Tustain}, {Tyra}, {Ubeda}, {Underwood}, {Uzzo}, {Van Campen}, {Vandal},
  {Vandenbussche}, {Vila}, {Volk}, {Wahlgren}, {Waldman}, {Walker}, {Wander},
  {Warfield}, {Warner}, {Wasiak}, {Watkins}, {Weaver}, {Weilert}, {Weiser},
  {Weiss}, {Weissman}, {Welty}, {West}, {Wheate}, {Wheatley}, {Wheeler},
  {White}, {Whiteaker}, {Whitehouse}, {Whiteleather}, {Whitman}, {Williams},
  {Willmer}, {Willoughby}, {Wilson}, {Wirth}, {Wislowski}, {Wolf}, {Wolfe},
  {Wolff}, {Workman}, {Wright}, {Wu}, {Wu}, {Wymer}, {Yates}, {Yeager},
  {Yeates}, {Yerger}, {Yoon}, {Young}, {Yu}, {Zak}, {Zeidler}, {Zhou},
  {Zielinski}, {Zincke}, \& {Zonak}}]{rigby_commissioning22}
{Rigby}, J., {Perrin}, M., {McElwain}, M., {et~al.} 2022, arXiv e-prints,
  arXiv:2207.05632.
\newblock \doarXiv{2207.05632}

\bibitem[{{Rooney} {et~al.}(2022){Rooney}, {Batalha}, {Gao}, \&
  {Marley}}]{Rooney_2022_virgaupdate}
{Rooney}, C.~M., {Batalha}, N.~E., {Gao}, P., \& {Marley}, M.~S. 2022, \apj,
  925, 33, \dodoi{10.3847/1538-4357/ac307a}

\bibitem[{{Rothman} {et~al.}(2010){Rothman}, {Gordon}, {Barber}, {Dothe},
  {Gamache}, {Goldman}, {Perevalov}, {Tashkun}, \& {Tennyson}}]{Rothman+10}
{Rothman}, L.~S., {Gordon}, I.~E., {Barber}, R.~J., {et~al.} 2010, \jqsrt, 111,
  2139, \dodoi{10.1016/j.jqsrt.2010.05.001}

\bibitem[{Rustamkulov {et~al.}(2023)Rustamkulov, Sing, Mukherjee, May, Kirk,
  Schlawin, Line, Piaulet, Carter, Batalha, Goyal, López-Morales, Lothringer,
  MacDonald, Moran, Stevenson, Wakeford, Espinoza, Bean, Batalha, Benneke,
  Berta-Thompson, Crossfield, Gao, Kreidberg, Powell, Cubillos, Gibson,
  Leconte, Molaverdikhani, Nikolov, Parmentier, Roy, Taylor, Turner, Wheatley,
  Aggarwal, Ahrer, Alam, Alderson, Allen, Banerjee, Barat, Barrado, Barstow,
  Bell, Blecic, Brande, Casewell, Changeat, Chubb, Crouzet, Daylan, Decin,
  Désert, Mikal-Evans, Feinstein, Flagg, Fortney, Harrington, Heng, Hong, Hu,
  Iro, Kataria, Kempton, Krick, Lendl, Lillo-Box, Louca, Lustig-Yaeger,
  Mancini, Mansfield, Mayne, Miguel, Morello, Ohno, Palle, Petit dit de~la
  Roche, Rackham, Radica, Ramos-Rosado, Redfield, Rogers, Shkolnik, Southworth,
  Teske, Tremblin, Tucker, Venot, Waalkes, Welbanks, Zhang, \&
  Zieba}]{rustamkulov_early_2023}
Rustamkulov, Z., Sing, D.~K., Mukherjee, S., {et~al.} 2023, Nature,
  \dodoi{10.1038/s41586-022-05677-y}

\bibitem[{Samra {et~al.}(2023)Samra, Helling, Chubb, Min, Carone, \&
  Schneider}]{samra_clouds_2023}
Samra, D., Helling, C., Chubb, K.~L., {et~al.} 2023, A\&A, 669, A142,
  \dodoi{10.1051/0004-6361/202244939}

\bibitem[{Seager \& Sasselov(2000)}]{seager_theoretical_2000}
Seager, S., \& Sasselov, D.~D. 2000, ApJ, 537, 916, \dodoi{10.1086/309088}

\bibitem[{Sing(2010)}]{sing_stellar_2010}
Sing, D.~K. 2010, A\&A, 510, A21, \dodoi{10.1051/0004-6361/200913675}

\bibitem[{Sing {et~al.}(2016)Sing, Fortney, Nikolov, Wakeford, Kataria, Evans,
  Aigrain, Ballester, Burrows, Deming, Désert, Gibson, Henry, Huitson,
  Knutson, Etangs, Pont, Showman, Vidal-Madjar, Williamson, \&
  Wilson}]{sing_continuum_2016}
Sing, D.~K., Fortney, J.~J., Nikolov, N., {et~al.} 2016, Nature, 529, 59,
  \dodoi{10.1038/nature16068}

\bibitem[{Spake {et~al.}(2020)Spake, Sing, Wakeford, Nikolov, Mikal-Evans,
  Deming, Barstow, Anderson, Carter, Gillon, Goyal, Hebrard, Hellier, Kataria,
  Lam, Triaud, \& Wheatley}]{spake_abundance_2020}
Spake, J.~J., Sing, D.~K., Wakeford, H.~R., {et~al.} 2020, Monthly Notices of
  the Royal Astronomical Society, 500, 4042, \dodoi{10.1093/mnras/staa3116}

\bibitem[{Speagle(2020)}]{speagle_dynesty_2020}
Speagle, J.~S. 2020, Monthly Notices of the Royal Astronomical Society, 493,
  3132, \dodoi{10.1093/mnras/staa278}

\bibitem[{Stevenson {et~al.}(2016)Stevenson, Lewis, Bean, Beichman, Fraine,
  Kilpatrick, Krick, Lothringer, Mandell, Valenti, Agol, Angerhausen, Barstow,
  Birkmann, Burrows, Charbonneau, Cowan, Crouzet, Cubillos, Curry, Dalba,
  de~Wit, Deming, Désert, Doyon, Dragomir, Ehrenreich, Fortney,
  García~Muñoz, Gibson, Gizis, Greene, Harrington, Heng, Kataria, Kempton,
  Knutson, Kreidberg, Lafrenière, Lagage, Line, Lopez-Morales, Madhusudhan,
  Morley, Rocchetto, Schlawin, Shkolnik, Shporer, Sing, Todorov, Tucker, \&
  Wakeford}]{stevenson_transiting_2016}
Stevenson, K.~B., Lewis, N.~K., Bean, J.~L., {et~al.} 2016, PASP, 128, 094401,
  \dodoi{10.1088/1538-3873/128/967/094401}

\bibitem[{Taylor {et~al.}(2023)Taylor, Radica, Welbanks, MacDonald, Blecic,
  Zamyatina, Roth, Bean, Parmentier, Coulombe, Feinstein, Espinoza, Benneke,
  Lafrenière, Doyon, \& Ahrer}]{Taylor_w96_2023}
Taylor, J., Radica, M., Welbanks, L., {et~al.} 2023, Monthly Notices of the
  Royal Astronomical Society, \dodoi{10.1093/mnras/stad1547}

\bibitem[{Thorngren {et~al.}(2016)Thorngren, Fortney, Murray-Clay, \&
  Lopez}]{thorngren_massmetallicity_2016}
Thorngren, D.~P., Fortney, J.~J., Murray-Clay, R.~A., \& Lopez, E.~D. 2016,
  ApJ, 831, 64, \dodoi{10.3847/0004-637X/831/1/64}

\bibitem[{{Toon} {et~al.}(1989){Toon}, {McKay}, {Ackerman}, \&
  {Santhanam}}]{Toon1989}
{Toon}, O.~B., {McKay}, C.~P., {Ackerman}, T.~P., \& {Santhanam}, K. 1989,
  \jgr, 94, 16287, \dodoi{10.1029/JD094iD13p16287}

\bibitem[{{Tremblin} {et~al.}(2015){Tremblin}, {Amundsen}, {Mourier},
  {Baraffe}, {Chabrier}, {Drummond}, {Homeier}, \& {Venot}}]{Tremblin2015}
{Tremblin}, P., {Amundsen}, D.~S., {Mourier}, P., {et~al.} 2015, \apjl, 804,
  L17, \dodoi{10.1088/2041-8205/804/1/L17}

\bibitem[{{Tsai} {et~al.}(2023){Tsai}, {Lee}, {Powell}, {Gao}, {Zhang},
  {Moses}, {H{\'e}brard}, {Venot}, {Parmentier}, {Jordan}, {Hu}, {Alam},
  {Alderson}, {Batalha}, {Bean}, {Benneke}, {Bierson}, {Brady}, {Carone},
  {Carter}, {Chubb}, {Inglis}, {Leconte}, {Line}, {L{\'o}pez-Morales},
  {Miguel}, {Molaverdikhani}, {Rustamkulov}, {Sing}, {Stevenson}, {Wakeford},
  {Yang}, {Aggarwal}, {Baeyens}, {Barat}, {de Val-Borro}, {Daylan}, {Fortney},
  {France}, {Goyal}, {Grant}, {Kirk}, {Kreidberg}, {Louca}, {Moran},
  {Mukherjee}, {Nasedkin}, {Ohno}, {Rackham}, {Redfield}, {Taylor}, {Tremblin},
  {Visscher}, {Wallack}, {Welbanks}, {Youngblood}, {Ahrer}, {Batalha}, {Behr},
  {Berta-Thompson}, {Blecic}, {Casewell}, {Crossfield}, {Crouzet}, {Cubillos},
  {Decin}, {D{\'e}sert}, {Feinstein}, {Gibson}, {Harrington}, {Heng},
  {Henning}, {Kempton}, {Krick}, {Lagage}, {Lendl}, {Lothringer}, {Mansfield},
  {Mayne}, {Mikal-Evans}, {Palle}, {Schlawin}, {Shorttle}, {Wheatley}, \&
  {Yurchenko}}]{tsai_direct_2022}
{Tsai}, S.-M., {Lee}, E. K.~H., {Powell}, D., {et~al.} 2023, \nat, 617, 483,
  \dodoi{10.1038/s41586-023-05902-2}

\bibitem[{Turrini {et~al.}(2021)Turrini, Schisano, Fonte, Molinari, Politi,
  Fedele, Panic, Kama, Changeat, \& Tinetti}]{turrini_tracing_2021}
Turrini, D., Schisano, E., Fonte, S., {et~al.} 2021, ApJ, 909, 40,
  \dodoi{10.3847/1538-4357/abd6e5}

\bibitem[{Virtanen {et~al.}(2020)Virtanen, Gommers, Oliphant, Haberland, Reddy,
  Cournapeau, Burovski, Peterson, Weckesser, Bright, {van der Walt}, Brett,
  Wilson, Millman, Mayorov, Nelson, Jones, Kern, Larson, Carey, Polat, Feng,
  Moore, {VanderPlas}, Laxalde, Perktold, Cimrman, Henriksen, Quintero, Harris,
  Archibald, Ribeiro, Pedregosa, {van Mulbregt}, \& {SciPy 1.0
  Contributors}}]{2020SciPy-NMeth}
Virtanen, P., Gommers, R., Oliphant, T.~E., {et~al.} 2020, Nature Methods, 17,
  261, \dodoi{10.1038/s41592-019-0686-2}

\bibitem[{{Visscher} {et~al.}(2010){Visscher}, {Lodders}, \&
  {Fegley}}]{Visscher2010}
{Visscher}, C., {Lodders}, K., \& {Fegley}, Bruce, J. 2010, \apj, 716, 1060,
  \dodoi{10.1088/0004-637X/716/2/1060}

\bibitem[{Wakeford {et~al.}(2017)Wakeford, Visscher, Lewis, Kataria, Marley,
  Fortney, \& Mandell}]{wakeford_high-temperature_2017}
Wakeford, H.~R., Visscher, C., Lewis, N.~K., {et~al.} 2017, Mon. Not. R.
  Astron. Soc., 464, 4247, \dodoi{10.1093/mnras/stw2639}

\bibitem[{{Welbanks} \& {Madhusudhan}(2019)}]{Welbanks2019}
{Welbanks}, L., \& {Madhusudhan}, N. 2019, \aj, 157, 206,
  \dodoi{10.3847/1538-3881/ab14de}

\bibitem[{{Welbanks} \& {Madhusudhan}(2021)}]{Welbanks2021}
---. 2021, \apj, 913, 114, \dodoi{10.3847/1538-4357/abee94}

\bibitem[{{Welbanks} \& {Madhusudhan}(2022)}]{Welbanks2022}
---. 2022, \apj, 933, 79, \dodoi{10.3847/1538-4357/ac6df1}

\bibitem[{Welbanks {et~al.}(2019)Welbanks, Madhusudhan, Allard, Hubeny,
  Spiegelman, \& Leininger}]{welbanks_massmetallicity_2019}
Welbanks, L., Madhusudhan, N., Allard, N.~F., {et~al.} 2019, ApJ, 887, L20,
  \dodoi{10.3847/2041-8213/ab5a89}

\bibitem[{Yip {et~al.}(2021)Yip, Changeat, Edwards, Morvan, Chubb, Tsiaras,
  Waldmann, \& Tinetti}]{yip_compatibility_2021}
Yip, K.~H., Changeat, Q., Edwards, B., {et~al.} 2021, AJ, 161, 4,
  \dodoi{10.3847/1538-3881/abc179}

\bibitem[{Zhou {et~al.}(2017)Zhou, Apai, Lew, \&
  Schneider}]{zhou_physical_2017}
Zhou, Y., Apai, D., Lew, B. W.~P., \& Schneider, G. 2017, AJ, 153, 243,
  \dodoi{10.3847/1538-3881/aa6481}

\bibitem[{Öberg {et~al.}(2011)Öberg, Murray-Clay, \&
  Bergin}]{oberg_effects_2011}
Öberg, K.~I., Murray-Clay, R., \& Bergin, E.~A. 2011, ApJ, 743, L16,
  \dodoi{10.1088/2041-8205/743/1/L16}

\end{thebibliography}
\bibliographystyle{aasjournal}

\appendix
\section{Details of Additional Reductions}
\label{sec: Additional Reductions}

Here we provide details of the three independent reductions carried out on the WASP-96\,b SOSS TSO using the \texttt{nirHiss} (Section~\ref{sec: nirHiss}), \texttt{transitspectroscopy} (Section~\ref{sec: transitspectroscopy}), and \texttt{NAMELESS} (Section~\ref{sec: NAMELESS}) pipelines. Although each pipeline was already described in \citet{feinstein_early_2023}, below we provide a brief outline of each, especially noting any particular steps which differ from what was presented in that work. The final transmission spectrum for each case, along with that from the reference \texttt{supreme-SPOON} reduction described in the main text, are shown in Figure~\ref{fig:Compare Spectra}.

\subsection{\texttt{nirHiss}}
\label{sec: nirHiss}

As described in \citet{feinstein_early_2023}, for the \texttt{nirHiss} reduction, we first process the TSOs through Stages 1 and 2 of the \texttt{Eureka} pipeline \citep{bell_eureka_2022}. From these outputs, \texttt{nirHiss} then follows three steps to remove additional background noise. First, we calculate the average scaling of the STScI JDox background model to a small region of the detector ($x\in[190,250]$, $y\in[200,500]$), and subtract this scaled background model from all integrations. For these data, we find the average scaling factor to be 0.448. Secondly, we use the F277W exposure, taken after the main TSO, to mitigate the effects of 0\textsuperscript{th} order contaminants which are present in the data. The F277W exposure consists of 11 integrations, and 14 groups per integration for a total exposure time of 846\,s. We take the average in time of the F277W exposure and mask the trace. The background was then modelled as in \citet{feinstein_early_2023}, and cosmic rays and bad pixels were identified as to not induce additional noise into the data. We then scale two 0\textsuperscript{th} order contaminants to the TSO observations. These contaminants were located at $x_1 \in [700, 800]$, $y_1 \in [110, 160]$ and  $x_2 \in [1850, 1950]$, $y_2 \in [220, 250]$. The scaling values from each region were averaged and applied to all integrations; we find an average scaling of 2.04. Lastly, pixels with non-zero data quality flags are interpolated using the same method as \citet{feinstein_early_2023}. Unlike \citet{feinstein_early_2023}, after identification of the precise locations of all three diffraction orders, a simple box aperture extraction, as opposed to an optimal extraction routine, is performed on the first two orders using a width of 24 pixels.

A white light curve is constructed for both orders by summing the flux across all wavelengths (only wavelengths $<$0.85\,µm are considered for order 2). The white light curves are then fit following the same procedure described in the main text for the \texttt{supreme-SPOON} reduction; the best fitting parameters from the order 1 white light curve are listed in Table~\ref{tab: WLC Parameters}. The spectrophotometric light curves are then fit at the pixel level, again following the same procedure described in the main text. 

\subsection{\texttt{transitspectroscopy}}
\label{sec: transitspectroscopy}

The \texttt{transitspectroscopy} reduction follows the same steps as those in \citet{feinstein_early_2023}. We start with the \texttt{\_rateints.fits} files produced by the official STScI pipeline, and use the STScI background model to subtract the zodiacal background from each SOSS integration. The background scaling for these data was found to be 0.466. 1/$f$ noise is then corrected following the procedure outlined in \citet{feinstein_early_2023}, and the stellar spectra for the first two orders are extracted using the \texttt{transitspectroscopy.spectroscopy.getSimpleSpectrum} routine and a box aperture of 30 pixels. 

The white light curves for each order are fit with \texttt{juliet} using the same prior setup as described in \citet{feinstein_early_2023}, except that the period is fixed to the 3.4252602\,d from \citet{nikolov_solar--supersolar_2022}. The best fitting values from the white light curve fit are shown in Table~\ref{tab: WLC Parameters}. For the spectrophotometric light curve fits, the orbital parameters are fixed to the \citet{nikolov_absolute_2018} values and coefficients for the square-root limb-darkening law are calculated following the method described in \citet{feinstein_early_2023}. The spectrophotometric fits are carried out at the pixel level. 

\subsection{\texttt{NAMELESS}}
\label{sec: NAMELESS}

All steps of the \texttt{NAMELESS} are followed in an identical manner to those presented in \citet{feinstein_early_2023} except for the 1/$f$ noise correction, for which we use a new method developed in \citet{coulombe_broadband_2023}. This method is essentially similar in spirit to that described in \ref{sec: Stage 1}, except it is applied at the integration level, and that instead of scaling the median image by an estimate of the white light curve to create the difference images, we allow each column to scale independently and simultaneously calculate these scaling factors with the 1/$f$ noise. For a more in-depth description of the algorithm, see \citet{coulombe_broadband_2023}. Stellar spectra are extracted from the corrected frames using the \texttt{transitspectroscopy.spectroscopy.getSimpleSpectrum} routine with a box width of 30 pixels.

We first fit for the white light curves of both orders 1 and 2 separately using the ExoTEP framework \citep{benneke_water_2019}. We fit for the mid-transit time $t_0$, the planet-to-star radius ratio $R_p/R_*$, impact parameter $b$, semi-major axis $a/R_*$, and quadratic limb-darkening coefficients ($u_1$, $u_2$; \citealp{mandel_analytic_2002, kreidberg_batman_2015}). We also fit for the scatter $\sigma$, as well as a linear systematics model with an offset $c$ and slope $v$. Uniform priors are considered for all parameters. For the spectrophotometric light curves, we follow the same process, but fix the impact parameter and semi-major axis to the \citet{nikolov_absolute_2018} values, and the time of mid-transit to its white light curve best fit value. We fit 610 bins for order 1 and 161 for order 2.

\section{Comparison of Group vs Integration Level 1/f Corrections}
\label{sec: Group vs Integration}

In order to assess any potential biases introduced into a transmission (or emission) spectrum resulting from performing the 1/$f$ noise correction at the group or integration level (i.e., before or after the non-linearity correction), we simulated TSO of WASP-96\,b, analogous to those presented here, with the IDTSOSS simulator \citep{radica_applesoss_2022, Albert2023SOSS}. The simulated TSO consisted of the same number of groups as integrations as the real TSO, and was seeded with a cloud-free, 10$\times$ solar metallicity, C/O=0.25 atmosphere model generated with the SCARLET framework \citep{benneke_strict_2015} under the assumption of chemical equilibrium. We then processed the simulated TSO through the \texttt{supreme-SPOON} pipeline in three different ways. 

\begin{itemize}
    \item Case 1: Group-level 1/$f$ correction as described in the text. 
    \item Case 2: Correct 1/$f$ at the group level. Subtract the background beforehand, but do not add it back after the \texttt{OneOverFStep}.
    \item Case 3: Subtract the background and correct 1/$f$ noise at the integration level. 
\end{itemize}

These three cases allowed us to test the interplay between the background correction, 1/$f$ correction and the non-linearity. After extracting the stellar spectra for each case, we fit the spectroscopic light curves at the pixel level, fixing the orbital and limb-darkening parameters to the same values as were input to the simulation. The resulting transmission spectra, binned to a resolution of $R=50$ are shown in Figure~\ref{fig:Groups vs Ints}. All three cases result in excellent agreement with the input spectrum ($\chi^2_\nu$ = 0.93, 0.99, and 1.31 for Cases 1, 2, and 3 respectively), with no systematic biases resulting from either treating the background before, or the 1/$f$ noise after the non-linearity correction. The integration level correction though does result in less precise transit depths (mean error bar of 144\,ppm vs 193\,ppm for Case 1), as well as a higher RMS residual scatter than the other two cases (139, 147, and 207\,ppm for Cases 1, 2, and 3 respectively). 

The importance of the non-linearity correction scales with the brightness of the target. Since these WASP-96\,b TSOs remain well below the 35000 counts threshold, it is possible that the excellent agreement of all three cases may stem from the relative unimportance of the non-linearity correction. We therefore simulated a second WASP-96\,b TSO, but increased the brightness of the host star by 0.6\,mag. This TSO has peak counts $\sim$25000, so non-linearity effects will be more important. We processed this simulation following the three methodologies described above, and once again found little difference between the three cases. Case 1 again yielded the lowest RMS scatter, most precise transit depths, and best $\chi^2_\nu$, followed by Case 2 and then Case 3, with values for all metrics similar to those calculated for the normal brightness case. We therefore conclude that, even for significantly brighter targets, no biases result from a non-optimal treatment of the background-1/$f$-non-linearity coupling. This is likely due to the fact that, although non-linearity effects become more prominent for brighter targets, the relative importance of the background and 1/$f$-noise decrease correspondingly, and in the end the two effects effectively cancel out. We note here as well that the above discussion assumes that the SOSS non-linearity effects are perfectly characterized, which is not entirely true in reality.

\section{Additional Figures and Tables}

\begin{figure*}
	\centering
	\includegraphics[width=\textwidth]{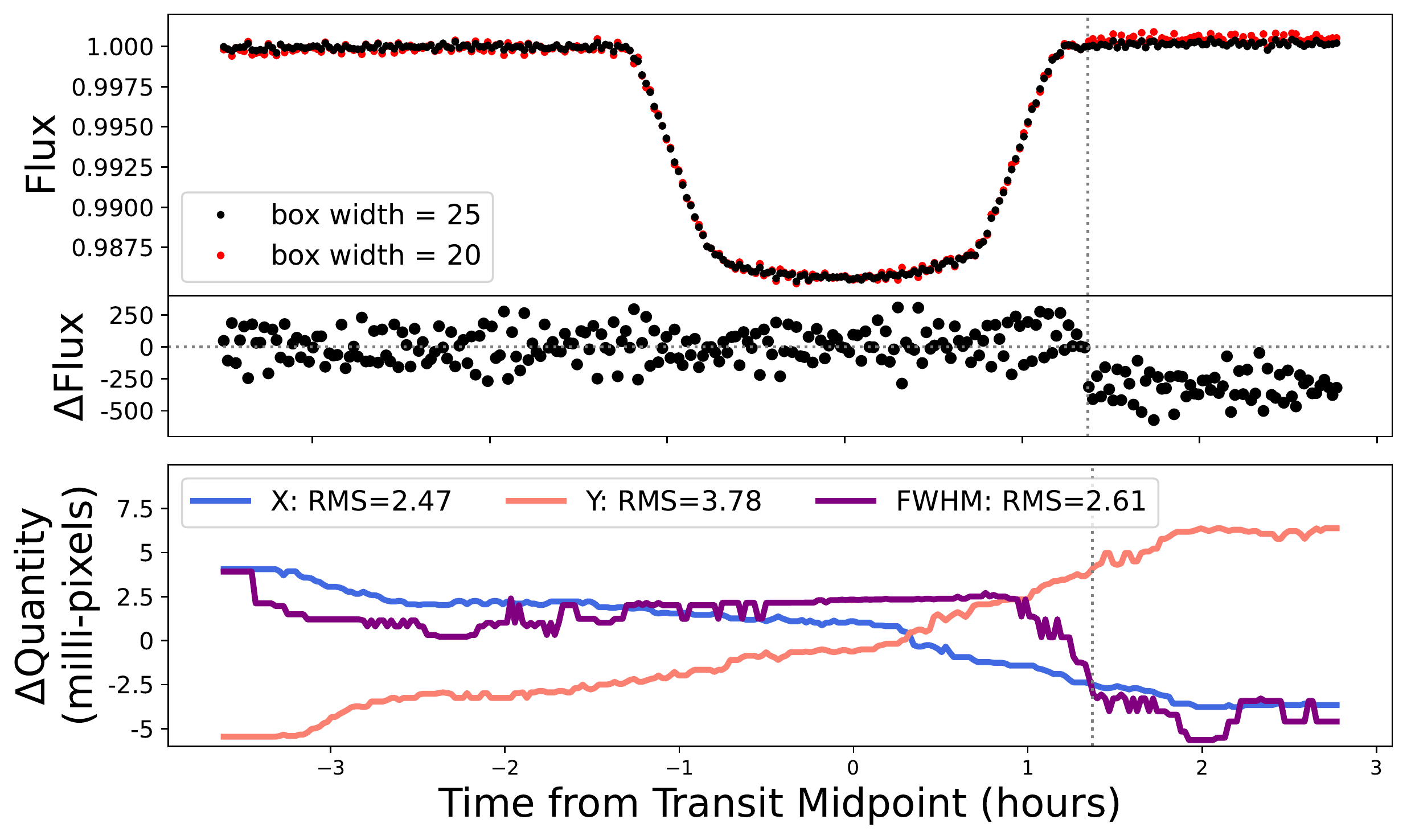}
    \caption{Detector level trends in the WASP-96\,b SOSS TSO. 
    \emph{Top}: Order 1 white light curve extracted with a box aperture of 25 pixels (black) and 20 pixels (red). The light curves are identical until the tilt event $\sim$1.4\,hr after the transit midpoint (grey vertical line). 
    \emph{Middle}: Difference in white light flux between the 25 and 20 box aperture extractions.
    \emph{Bottom}: Temporal trends in the X-position (blue), Y-position (red) and FWHM (purple) of the SOSS trace relative to the median stack through the TSO. The trace position is incredibly stable with RMS shifts in X and Y positions of $<$5\,milli-pixels. The FWHM is also generally stable, except during the tilt event where there is an abrupt decrease of $\sim$6\,milli-pixels. 
    \label{fig:SOSS Stability}}
\end{figure*}

\begin{figure}
	\centering
	\includegraphics[width=\columnwidth]{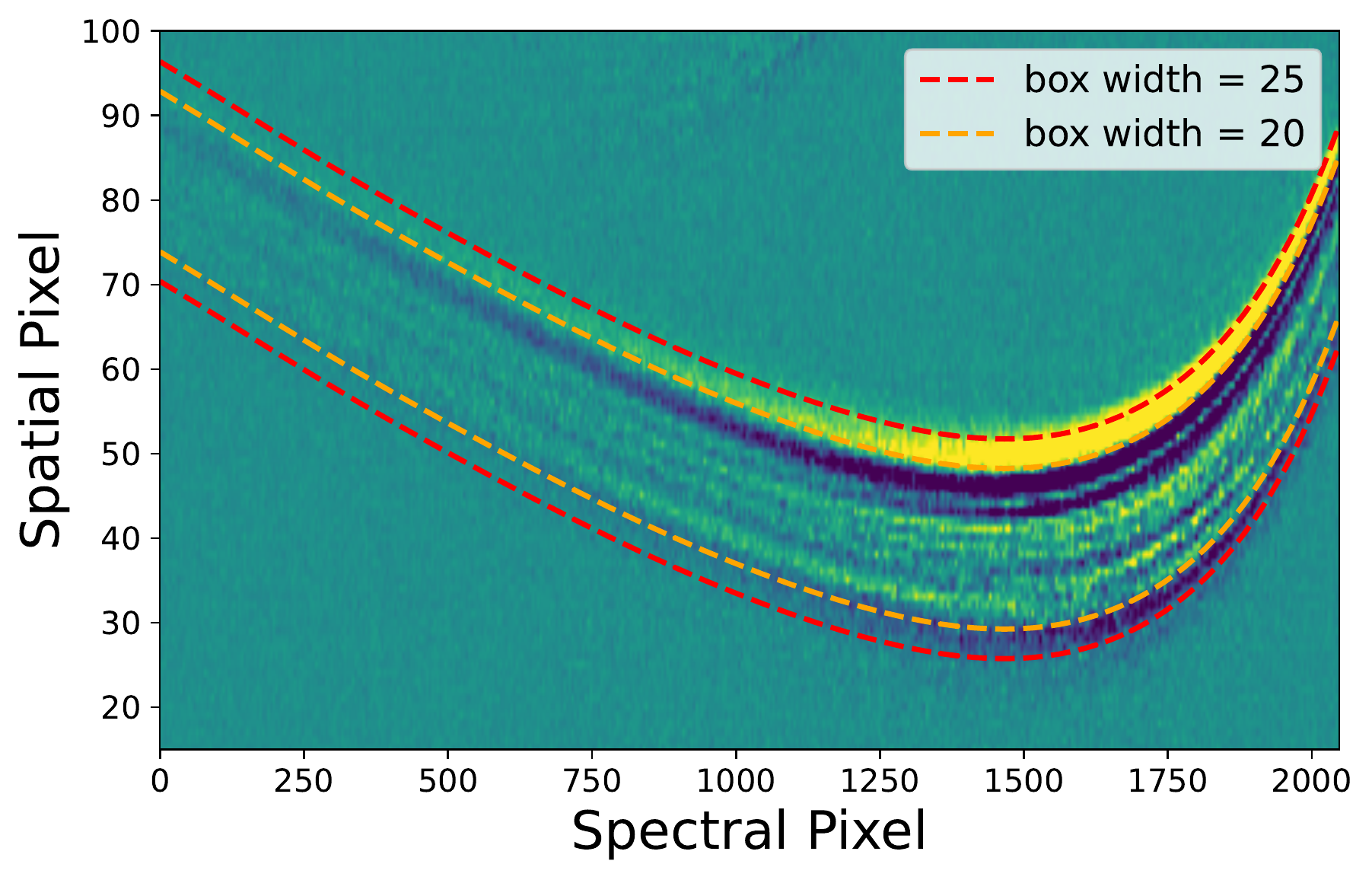}
    \caption{Visualization of the tilt event in 2D. Shown is the difference between a median stack of the first 50, and median stack of the last 50 integrations of the TSO, zoomed-in on the order 1 trace. The morphology change during the tilt event can be clearly seen at the upper edge of the trace. The width of a 25, as well as 20 pixel extraction box are shown in red and orange, respectively. The morphological change is entirely contained within the 25 pixel box aperture, explaining why we do not see any evidence for the tilt event in our analysis. However, when using a 20 pixel-wide box, some additional flux falls into the aperture after the tilt event, resulting in a discontinuity in the light curve (e.g., Figure~\ref{fig:SOSS Stability}).
    \label{fig:2D tilt}}
\end{figure}

\begin{figure*}
	\centering
	\includegraphics[width=\textwidth]{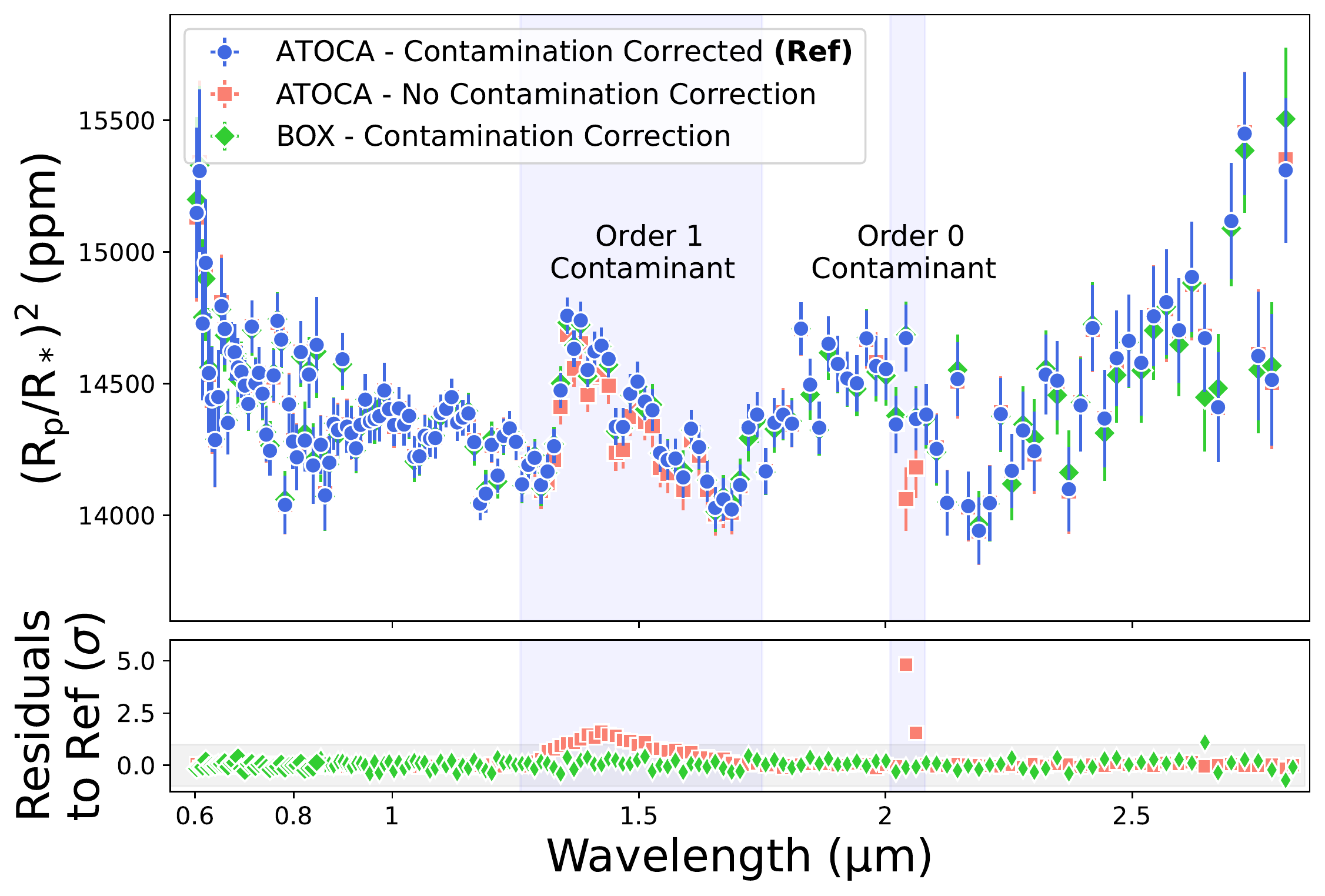}
    \caption{Comparison of WASP-96\,b transmission spectra obtained with different methodologies. \emph{Top}: The reference transmission spectrum extracted with \texttt{ATOCA} and corrected for dilution from background sources (blue) compared to an \texttt{ATOCA} spectrum without dilution correction (red), and a box-extracted spectrum with dilution correction (green). All spectra have been binned to R=150 here for visual clarity. The wavelength regimes affected by each background contaminant are denoted with faded blue boxes. 
    \emph{Bottom}: Residuals between each spectrum shown above and the reference spectrum, normalized by the reference spectrum error bars. The $\pm$1$\sigma$ range is shaded in grey. 
    \label{fig:Compare Spectra Contamination}}
\end{figure*}

\begin{figure}
	\centering
	\includegraphics[width=\columnwidth]{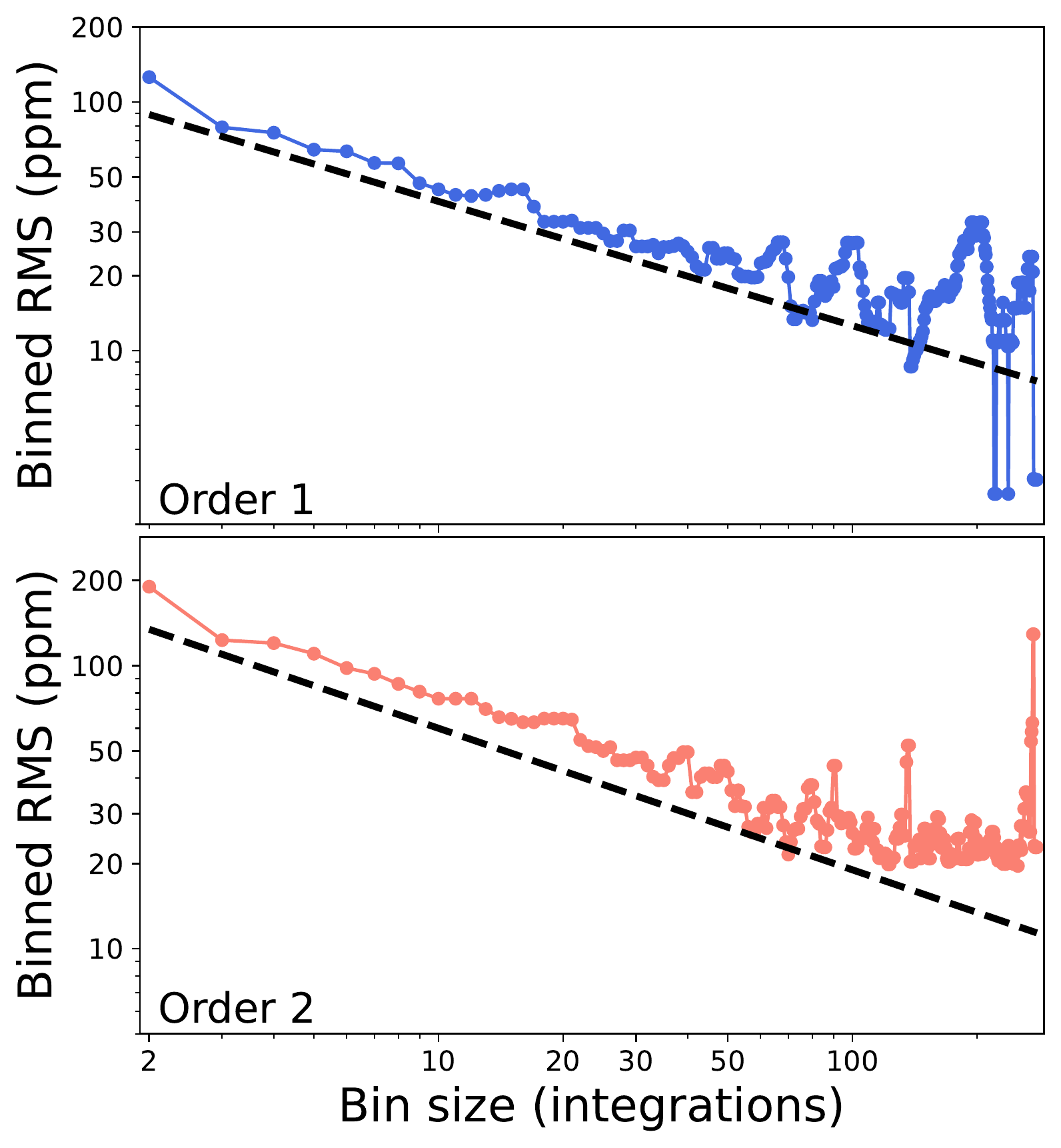}
    \caption{Allan variance plots for order 1 (\emph{top}) and order 2 (\emph{bottom}). The coloured lines in each panel are the white light curve residuals binned to different bin widths. The black dashed lines represent the trends for pure photon noise. In general, the binned residuals trace well the pure photon noise trend.
    \label{fig:Allan Variance}}
\end{figure}

\begin{figure*}
	\centering
	\includegraphics[width=\textwidth]{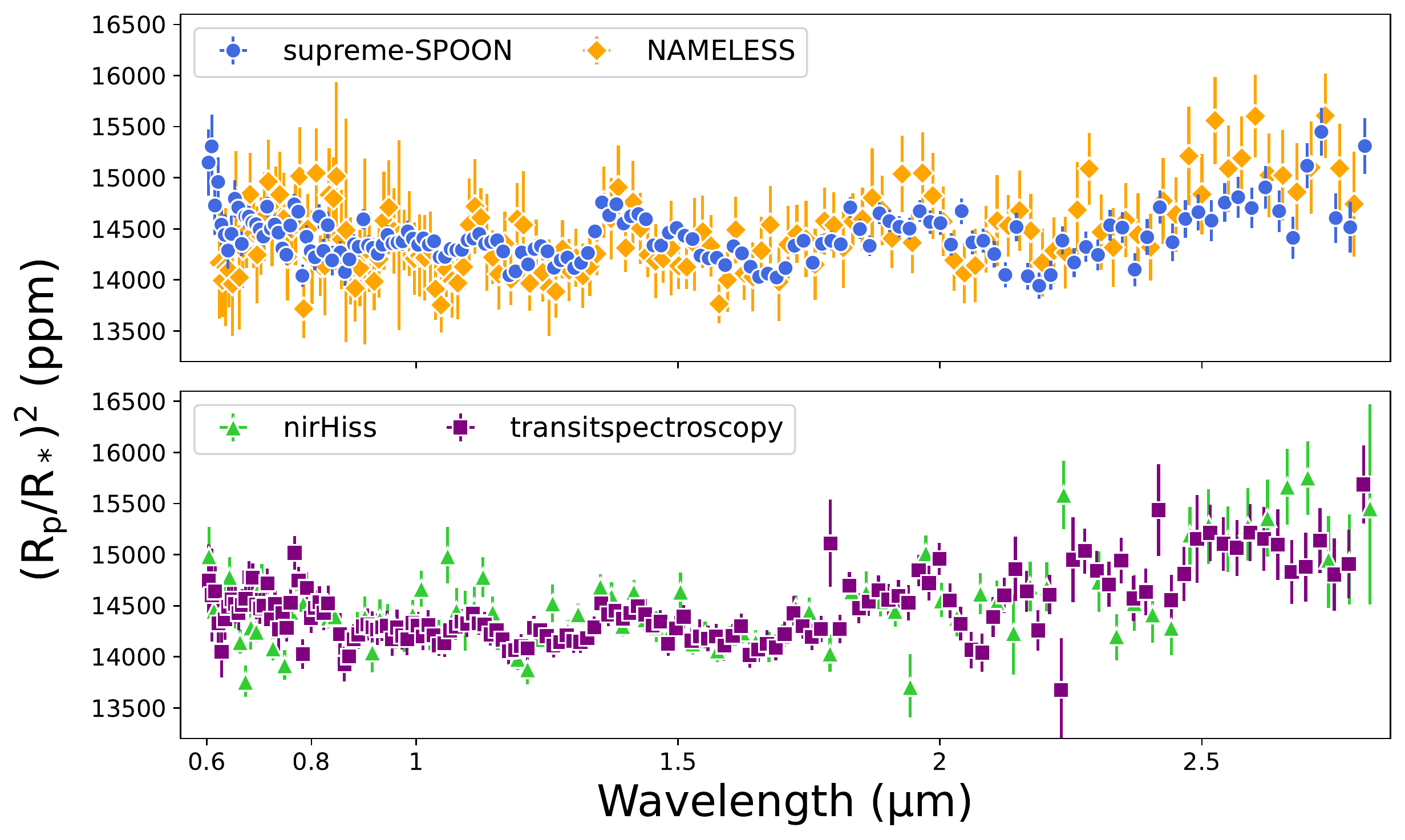}
    \caption{Comparison of transmission spectra for WASP-96\,b obtained through four different pipelines: \texttt{supreme-SPOON} (blue), \texttt{nirHiss} (green), \texttt{transitspectroscopy} (purple), and \texttt{NAMELESS} (orange). All transmission spectra here are binned to R=100. The four independent spectra are in good agreement, showing consistent transit depths and features across the full wavelength range of SOSS. Note that only the \texttt{supreme-SPOON} reduction is completely corrected for contamination from the background order 1 and order 0 contaminants, and thus shows a slightly larger 1.4\,µm water feature.
    \label{fig:Compare Spectra}}
\end{figure*}

\begin{figure*}
	\centering
	\includegraphics[width=\textwidth]{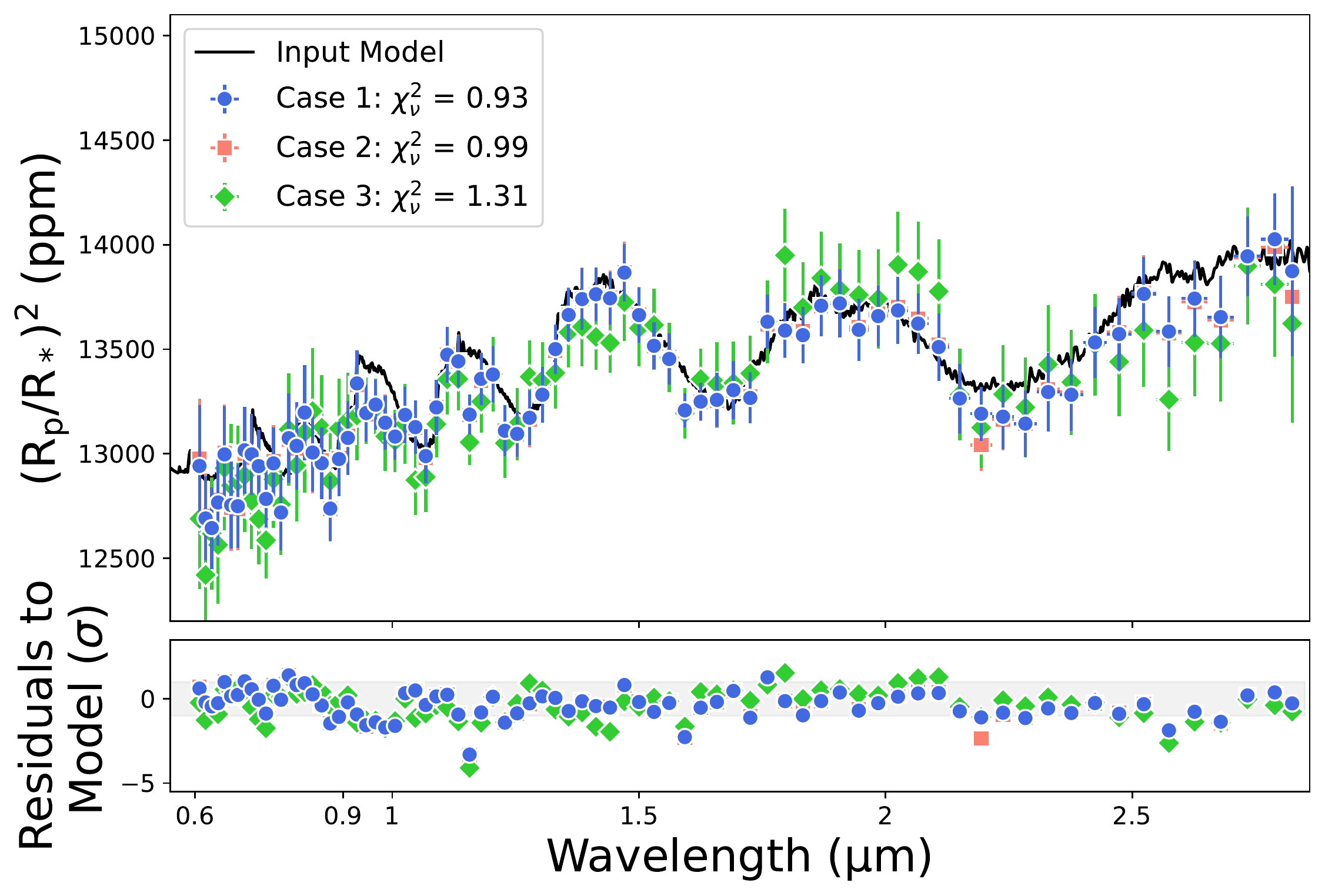}
    \caption{\emph{Top}: Transmission spectra resulting from three different reductions of a simulated WASP-96\,b SOSS TSO to test different 1/$f$ noise correction methodologies. Case 1 represents the reduction described in the main text body, which is a group-level 1/$f$ correction where the background is re-added after the 1/$f$ correction is performed. Case 2 is the reduction where the background was not re-added after the 1/$f$ correction, and Case 3 represents an integration-level background and 1/$f$ noise correction. The input atmosphere model is shown in black. Case 1 results in the best $\chi^2_\nu$, as well as the lowest residual RMS and most precise transit depths. However, no systematic biases result from the other two cases. 
    \emph{Bottom}: Residuals divided by the error bar on each point between each transmission spectrum and the input atmosphere model. 
    \label{fig:Groups vs Ints}}
\end{figure*}

\bsp	
\label{lastpage}
\end{document}